\documentclass[a4paper,11pt]{article}
\usepackage{jheppub} 
\usepackage{amsmath, amssymb, amsthm, xcolor, physics, graphicx, caption, cancel}
\usepackage{float}

\newcommand{\Eval}[1]{\langle #1 \rangle} 
\newcommand{\RM}[1]{\mathrm{#1}}

\allowdisplaybreaks[4]
\title{Effective dynamics of quantum fluctuations in field theory: with applications to cosmology}
\date{}
\author[a,b]{Ding Ding,\footnote{Corresponding author.}}
\author[a,b]{Yu Zhao,}
\author[a,b]{and Yidun Wan\footnote{Corresponding author.}}
\affiliation[a]{State Key Laboratory of Surface Physics, Center for Astronomy and Astrophysics,
Department of Physics, Center for Field Theory and Particle Physics, and Institute for Nanoelectronic devices and Quantum computing, Fudan University,\\2005 Songhu Road, Shanghai 200433, China}
\affiliation[b]{Shanghai Research Center for Quantum Sciences,\\99 Xiupu Road, Shanghai 201315, China}

\emailAdd{ding\_ding@fudan.edu.cn}
\emailAdd{yuzhao20@fudan.edu.cn}
\emailAdd{ydwan@fudan.edu.cn}

\abstract{We develop a novel framework for describing quantum fluctuations in field theory, with a focus on cosmological applications. Our method uniquely circumvents the use of operator/Hilbert-space formalism, instead relying on a systematic treatment of classical variables, quantum fluctuations, and an effective Hamiltonian. Our framework not only
aligns with standard formalisms in flat and de Sitter spacetimes,
which assumes no backreaction, demonstrated through the $\varphi^3$-model,
but also adeptly handles time-dependent backreaction in more general cases. The uncertainty principle and spatial symmetry emerge as critical tools for selecting initial conditions and understanding effective potentials. We discover that modes inside
the Hubble horizon \emph{do not}
necessarily feel an initial Minkowski vacuum,
as is commonly assumed.
Our findings offer fresh insights into the early universe's quantum fluctuations
and potential explanations to
    large-scale CMB anomalies.}

\begin{document}
\maketitle
\flushbottom

\clearpage

\section{Introduction}
The framework of inflation explains elegantly 
how quantum fluctuations of the early universe can seed
a Gaussian primordial power spectrum.
This primordial power spectrum would then grow to form the
inhomogeneous structures of our universe
--- the temperature correlations of the cosmic microwave background
(CMB)
and the Large Scale Structure (LSS) of galaxies.
Therefore, quantum fluctuations are crucial for interpreting observations in contemporary cosmology, necessitating a systematic framework for their study.

\subsection{Key results}
In this paper, we develop a framework describing quantum fluctuations
and its backreaction on classical variables.
It is an extension of \cite{bojowald2006effective}---which mainly considers quantum mechanical systems---
to the case of generic interacting fields.

Our framework is novel in that it consistently describes
evolution and initial conditions without resorting to any operator/Hilbert-space language.

Our framework leads to the following results:
\begin{itemize}
	\item 
		Prescriptions on how to describe the evolution of quantum
		fluctuations to arbitrary order in interaction strength,
		how to find suitable initial conditions, and how to utilize
		spatial symmetries to solve the equations
  of motion by explicitly decoupling the irrelevant quantum fluctuations.
	\item A new way of understanding effective potentials and selecting initial conditions by means of the uncertainty principle and symmetries in position-space.
\end{itemize}

Albeit devoid of operators and Hilbert spaces,
our framework aligns with the standard operator/Hilbert-space formalism.
We demonstrate this in cases free of backreaction,
as is required for a fixed background metric,
using the $\varphi^3$-model on both Minkowski and de Sitter spacetimes as an example.
Nevertheless, our framework can coherently encompass time-dependent backreaction,
as to be discussed in 
Section \ref{sec:backreact}.
In the context of cosmology,
non-vanishing backreaction
will affect the preservation
of background homogeneity,
which can be analyzed adeptly in our
framework.

\subsection{Motivations and background}
The improved precision of cosmological measurements
\cite{planck2018const} in the past
two decades has allowed physicists to peek into the early conditions
of our universe and understand their subsequent evolution.
This in turn helped theorists establish cosmological perturbation theory
as the preferred description for the observed Large Scale Structure and 
cosmic microwave background.
In essence, the theory tells us how small inhomogeneous perturbations
evolve on a homogeneous background,
assuming some initial conditions for the inhomogeneities.
Based on the temperature correlations in the CMB,
the initial condition is thought to obey (classical) Gaussian statistics,
from which there are no strong hints of deviation.

But this is not the end of the story, as our universe is unlikely to have started
with purely classical initial conditions.
Evolving our universe backward in time, energy will increase to the point where quantum effects deserve serious consideration. We must then ask: What are the quantum conditions that seeded the classical
initial inhomogeneities?
The theory of inflation \cite{guth1981inflationary},
which was originally devised to solve cosmological puzzles largely in the homogeneous
sector, answers this question elegantly
\cite{mukhanov1992theory}.
In the inflationary framework,
the universe exists in a quasi-de Sitter background
and (in most models)
undergoes roughly 70 $e$-folds of rapid expansion before transferring all
its energy into the preheating phase
\cite{kofman1994reheating}, giving rise to the ``hot big bang''.
During this quasi-de Sitter phase,
the quantum fluctuations of inhomogeneities exit the Hubble horizon
and classicalize (through some yet-to-be known mechanism).
They re-enter the Hubble horizon post-inflation as classical inhomogeneities
to seed the Gaussian statistics that is required to explain our CMB observations.
In other words, inflation builds a coherent picture of the origin of structure
based on early quantum fluctuations.

The inflationary framework thus offers an opportunity to bridge quantum theory and classical physics, incorporating the interplay between matter fields and gravity. Unfortunately, since we do not have a full theory of quantum gravity, how one should translate theory to observation via first principle is easier said than done. Even the compromise of using the framework of quantum field theory in curved spacetime entails major obstacles. (For de Sitter spacetime, see for example \cite{krotov2011infrared,polyakov2012infrared,danielsson2019quantum}.) Such operator/Hilbert-space formalisms therefore suffer from both conceptual and practical difficulties. Fortunately, we can turn to effective methods that aim to approximate
quantum effects with a (modified) classical system.

In fact, since the infancy of the inflationary paradigm, the use of effective potentials has already proved to be successful in fixing the inflation potential in model-building. The idea is that instead of analyzing the full quantum system, one integrates out quantum fluctuations in the path-integral to obtain a modified classical potential
\cite{jona1964relativistic,coleman1973radiative,jackiw1979time}. In recent years further development in effective methods---using classical variables
and effective classical dynamics for expectation values---has experienced a modest resurgence.
(As an incomplete list, see for example
\cite{mulryne2010moment, mulryne2011moment,
vachaspati2018classical,vachaspati2019classical,
bojowald2021canonical, bojowald2021multi}).

In this paper, we introduce a systematic way to effectively describe an interacting quantum field
through three elements: ``classical'' variables, quantum fluctuations, and an effective Hamiltonian to generate their dynamics. It is found that the uncertainty principle and position-space symmetries allow us to solve (in some cases analytically) the equations of motion for quantum fluctuations.
The quantum mechanical version of the method was conceptualized in \cite{bojowald2006effective}.
In field theory context, attempts to derive effective potentials were made in \cite{bojowald2014canonical}.
The Coleman-Weinberg potential was obtained
via the \emph{assumptions} of an adiabatic approximation and the saturation of the uncertainty principle. Nevertheless, the key role of position-space symmetries was yet to be uncovered. In our present work, we will argue that the preservation of position-space
symmetries is a consequence of minimal backreaction, as is required for cosmology if the background metric is to stay fixed.
On the other hand,
initial position-space symmetries will decouple the evolution of quantum fluctuation from each other
and from the ``classical" variables 
to an extent that allows us to find solutions either analytically (in flat space) or numerically.
In this sense, spatial-symmetries and the suppression of backreaction reinforce each other.
Additionally, saturation of the uncertainty principle for systems dictated by low-energy effective potentials will arise as a \emph{consequence} rather than an assumption.

Our framework is shown \cite{bojowald2021multi}
to share connections with the variational interpretation of the effective action approach developed by Jackiw \cite{jackiw1979time}. At its Gaussian level, it coincides with the method of classical-quantum correspondence \cite{vachaspati2019classical}. But in \cite{bojowald2021canonical} it was argued that our method has the advantage of being more systematic and works for more generic situations. The dynamics of our method at leading order also share similarities with the well-known transport method \cite{mulryne2010moment}. As we will show later, however, the two differ both conceptually and practically. Furthermore, our method also provides a natural way of treating quantum backreaction because the mixed dynamics between (both homogeneous and inhomogeneous) classical variables and quantum fluctuations are naturally embedded in our framework, as was shown in \cite{bojowald2021canonical}.
Finally, attempts to extend \cite{bojowald2006effective}
to inflation also include 
\cite{brizuela2019moment,brizuela2021mode,brizuela2022quantum},
which uses quantum constraints (Wheeler-DeWitt equations)
as a starting point for discussing quantum effects.
In these works, the inhomogeneous part of the Hamiltonian 
is kept to second order. In particular, the effects of
higher-power interaction terms are not discussed.
While their results shed light on leading-order quantum backreaction
effects for constrained systems,
our present work provides a systematical prescription
for dealing with quantum fluctuations of generic field systems.

The paper is organized as follows: Section \ref{QM-sec} reviews our effective framework in a quantum mechanical setting and makes contact with the variational approach to effective potentials. In particular, we will show how one could arrive at the celebrated Coleman-Weinberg potential as a leading-order result. Section \ref{sec-Mink-fields} demonstrates the field version of our method in flat spacetime. We discover that the uncertainty principle and symmetries are pivotal in selecting minimal-energy Hamiltonians. Section \ref{sec-dS-fields} applies our method to cosmological spacetimes and computes the power spectrum for quadratic systems. We find, with the aid of the uncertainty principle, that the large-wavelength modes no longer see a Minkowski vacuum when initial conditions are set at finite times in the past. In particular, initial conditions differ physically for different choices of variables related to canonical transformations. (A similar conclusion was argued for with operator methods in \cite{grain2020canonical}.) Section \ref{sec-interacting-field} then generalizes our method to interacting fields and discusses how one could compute non-Gaussianities. For a $\varphi^3$-model, we obtain the same superhorizon limit for the bispectrum as that derived by standard operator methods. Section \ref{sec-conclusions} summarizes the paper. The paper ends with discussing the challenges and potential directions for future work.

Unless explicitly stated otherwise, natural units where $\hbar=c=G_N=1$ are assumed throughout the paper.

\section{Review of effective dynamics in quantum mechanics}
\label{QM-sec}
To understand the philosophy behind our effective framework, consider
first the classical case.
Given a classical physical system governed by
a Hamiltonian $H[x,p]$
with the canonical pair $(x,p)$, one can describe its evolution completely
with two functions of time $x(t)$ and $p(t)$,
or equivalently, two degrees of freedom.

The quantum counterpart of such a classical system however requires infinitely many complex degrees of freedom.
In textbook quantum mechanics (Schrodinger picture), these are the time-dependent expansion coefficients
of a generic state when expanded in a basis of the typically infinite-dimensional Hilbert space.

One could also choose a different parameterization for quantum states
by opting for fluctuations (also known as moments of a wavefunction)
as opposed to these expansion coefficients \cite{bojowald2006effective}.
If a state is ``semi-classical'',
or if one is only interested in leading order quantum effects,
the quantum system can often be consistently approximated
with finitely many fluctuation parameters.
For Gaussian-type states,
early attempts to apply this philosophy include
going beyond the 1PI effective potential
\cite{jackiw1979time}
and
describing quantum tunneling in quantum chemistry
\cite{arickx1986gaussian,prezhdo2006quantized}.
These works typically deal with second-order fluctuations
and, when extended to cosmology,
are directly relevant to the production of
the primordial power spectrum.
Despite differences in appearance,
this class of methods was shown
\cite{bojowald2021canonical}
to be equivalent to the more recently proposed
``quantum-classical correspondence'',
which can also be used to describe inhomogeneities generated by backreaction
in cosmology and black holes 
\cite{vachaspati2019classical,bojowald2021canonical,
mukhopadhyay2019rolling,vachaspati2019HR}.

Going beyond Gaussian systems,
a systematic prescription of using quantum fluctuations to approximate a state
was developed by Bojowald and collaborators
\cite{bojowald2006effective,baytacs2019effective,baytacs2020faithful}.
A key input that was not previously discovered was how one could construct
an effective Hamiltonian in such a way that it drove the dynamics of 
arbitrary-order fluctuations consistently.
Once such a prescription was found,
the parameterization of quantum dynamics using fluctuations
could be done to arbitrary orders for generic Hamiltonians,
at least in principle.

In practice, one cannot afford to keep track of arbitrary orders of fluctuations,
nor is it needed for describing observationally relevant quantum fluctuations
in cosmology.
A truncation or hierarchy
\footnote{In this paper, we will use the term ``hierarchy"
to refer to any relation among fluctuations of different types.
Finding a hierarchy 
is intended to help reduce the number of fluctuations relevant to some physical process,
such that calculations are tractable.
For example, for a set of fluctuations obeying a Gaussian hierarchy, all
even-order fluctuations are related to (products of) quadratic-order ones,
whereas all odd-order fluctuations vanish.
In this case, we only need to keep track of the quadratic-order fluctuations.}
among different orders of fluctuations 
can be assumed.
For example, one could assume on physical grounds that higher-order fluctuations
are suppressed by $O(\hbar)$ and thus negligible.
One could, on the other hand, also assume a Gaussian hierarchy, where
Wick's theorem reduces higher-order fluctuations to products of quadratic fluctuations.
In both cases,
only a finite number of fluctuations is needed to describe the system consistently.

In the remainder of this paper,
we shall first illustrate how to implement our effective framework
for generic quantum mechanical Hamiltonians.
Then we restrict to a system with standard kinetic terms
to acquire some physical intuition.
We will show that the well-known Coleman-Weinberg effective potential
can be thought of as a leading-order result in our effective framework.
This part of the paper is largely a review of
\cite{bojowald2006effective,bojowald2014canonical,
baytacs2019effective}.

The contribution of our paper to the field version of the effective framework will be presented in
Sections \ref{sec-Mink-fields}
and \ref{sec-dS-fields}.
These two sections will show that one 
could describe the quantum fluctuations systematically
for an interacting quantum field 
on both flat and cosmological spacetimes.
We also highlight the role of spatial symmetries,
which is absent in the quantum mechanical case.

\subsection{Basic elements of fluctuation dynamics}
To showcase,
let us first restrict to a system of one canonical pair
$(x,p)$.
Given its quantum Hamiltonian
$H[\hat{x},\hat{p}]$,
one can construct three objects:
(1) fluctuation parameters
(or ``fluctuations" for short),
denoted as $\Delta(\dots)$,
(2) an effective Hamiltonian $H_{\Delta}$
associated with these fluctuations,
and 
(3) an effective Poisson bracket $\{\dots,\dots\}$
associated with the phase space of fluctuations
and classical variables as $\Eval{\hat{x}}$
and $\Eval{\hat{p}}$.

Mathematically, for some arbitrary state
$|\psi\rangle$,
the three objects are defined as
\begin{align}
	\Delta(x^{n-m}p^m)
	\equiv&\Eval{(\hat{x}-\Eval{\hat{x}})^{n-m}(\hat{p}-\Eval{\hat{p}})^{m}}_W
	\nonumber\,,\\
	H_{\Delta}
	\equiv&\Eval{H[\hat{x},\hat{p}]}_W-H[\Eval{\hat{x}},\Eval{\hat{p}}]
	\nonumber\,,\\
	\{\Eval{\hat{A}},\Eval{\hat{B}}\}
	\equiv& \frac{1}{i\hbar}\Eval{[\hat{A},\hat{B}]}\,.
	\label{method-rules}
\end{align}
In the first line, the subscript $W$
refers to Weyl-ordering: an operation that symmetrizes the operators
and then divides the result by the number of possible permutations
\footnote{For example,
$\Eval{x^2p}_W=\Eval{xxp}_W=\Eval{xxp+xpx+pxx+xxp+xpx+pxx}/3!
=\Eval{x^2p+xpx+px^2}/3$,
where we have viewed $x^2$ as the product of two distinct operators $xx$.
Here, we have used the shorthand $x=\hat{x},p=\hat{p}$.}.
In particular, the ordering of operators in the
$\Delta(\dots)$
notation does not matter.
In the last line of \eqref{method-rules},
$\hat{A}$ and $\hat{B}$ can be arbitrary powers of operators.
We also demand that the Poisson brackets follow the Leibniz rule when
one of its slots has products of expectation values.

Our effective phase space associated with the Poisson brackets defined above
contains only functions of operator expectation 
values\footnote{There may be instances where there are $c$-numbers that have their own time evolution
but might not explicitly correspond to known/well-defined operators, as is the case for, say, the background metric
$\bar{g}_{\mu\nu}$
in an inflationary setting.
In these cases, one can either treat these $c$-numbers as time-dependent parameters or
chose to append their dynamics to our framework by introducing purely classical canonical variables to derive
their equations of motion.
These classical canonical variables will only have non-vanishing Poisson brackets among themselves
but may still couple to quantum fluctuations if the total effective Hamiltonian contains
terms like $\bar{g}_{\mu\nu}\Delta(\varphi^2)$ with inflaton field $\varphi$.}---i.e. complex numbers (or $c$-numbers).
So before calculating the Poisson bracket between any 
$c$-number function using \eqref{method-rules}, 
one should always make explicit the operators $\hat{A}$ and $\hat{B}$
that give rise to the $c$-numbers $\Eval{\hat{A}}$ and $\Eval{\hat{B}}$.
For example, consider the bracket between the following $c$-number functions
\begin{align*}
     \{\Eval{\hat{A}}\exp(\Eval{\hat{A}}),\Eval{\hat{B}}^2\}=&
      \{\Eval{\hat{A}}\sum_{n=0}\frac{1}{n!}\Eval{\hat{A}}^n,\Eval{B}^2\}\\
      =&\{\Eval{\hat{A}},\Eval{\hat{B}}^2\}\sum_{n=0}\frac{1}{n!}\Eval{\hat{A}}^n
      +\Eval{\hat{A}}\sum_{n=0}\frac{1}{n!}\{\Eval{\hat{A}}^n,\Eval{\hat{B}}^2\}\\
      =&2\{\Eval{\hat{A}},\Eval{\hat{B}}\}
      \Eval{\hat{B}}\sum_{n=0}\frac{1}{n!}\Eval{\hat{A}}^n
      +\sum_{n=0}\frac{2n}{n!}\Eval{\hat{A}}^n\{\Eval{\hat{A}},\Eval{\hat{B}}\}\Eval{\hat{B}}\,.
\end{align*}
One can then directly apply the Poisson bracket rule of \eqref{method-rules} to the last line.

For brevity of notation, in the following, we will refer to the power $n$ of product-operators
in $\Delta(\dots)$ as the order of the fluctuation,
denoted as $\Delta^{(n)}$
\footnote{For example, both
$\Delta(x^2)$ and $\Delta(xp)$ are
quadratic-order fluctuations, which can be collectively denoted as
$\Delta^{(2)}$.}.
Hereafter,
we shall also drop the hats on operators when there is no confusion.

Let us obtain some intuition for the dynamics of a system
by restricting to the Hamiltonian
\begin{equation}
	H[x,p]=
	\frac{1}{2}p^2+V(x)\,.
	\label{}
\end{equation}
To obtain $H_{\Delta}$, we take an expectation value and expand the operators
around their averages
\begin{equation}
	\Eval{H[x,p]}_W=\frac{1}{2}\Eval{p}^2+V(\Eval{x})
	+\underbrace{\frac{1}{2}\Delta(p^2)
	+\sum_{n=2}\frac{1}{n!}\partial_x^nV|_{x=\Eval{x}}\Delta(x^n)}_{\textstyle H_{\Delta}}
	\,.
	\label{}
\end{equation}

The quantum effects of the system are captured by $H_{\Delta}$
in the sense that it augments the classical equations of motion
through fluctuations $\Delta$
\begin{align}
	\dot{\Eval{x}}=&\{\Eval{x},\Eval{H[x,p]}_W\}
	=\Eval{p}\,,\nonumber\\
	\dot{\Eval{p}}=&
	\{\Eval{p},\Eval{H[x,p]}_W\}
	=-\partial_{x}V(x)|_{x=\Eval{x}}+
	{\{\Eval{p},H_{\Delta}\}}\,.
	\label{QM-EoM}
\end{align}
The second term on the RHS of the second line represents additional quantum evolution
sourced by the quantum fluctuations $\Delta$.
These terms, which couple classical variables
to quantum fluctuations, can be interpreted
as backreaction on the classical system
dictated by
$\Eval{x}$ and $\Eval{p}$.

The backreaction in our framework
comes from terms in $H_{\Delta}$
that are products between classical variables
and quantum fluctuations.
Note that however, backreaction is not induced
by the effective Poisson bracket between
classical variables and fluctuations.
Indeed, one can show
\cite{bojowald2006effective}
that at any order $n$ of fluctuations 
$\Delta^{(n)}$,
one has $\{\Eval{x},\Delta^{(n)}\}=0$
and similarly for $\Eval{p}$.
This implies that fluctuations $\Delta$ make up a subspace that is
symplectically orthogonal
to the classical phase space spanned by $\Eval{x}$ and $\Eval{p}$.
Therefore by including the fluctuation parameters $\Delta$,
we obtain an extended phase space on top of the classical one,
via definitions
\eqref{method-rules}.
This extension is what allows us to include quantum
degrees of freedom in our effective framework.

\subsection{Quantum backreaction and effective potentials}
\label{sec:backreact}

Given a classical inflation (scalar) field,
described by $\varphi$,
a full quantum analysis should quantize the full field: both the inhomogeneous and the homogeneous parts of the scalar.
But,
since the homogeneous part of the inflation couples to the Hubble parameter
via the Einstein equations,
the quantization of the homogeneous sector involves the gravitational background,
making it extremely difficult.
In practice,
one could compromise by including quantum backreaction on the classical background
field, described by $\Eval{\varphi}$,
with the use of effective potentials.
This allows us to integrate out fluctuations like $\Delta(\varphi\varphi)$ to effectively couple them to
$\Eval{\varphi}$ with a modification to the original classical
potential of $\varphi$.

The price to pay for such a compromise is at least twofold.
First, the time dependence of the backreaction is either partially lost or, at the very least,
made implicit.
One uses the same function(al)
--- the effective potential ---
to describe backreaction at all times.
Second, the dynamics of quantum fluctuations are lost, too,
as they are integrated into the path-integral.
But the evolution of these quantum fluctuations is precisely
what generates the primordial power spectrum
in the inflationary framework.
One would ultimately still need to analyze them.
In the process,
one runs into the danger of ``re-quantizing'' the inflaton,
as the quantum fluctuations now evolve on an effective background
that was obtained via their ``initial'' quantization.

Our effective framework based on
\eqref{method-rules}
can avoid making these compromises.
The ``classical'' variables
and quantum fluctuations are described consistently in a single
framework.
From \eqref{QM-EoM},
we see how the extended part of the phase space provides quantum backreaction
to the classical variables $\Eval{x}$ and $\Eval{p}$
through $H_{\Delta}$.
The quantum backreaction is also dynamic in the sense that the fluctuations
evolve according to
\begin{equation}
	\dot{\Delta}(\dots)=\{\Delta(\dots),\Eval{H[x,p]}\}
	=\{\Delta(\dots),H_{\Delta}\}\nonumber\,.
	\label{}
\end{equation}
Notice that since $H_{\Delta}$ may contain classical variables
$\Eval{x}$ and $\Eval{p}$ through
$\partial_{x}^nV(x)|_{x=\Eval{x}}$,
the evolution of $\Delta$\ is also affected by the classical background.
Our framework captures the two-way communication between classical variables
and quantum effects in a time-dependent manner.

When applying cosmological perturbation theory
to inflation, one often sets apart the classical portion of the full
system from the sub-system we wish to quantize.
The legitimacy of this operation relies on 
the assumption of minimal quantum backreaction
\cite{mukhanov1992theory}.
So when is quantum backreaction --- loosely speaking --- minimal?
It is minimal in the case of quadratic systems,
which is incidentally how one describes cosmological perturbations
in inflation at the leading order.
This answer suggests that one needs to be very
careful when treating non-Gaussian systems
during inflation,
for the sake of keeping backreaction small.
We will show in Section
\ref{sec-Mink-fields}
how spatial symmetries help us minimize
backreaction and control the growth of inhomogeneity.

That quadratic systems lead to minimal backreaction
is most easily seen in a quantum mechanical toy model,
with a potential $V(x)=m^2x^2/2$ for some parameter $m$.
one finds that quantum fluctuations and classical variables are decoupled:
\begin{align*}
	\Eval{H[x,p]}
	=&H[\Eval{x},\Eval{p}]+\frac{1}{2}\Delta(p^2)
	+\frac{1}{2}m^2\Delta(x^2)\\
	\dot{\Eval{x}}
	=&\Eval{p}\\
	\dot{\Eval{p}}
	=&-\partial_xV(x)|_{x=\Eval{x}}\,.
\end{align*}
The evolution of the quadratic fluctuation
$\Delta(x^2)\,,\;\Delta(xp)$, and $\Delta(p^2)$
also decouples from the classical variables
due to the form of $\Eval{H[x,p]}$.

For anharmonic systems,
it is possible to capture leading-order quantum effects by
truncating $\Eval{H[x,p]}$
beyond the quadratic order $\Delta^{(2)}$ in fluctuations.
\begin{equation}
	\Eval{H[x,p]}
	=H[\Eval{x},\Eval{p}]+\frac{1}{2}
 \Delta(p^2)
	+\frac{1}{2}V''(\Eval{x})\Delta(x^2)
	+\cancel{\sum_{n=3}\partial_x^nV(x)|_{x=\Eval{x}}\Delta(x^n)}
	\,.
	\label{}
\end{equation}
With such a truncated Hamiltonian,
denoted as
\begin{equation}
	\Eval{H[x,p]}_2
	=H[\Eval{x},\Eval{p}]
 +\frac{1}{2}\Delta(p^2)
	+\frac{1}{2}V''(\Eval{x})\Delta(x^2)\,,
	\label{}
\end{equation}
one can connect our effective framework to the conventional description of quantum backreaction that is based on the idea of effective actions and time-dependent variational principle
\cite{jackiw1979time}.
At the leading order (where non-local terms are absent),
the conventional result is commonly referred to as the
Coleman-Weinberg effective potential
\cite{coleman1973radiative}.

Specifically, our effective system can be reduced to one with
a Coleman-Weinberg effective potential by varying $\Eval{H[x,p]}$
with respect to the fluctuations.
This is a manifestation of the variational interpretation of effective potentials.
In this procedure, 
an obstacle to overcome is implementing the
consistency with the Heisenberg uncertainty principle,
which dictates the quantum fluctuations.
To this end, it is useful to introduce a pair of canonical variables
--- constructed from the fluctuation parameters ---
that automatically satisfies the uncertainty principle
\cite{baytacs2020faithful}:
\begin{equation}\label{canonical-variables}
\Delta(x^2)=Q^2,\quad \Delta(xp)=QP,\quad \Delta(p^2)=P^2+\frac{U}{Q^2},\quad U=\Delta(x^2)\Delta(p^2)-\Delta^2(xp), 
\end{equation}
where $Q$ and $P$ are canonical pairs satisfying
$\{Q,P\}=1$.
The parameter $U$ is time-independent
when one restricts to systems that contain up to second-order fluctuations
--- the parameter $U$ is the Casimir of the Poisson algebra
of fluctuations $(\Delta(xp),\Delta(x^2),\Delta(p^2))$.
For minimal-uncertainty states, it is simply
$U=\frac{\hbar^2}{4}$.
One can check that the map \eqref{canonical-variables}
is consistent with the Poisson bracket algebra
defined by \eqref{method-rules}.

With such a change of variables,
the effective Hamiltonian we wish to extremize becomes
\begin{equation}
	\Eval{H[x,p]}=
	H[\Eval{x},\Eval{p}]
	+\frac{1}{2}P^2+\frac{1}{2}V''(\Eval{x})Q^2+\frac{U}{2Q^2}\,.
	\label{}
\end{equation}
Note that due to the uncertainty principle,
$\Delta(p^2)$ contributes a $U/Q^2$ term that acts as a
``repulsive'' potential stopping the variance
of position
$\Delta(x^2)=Q^2$ from reaching zero.

By extremizing the effective Hamiltonian for $Q, P$, which represents the fluctuation parameters, we obtain the minimized Hamiltonian
\begin{align}
	&\RM{Extr}[\Eval{H[x,p]}]=H[\Eval{x},\Eval{p}]+\sqrt{UV''(\Eval{x})}
	\nonumber\\
	&\text{when } P^2=0\,,\; Q^2=\sqrt{\frac{U}{V''(\Eval{x})}}
	\Leftrightarrow \Delta(p^2)=V''(\Eval{x})\Delta(x^2)\,,\;
	\Delta(xp)=0\,.
	\label{QM-min-EP}
\end{align}
Setting $U=\hbar^2/4$ leads to the Coleman-Weinberg result in quantum mechanics
\cite{cametti1999comparison}.

Let us recap what we did.
We arrived at the Coleman-Weinberg effective potential by applying two
principles:
(1) Extremization of our effective Hamiltonian $\Eval{H}$ with respect to
the quantum fluctuations
and (2) quantum fluctuations complying with the uncertainty principle.
The two principles entail a variational
problem with inequality constraints.
For a quantum mechanical system with one canonical pair,
it is easy enough to find a canonical map \eqref{canonical-variables}
that embodies the uncertainty relation,
making the constraints trivial.
Direct extremization of $\Eval{H}$ with respect to the canonical variables $Q$ and $P$
gives the Coleman-Weinberg potential.

In field theory,
there are infinitely many canonical pairs.
Finding a map from 
fluctuations to canonical variables 
$\{\Delta\} \rightarrow \{Q_1,Q_2,\dots,P_1,P_2,\dots\}$
is all but impossible.
Nevertheless, one could still apply principles (1) and (2)
to obtain the Coleman-Weinberg potential, as we will show in
Section \ref{UP-IV-phi3}.

To summarize,
we have shown that our effective framework centered around quantum fluctuations
can recover the results in an operator formalism.
The leading order of the effective action, viz the Coleman-Weinberg effective potential,
can be recovered by keeping up to second order in fluctuation parameters
$\Delta^{(2)}$ and demanding extremization
of the effective Hamiltonian.
This allows us to interpret the Coleman-Weinberg result
physically as a low-energy effective system that is stable to quantum fluctuations.
In situations beyond such an approximation,
our method provides a systematic way to include higher-order
quantum fluctuations in a time-dependent way.
In particular, our method applies to the case of interacting fields,
which we shall discuss in Sections \ref{sec-Mink-fields} and \ref{sec-dS-fields}.

\section{Fields on a flat spacetime}\label{sec-Mink-fields}

Our effective framework also applies to systems with infinite degrees of freedom, namely fields.
Nevertheless, compared with quantum mechanics, 
three key aspects become more consequential:
(1) calculating Poisson brackets between various fluctuations for fields, 
(2) global symmetries of the system, 
especially the translational and rotational invariance of expectation values and fluctuations, and
(3) determining initial values based on the uncertainty principle and an assumption of initial Gaussianity
\footnote{One could forgo the Gaussianity assumption
and replace it with some other hierarchy
between higher-order fluctuations and lower-order
ones.
The underlying idea is that we want a way to
fix the initial values for all the fluctuations
by specifying
only a finite number of them
(for example the quadratic-order ones).
To this end, Gaussianity --- or the hierarchy behind
Wick's theorem --- is the most common realization
in traditional quantum field theory.}.

We will demonstrate these aspects through a massive real 
\(\varphi^3\)-model in the Minkowski spacetime. 
Nevertheless, almost all of our conclusions in this section
also hold for more general cosmological spacetimes. 
We will highlight the exceptions as they appear. 
Our present example will also help clarify the notations 
used in our effective framework for field applications.

The action of a real scalar field \(\varphi(t, \mathbf{x})\) with mass \(m\) is
\begin{equation}
S_{\text{Mink.}}[\varphi] \equiv \int dt d^3x \left[\frac{1}{2}(\dot{\varphi}^2 
-\delta^{ij}\partial_i\varphi\partial_j\varphi) 
- \frac{1}{2}m^2\varphi^2 - \frac{\mu}{3}\varphi^3 \right]\,.
\end{equation}
Here, \(\mu \ne 0\) is the coupling constant of 
the $\varphi^3$-interaction. 
The term ``Mink.'' refers to the Minkowski spacetime. 
The associated Hamiltonian is
\begin{align}
    H_{\text{Mink.}} &= \int d^3x\left[\frac{1}{2}\pi^2 
    + \frac{1}{2}\delta^{ij}\partial_i\varphi\partial_j\varphi 
    + \frac{1}{2}m^2\varphi^2 + \frac{\mu}{3}\varphi^3 \right] \label{eq:HamilReal}\\
    &= \int \frac{d^3k}{(2\pi)^3}\frac{1}{2}
    \left[ \pi_{\mathbf{k}}\pi_{\mathbf{-k}} 
    + \omega_k^2\varphi_{\mathbf{k}}\varphi_{\mathbf{-k}} \right] \nonumber\\
    &\quad + 
    \frac{\mu}{3}
    \int \frac{d^3k}{(2\pi)^3}\frac{d^3p}{(2\pi)^3}\frac{d^3q}{(2\pi)^3}(2\pi)^3
    \delta^3(\mathbf{k+p+q})
    \varphi_{\mathbf{k}}\varphi_{\mathbf{p}}\varphi_{\mathbf{q}}\,, \label{eq:HamilMoment}
\end{align}
where we have denoted the conjugate momentum
and effective frequency of the scalar field by
\[\pi(t, \mathbf{x}) \equiv \dot\varphi(t, \mathbf{x}),\qquad \omega_k^2 \equiv k^2 + m^2,\qquad k = \sqrt{\mathbf{k}\cdot\mathbf{k}}.\]
The commutator between the field operators \(\varphi(t, \mathbf{x})\) and \(\pi(t, \mathbf{x})\) is 
\[\left[\varphi(t, \mathbf{x}), \pi(t, \mathbf{y})\right] = i\hbar\delta^3(\mathbf{x - y})\,.\]

In Eq. \eqref{eq:HamilMoment}, 
we rewrite the Hamiltonian in terms of
the momentum space representations
of the field operators:
\begin{equation*}
	\varphi_\mathbf{k}(t) \equiv
	\int d^3x\varphi(t, \mathbf{x})
	e^{-i\mathbf{k}\cdot \mathbf{x}},\qquad 
	\pi_{\mathbf{k}}(t) \equiv
	\int d^3x\pi(t,\mathbf{x})
	e^{-i\mathbf{k}\cdot \mathbf{x}},
\end{equation*}
satisfying
\begin{equation*}
	\left[\varphi_{\mathbf{k}},\pi_{\mathbf{p}}\right] =
	i\hbar(2\pi)^3\delta^3(\mathbf{k+p}).
\end{equation*}
To simplify the notations, 
we omit the time argument when there is no risk of confusion:
\(\varphi_{\mathbf{k}}\equiv\varphi_{\mathbf{k}}(t), \pi_{\mathbf{p}}\equiv\pi_{\mathbf{p}}(t)\).

In the effective dynamics of quantum fields, 
calculations often involve numerous 
integrals and delta functions of momenta-sums. 
For brevity, 
we introduce the following conventions:
\begin{equation*}
    \int_{k,p,q,\dots} \equiv \int\frac{d^3k}{(2\pi)^3}\frac{d^3p}{(2\pi)^3}\frac{d^3q}{(2\pi)^3}\dots\qquad
    \tilde{\delta}_{\mathbf{k,p,q,}\dots} \equiv (2\pi)^3\delta^3(\mathbf{k+p+q}+\dots)
    \,.
\end{equation*}
Using these notations, the Hamiltonian simplifies to
\begin{equation}
H_{\text{Mink.}}[\varphi, \pi] = 
\int_k \frac{1}{2}\left[\pi_{\mathbf{k}}\pi_{\mathbf{-k}} + \omega_k^2\varphi_{\mathbf{k}}\varphi_{\mathbf{-k}}\right]
+ \frac{\mu}{3}\int_{k, p, q}\tilde\delta_\mathbf{k, p, q}\varphi_\mathbf{k}\varphi_\mathbf{p}\varphi_\mathbf{q}.
\end{equation}

For any state \(\ket\psi\) in Fock space, 
the effective Hamiltonian is
\begin{equation} \label{eq:HamilQ}
	H_{\text{Mink.},Q}
\equiv \Eval{H_{\RM{Mink.}}[\varphi,\pi]}
= H_{\text{Mink.}}[\Eval{\varphi}, \Eval{\pi}] + H_{\text{Mink.},\Delta}\,.
\end{equation}
The subscript ``Q'' denotes the full effective Hamiltonian
of the quantum theory, where only $H_{\text{Mink.},\Delta}$
drives the evolution of the fluctuations
and is given by
\begin{align}
H_{\text{Mink.},\Delta}
= &\int_k
\frac{1}{2}\left[\Delta(\pi_{\mathbf{k}}\pi_{\mathbf{-k}})
+\omega_k^2\Delta(\varphi_{\mathbf{k}}\varphi_{\mathbf{-k}})\right]\nonumber\\
&\ + \mu\int_{k,p,q}\tilde{\delta}_{\mathbf{k,p,q}}
\left[\Delta(\varphi_{\mathbf{k}}\varphi_{\mathbf{p}})
\Eval{\varphi_{\mathbf{q}}}
+\frac{1}{3}\Delta(\varphi_{\mathbf{k}}\varphi_{\mathbf{p}}\varphi_{\mathbf{q}})\right].\label{eq:HamilDelta}
\end{align}

\subsection{Effective Poisson brackets and evolution}

One needs to calculate the Poisson brackets between different fluctuations
before one can obtain their evolution.
Given that \(\varphi_\mathbf{k}, \pi_\mathbf{p}\) with different $\mathbf{k}, \mathbf{p}$
are distinct operators, 
the direct calculation of higher-order Poisson brackets 
for Weyl-ordered expectation values is a complex task. 
Nevertheless, inspired by \cite{bojowald2006effective},
we have developed a less tedious method for calculating the Poisson bracket
between any two fluctuations via
the Baker-Campbell-Hausdorff formula.
(See Appendix \ref{BCH-PB-appendix}.)

Specifically, the equations of motion for fluctuations 
in the \(\varphi^3\)-model only involve
Poisson brackets containing fluctuations in \(H_{\text{Mink.},\Delta}\), namely,
\(\Delta(\varphi_\mathbf{k}\varphi_\mathbf{p}), \Delta(\pi_\mathbf{k}\pi_\mathbf{p})\), 
and \(\Delta(\varphi_\mathbf{k}\varphi_\mathbf{p}\varphi_\mathbf{q})\). 
The relevant nonvanishing Poisson brackets are
\begin{align}
&\Big\{\Delta(\varphi_\mathbf{k}\varphi_\mathbf{p}),\  \Delta(\varphi_{\mathbf{a}_1}\cdots\varphi_{\mathbf{a}_m}\ \pi_{\mathbf{b}_1}\cdots\pi_{\mathbf{b}_n})\Big\}\nonumber\\
=\ &\Bigg[\sum_{j = 1}^n\tilde\delta_{\mathbf{k,b}_j}\Delta(\varphi_\mathbf{p}\ \varphi_{\mathbf{a}_1}\cdots\varphi_{\mathbf{a}_m}\ \pi_{\mathbf{b}_1}\cdots\pi_{\mathbf{b}_{j-1}}\pi_{\mathbf{b}_{j+1}}\cdots\pi_{\mathbf{b}_n})\Bigg] + (\mathbf{k\leftrightarrow p}),\label{eq:matchphiphipi}\\\nonumber\\
&\Big\{\Delta(\pi_\mathbf{k}\pi_\mathbf{p}),\ \Delta(\varphi_{\mathbf{a}_1}\cdots\varphi_{\mathbf{a}_m}\ \pi_{\mathbf{b}_1}\cdots\pi_{\mathbf{b}_n})\Big\}\nonumber\\
=\ &\Bigg[-\sum_{i = 1}^m\tilde\delta_{\mathbf{k,a}_i}\Delta(\pi_\mathbf{p}\ \varphi_{\mathbf{a}_1}\cdots\varphi_{\mathbf{a}_{i-1}}\varphi_{\mathbf{a}_{i+1}}\cdots\varphi_{\mathbf{a}_m}\ \pi_{\mathbf{b}_1}\cdots\pi_{\mathbf{b}_n}) \Bigg] + (\mathbf{k\leftrightarrow p}),\label{eq:matchpipiphi}\\\nonumber\\
&\Big\{\Delta(\varphi_\mathbf{k}\varphi_\mathbf{p}\varphi_\mathbf{q}),\ \Delta(\varphi_{\mathbf{a}_1}\cdots\varphi_{\mathbf{a}_m}\ \pi_{\mathbf{b}_1}\cdots\pi_{\mathbf{b}_n})\Big\}\nonumber\\
=\ &\sum_{j = 1}^n\tilde\delta_{\mathbf{k, b}_j}\Bigg[\Delta(\varphi_\mathbf{p}\varphi_\mathbf{q}\ \varphi_{\mathbf{a}_1}\cdots\varphi_{\mathbf{a}_m}\ \pi_{\mathbf{b}_1}\cdots\pi_{\mathbf{b}_{j-1}}\pi_{\mathbf{b}_{j+1}}\cdots\pi_{\mathbf{b}_n}) - \Delta(\varphi_\mathbf{p}\varphi_\mathbf{q})\times\nonumber\\
&\Delta(\varphi_{\mathbf{a}_1}\cdots\varphi_{\mathbf{a}_m}\ \pi_{\mathbf{b}_1}\cdots\pi_{\mathbf{b}_{j-1}}\pi_{\mathbf{b}_{j+1}}\cdots\pi_{\mathbf{b}_n})\Bigg] + (\mathbf{k\to p\to q})\nonumber\\
&-\frac{\hbar^2}{4}\sum_{1\le j_1 < j_2 < j_3 \le n}\Bigg[\tilde\delta_{\mathbf{k, b}_{j_1}}\Big(\tilde\delta_{\mathbf{p, b}_{j_2}}\tilde\delta_{\mathbf{q, b}_{j_3}} + \tilde\delta_{\mathbf{q, b}_{j_2}}\tilde\delta_{\mathbf{p, b}_{j_3}} \Big) +(\mathbf{k\to p\to q})\Bigg]\times\nonumber\\
& \Delta(\varphi_{\mathbf{a}_1}\cdots\varphi_{\mathbf{a}_m}\ \pi_{\mathbf{b}_1}\cdots\pi_{\mathbf{b}_{j_1-1}}\pi_{\mathbf{b}_{j_1+1}}\cdots\pi_{\mathbf{b}_{j_2-1}}\pi_{\mathbf{b}_{j_2+1}}\cdots\pi_{\mathbf{b}_{j_3-1}}\pi_{\mathbf{b}_{j_3+1}}\cdots\pi_{\mathbf{b}_n}),\label{eq:matchphiphiphipipipi}
\end{align}
Here, the notation \((\mathbf{a \leftrightarrow b})\) denotes a term identical to the one preceding it but with the momenta \(\mathbf{a}\) and \(\mathbf{b}\) interchanged. Similarly, \((\mathbf{k \rightarrow p\rightarrow q})\) 
indicates two additional terms, obtained from the preceding term by cyclically permuting the momenta \(\mathbf{k}, \mathbf{p}\), and \( \mathbf{q} \). Especially, we define $\Delta(1)=1$, $\Delta(\varphi_\mathbf{p}) = \Delta(\pi_\mathbf{p}) = 0$.
For instance,
\begin{align}
	&\{\Delta(\varphi_{\mathbf{k}}\varphi_{\mathbf{p}}),
	\Delta(\pi_{\mathbf{a}}\pi_{\mathbf{b}})\}
	= \tilde{\delta}_{\mathbf{k,a}}\Delta(\varphi_{\mathbf{p}}\pi_{\mathbf{b}}) 
	+ \tilde{\delta}_{\mathbf{k,b}}\Delta(\varphi_{\mathbf{p}}\pi_{\mathbf{a}}) 
	+ \tilde{\delta}_{\mathbf{p,a}}\Delta(\varphi_{\mathbf{k}}\pi_{\mathbf{b}}) 
	+ \tilde{\delta}_{\mathbf{p,b}}\Delta(\varphi_{\mathbf{k}}\pi_{\mathbf{a}}),\nonumber\\ 
	&\{\Delta(\varphi_{\mathbf{k}}\varphi_{\mathbf{p}}\varphi_{\mathbf{q}}),
	\Delta(\pi_{\mathbf{a}}\pi_{\mathbf{b}}\pi_\mathbf{c})\}
	= \tilde{\delta}_{\mathbf{k,a}} 
	\Big[\Delta(\varphi_{\mathbf{p}}\varphi_{\mathbf{q}}\pi_{\mathbf{b}}\pi_{\mathbf{c}}) 
	- \Delta(\varphi_\mathbf{p}\varphi_\mathbf{q})\Delta(\pi_\mathbf{b}\pi_\mathbf{c})\Big]\nonumber\\ 
	&+ \tilde{\delta}_{\mathbf{k,b}} 
	\Big[\Delta(\varphi_{\mathbf{p}}\varphi_{\mathbf{q}}\pi_{\mathbf{c}}\pi_{\mathbf{a}}) 
	- \Delta(\varphi_\mathbf{p}\varphi_\mathbf{q})\Delta(\pi_\mathbf{c}\pi_\mathbf{a})\Big] 
	+ \tilde{\delta}_{\mathbf{k,c}} 
	\Big[\Delta(\varphi_{\mathbf{p}}\varphi_{\mathbf{q}}\pi_{\mathbf{a}}\pi_{\mathbf{b}}) 
	- \Delta(\varphi_\mathbf{p}\varphi_\mathbf{q})\Delta(\pi_\mathbf{a}\pi_\mathbf{b})\Big]\nonumber\\ 
	&+ \tilde{\delta}_{\mathbf{p,a}} 
	\Big[\Delta(\varphi_{\mathbf{q}}\varphi_{\mathbf{k}}\pi_{\mathbf{b}}\pi_{\mathbf{c}}) 
	- \Delta(\varphi_\mathbf{q}\varphi_\mathbf{k})\Delta(\pi_\mathbf{b}\pi_\mathbf{c})\Big] 
	+ \tilde{\delta}_{\mathbf{p,b}} 
	\Big[\Delta(\varphi_{\mathbf{q}}\varphi_{\mathbf{k}}\pi_{\mathbf{c}}\pi_{\mathbf{a}}) 
	- \Delta(\varphi_\mathbf{q}\varphi_\mathbf{k})\Delta(\pi_\mathbf{c}\pi_\mathbf{a})\Big]\nonumber\\
	&+ \tilde{\delta}_{\mathbf{p,c}} 
	\Big[\Delta(\varphi_{\mathbf{q}}\varphi_{\mathbf{k}}\pi_{\mathbf{a}}\pi_{\mathbf{b}}) 
	- \Delta(\varphi_\mathbf{q}\varphi_\mathbf{k})\Delta(\pi_\mathbf{a}\pi_\mathbf{b})\Big] 
	+ \tilde{\delta}_{\mathbf{q,a}} 
	\Big[\Delta(\varphi_{\mathbf{k}}\varphi_{\mathbf{p}}\pi_{\mathbf{b}}\pi_{\mathbf{c}}) 
	- \Delta(\varphi_\mathbf{k}\varphi_\mathbf{p})\Delta(\pi_\mathbf{b}\pi_\mathbf{c})\Big]\nonumber\\
	&+ \tilde{\delta}_{\mathbf{q,b}} 
	\Big[\Delta(\varphi_{\mathbf{k}}\varphi_{\mathbf{p}}\pi_{\mathbf{c}}\pi_{\mathbf{a}}) 
	- \Delta(\varphi_\mathbf{k}\varphi_\mathbf{p})\Delta(\pi_\mathbf{c}\pi_\mathbf{a})\Big] 
	+ \tilde{\delta}_{\mathbf{q,c}} 
	\Big[\Delta(\varphi_{\mathbf{k}}\varphi_{\mathbf{p}}\pi_{\mathbf{a}}\pi_{\mathbf{b}}) 
	- \Delta(\varphi_\mathbf{k}\varphi_\mathbf{p})\Delta(\pi_\mathbf{a}\pi_\mathbf{b})\Big]\nonumber\\ 
	&- \frac{\hbar^2}{4}\Big(
		\tilde\delta_\mathbf{k, a}\tilde\delta_\mathbf{p, b}\tilde\delta_\mathbf{q, c} 
		+ \tilde\delta_\mathbf{k, a}\tilde\delta_\mathbf{p, c}\tilde\delta_\mathbf{q, b} 
		+ \tilde\delta_\mathbf{k, b}\tilde\delta_\mathbf{p, a}\tilde\delta_\mathbf{q, c} \nonumber\\
	&\qquad\quad + \tilde\delta_\mathbf{k, b}\tilde\delta_\mathbf{p, c}\tilde\delta_\mathbf{q, a}
		+ \tilde\delta_\mathbf{k, c}\tilde\delta_\mathbf{p, a}\tilde\delta_\mathbf{q, b}
		+ \tilde\delta_\mathbf{k, c}\tilde\delta_\mathbf{p, b}\tilde\delta_\mathbf{q, a}\Big).
\label{phiphiphipipipi}\end{align}
These brackets are instrumental in deriving the dynamics of quadratic- and cubic-order fluctuations,
\(\Delta^{(2)}\) and $\Delta^{(3)}$,
within the \(\varphi^3\)-model.

For further clarity, a few comments are in order:
\begin{enumerate}
\item The Poisson structure of fluctuations is model-independent. 
The Poisson brackets we have listed above can be directly applied 
to a $\varphi^3$-model in a cosmological spacetime, 
or to any other real scalar field systems.

\item Calculations of effective Poisson brackets (via \eqref{method-rules}) can be tedious in general.
Fortunately, the Poisson brackets involving quadratic fluctuations
\(\Delta^{(2)}\) can be directly ``read out'' 
	using the Leibniz rule and intuitions from classical Poisson brackets, without explicitly calculating
 any commutators.
 For example:
	\[\{\Delta(\varphi_{\mathbf{k}}\varphi_{\mathbf{p}}),
	\Delta(\pi_{\mathbf{a}}\pi_{\mathbf{b}})\}
	= \tilde{\delta}_{\mathbf{k,a}}\Delta(\varphi_{\mathbf{p}}\pi_{\mathbf{b}}) 
	+ \tilde{\delta}_{\mathbf{k,b}}\Delta(\varphi_{\mathbf{p}}\pi_{\mathbf{a}}) 
	+ \tilde{\delta}_{\mathbf{p,a}}\Delta(\varphi_{\mathbf{k}}\pi_{\mathbf{b}}) 
	+ \tilde{\delta}_{\mathbf{p,b}}\Delta(\varphi_{\mathbf{k}}\pi_{\mathbf{a}})\]
can be read out by pairing a canonical variable (i.e. field $\varphi$) in the argument of
$\Delta(\varphi_{\mathbf{k}}\varphi_{\mathbf{p}})$ with its conjugate
in the argument of $\Delta(\pi_{\mathbf{a}}\pi_{\mathbf{b}})$.
We sum all possible pairings.
A given pairing produces a $\tilde{\delta}$-function multiplied with $\Delta(\varphi\pi)$ which contains
arguments of fields not involved in such a pairing. 

Nevertheless, when dealing with brackets involving cubic fluctuations
(or higher-order ones), 
such as the cubic-cubic bracket \(\{\Delta(\varphi_{\mathbf{k}}\varphi_{\mathbf{p}}\varphi_{\mathbf{q}}),
\Delta(\pi_{\mathbf{a}}\pi_{\mathbf{b}}\pi_\mathbf{c})\}\), ``reading out" does not work anymore because
unexpected terms may emerge, as seen in the last line of \eqref{phiphiphipipipi}. 
These terms are pivotal in distinguishing between classical and quantum theories
because they are the remnants of non-commuting variables.
When higher-order fluctuations are present,
it is usually better to use the Baker-Campbell-Hausdorff formula 
in Appendix \ref{BCH-PB-appendix} to calculate the brackets.

\item The Hamilton's Equation $\dot\Delta = \{\Delta, H_{\text{Mink.}, \Delta}\}$ gives the equations of motion:
\begin{align}
&\dot\Delta(\varphi_{\mathbf{a}_1}\cdots\varphi_{\mathbf{a}_m}\ \pi_{\mathbf{b}_1}\cdots\pi_{\mathbf{b}_n})\nonumber\\ 
&= \sum_{i = 1}^m\Delta(\varphi_{\mathbf{a}_1}\cdots\varphi_{\mathbf{a}_{i-1}}\pi_{\mathbf{a}_i}\varphi_{\mathbf{a}_{i+1}}\cdots\varphi_{\mathbf{a}_m}\ \pi_{\mathbf{b}_1}\cdots\pi_{\mathbf{b}_n})\nonumber\\
&-\sum_{j = 1}^n\omega_{b_j}^2\Delta(\varphi_{\mathbf{a}_1}\cdots\varphi_{\mathbf{a}_m}\ \pi_{\mathbf{b}_1}\cdots\pi_{\mathbf{b}_{j-1}}\varphi_{\mathbf{b}_j}\pi_{\mathbf{b}_{j+1}}\cdots\pi_{\mathbf{b}_n})\nonumber\\
&-\mu\sum_{j = 1}^n\int_p\Bigg[\Delta(\varphi_{\mathbf{a}_1}\cdots\varphi_{\mathbf{a}_m}\ \pi_{\mathbf{b}_1}\cdots\pi_{\mathbf{b}_{j-1}}\varphi_{\mathbf{p}}\varphi_{\mathbf{b}_j-\mathbf{p}}\pi_{\mathbf{b}_{j+1}}\cdots\pi_{\mathbf{b}_n}) -\Delta(\varphi_\mathbf{p}\varphi_{\mathbf{b}_j - \mathbf{p}})\times\nonumber\\
&\qquad\qquad\quad\ \Delta(\varphi_{\mathbf{a}_1}\cdots\varphi_{\mathbf{a}_m}\ \pi_{\mathbf{b}_1}\cdots\pi_{\mathbf{b}_{j-1}}\pi_{\mathbf{b}_{j+1}}\cdots\pi_{\mathbf{b}_n})\Bigg]\nonumber\\
&+ \frac{\mu\hbar^2}{2}\sum_{1\le j_1 < j_2 < j_3 \le n}\tilde\delta_{\mathbf{b}_{j_1}, \mathbf{b}_{j_2}, \mathbf{b}_{j_3}}\times\nonumber\\
&\Delta(\varphi_{\mathbf{a}_1}\cdots\varphi_{\mathbf{a}_m}\ \pi_{\mathbf{b}_1}\cdots\pi_{\mathbf{b}_{j_1-1}}\pi_{\mathbf{b}_{j_1+1}}\cdots\pi_{\mathbf{b}_{j_2-1}}\pi_{\mathbf{b}_{j_2+1}}\cdots\pi_{\mathbf{b}_{j_3-1}}\pi_{\mathbf{b}_{j_3+1}}\cdots\pi_{\mathbf{b}_n}).\label{eq:eom}
\end{align}
Here, we have assumed that \(\Eval{\varphi_\mathbf{k}} = \Eval{\pi_\mathbf{k}} = 0\) --- their proof will be discussed later in Section \ref{sec:quanteffect}.
 
\item When perturbatively expanding \(\Delta\)'s in terms of the powers of \(\mu\):
\[
\Delta(\cdots) = \Delta_0(\cdots) + \Delta_1(\cdots) + \cdots + \Delta_n(\cdots) + \cdots,
\]
where \(\Delta_n(\cdots)\) is at the \(\mu^n\) order, the equations of motion \eqref{eq:eom} suggest that the dynamics of \(\Delta^{(k)}_n\) are influenced by
fluctuations that are lower-order in either the
power of product-operators or
in the powers of $\mu$. Therefore, if all \(\Delta_0(\cdots)\)'s at any time are known, it is possible in principle to compute all \(\Delta^{(k)}_n\) sequentially, in the order of powers of \(\mu\hbar^2\). The zeroth order \(\Delta_0(\cdots)\)'s evolve as free fields:
\begin{align*}
\dot\Delta_0(\varphi_{\mathbf{a}_1}\cdots\varphi_{\mathbf{a}_m}\pi_{\mathbf{b}_1}\cdots\pi_{\mathbf{b}_n}) &= \sum_{i = 1}^m\Delta_0(\varphi_{\mathbf{a}_1}\cdots\varphi_{\mathbf{a}_{i-1}}\pi_{\mathbf{a}_i}\varphi_{\mathbf{a}_{i+1}}\cdots\varphi_{\mathbf{a}_m}\pi_{\mathbf{b}_1}\cdots\pi_{\mathbf{b}_n})\\
&-\sum_{j = 1}^n\omega_{b_j}^2\Delta_0(\varphi_{\mathbf{a}_1}\cdots\varphi_{\mathbf{a}_m}\pi_{\mathbf{b}_1}\cdots\pi_{\mathbf{b}_{j-1}}\varphi_{\mathbf{b}_j}\pi_{\mathbf{b}_{j+1}}\cdots\pi_{\mathbf{b}_n}),
\end{align*}
whose dynamics are exactly solvable using Wick's theorem. 
We will illustrate two applications of this in later sections.
	\item In the fourth and fifth line of Eqs. \eqref{eq:eom}, the interaction perturbation is represented as 
	\[-\mu\int_p\Big[\Delta(X\varphi_\mathbf{p}\varphi_{\mathbf{a-p}}) - \Delta(X)\Delta(\varphi_\mathbf{p}\varphi_\mathbf{a-p})\Big],\] 
 where \(X\) represents some products of $\varphi$ and $\pi$ field operators. When assuming that the state \(\ket{\psi}\) is a harmonic vacuum state \(\ket{0}\) where Wick's theorem applies, the ``bubble term'' in \(\int_p\Delta(X\varphi_\mathbf{p}\varphi_\mathbf{a-p})\) is canceled by \(-\int_p\Delta(X)\Delta(\varphi_\mathbf{p}\varphi_\mathbf{a-p})\), 
	leaving only ``connected contributions'' that include fluctuations with fields 
	involving the external momenta in \(X\) as well as the internal momenta \(\mathbf{p}\) and \(\mathbf{a-p}\). 
	For instance, with \(X \equiv \varphi_\mathbf{b}\varphi_\mathbf{c}\), Wick's theorem gives
	\begin{align*}
 &\Delta(\varphi_\mathbf{b}\varphi_\mathbf{c}\varphi_\mathbf{p}\varphi_\mathbf{a-p}) - \Delta(\varphi_\mathbf{b}\varphi_\mathbf{c})\Delta(\varphi_\mathbf{p}\varphi_\mathbf{a-p})
 \\
 =\ &\Delta(\varphi_\mathbf{b}\varphi_\mathbf{p})\Delta(\varphi_\mathbf{c}\varphi_\mathbf{a-p}) + \Delta(\varphi_\mathbf{b}\varphi_\mathbf{a-p})\Delta(\varphi_\mathbf{c}\varphi_\mathbf{p}).
 \end{align*}
\end{enumerate}

In this paper, for demonstration, we primarily focus on the dynamics of quadratic and cubic fluctuations. Our goal is to deduce the tree-level solutions at the $\mu^0$ and $\mu^1$ orders, and then compare our findings in Minkowski spacetime with those in de Sitter cosmological spacetime
in Section \ref{sec-dS-fields}. Incorporating Eqs. \eqref{eq:eom}, we present the equations of motion for quadratic fluctuations $\Delta^{(2)}$ as follows:
\begin{align}
	\dot{\Delta}(\varphi_{\mathbf{a}}\varphi_{\mathbf{b}})
	&= \Delta(\varphi_{\mathbf{a}}\pi_{\mathbf{b}}) + \Delta(\varphi_{\mathbf{b}}\pi_{\mathbf{a}}), \nonumber\\ 
	\dot{\Delta}(\varphi_{\mathbf{a}}\pi_{\mathbf{b}})
	&= \Delta(\pi_{\mathbf{a}}\pi_{\mathbf{b}}) - \omega_b^2\Delta(\varphi_{\mathbf{a}}\varphi_{\mathbf{b}})
	-\mu\int_{p}
	\Delta(\varphi_{\mathbf{a}}\varphi_{\mathbf{p}}\varphi_{\mathbf{b-p}}),\nonumber\\
	\dot{\Delta}(\pi_{\mathbf{a}}\pi_{\mathbf{b}})
	&= -\omega_a^2\Delta(\varphi_{\mathbf{a}}\pi_{\mathbf{b}}) -\omega_b^2\Delta(\varphi_{\mathbf{b}}\pi_{\mathbf{a}}) - \mu\int_{p}
	\Bigg[\Delta(\pi_{\mathbf{a}}\varphi_{\mathbf{p}}\varphi_{\mathbf{b-p}}) 
	+ (\mathbf{a\leftrightarrow b})\Bigg].
	\label{phi3-quad-EoM}
\end{align}

For cubic fluctuations \(\Delta^{(3)}\), the equations of motion are:
\begin{align}
	\dot{\Delta}(\varphi_{\mathbf{a}}\varphi_\mathbf{b}\varphi_{\mathbf{c}})
	=&
	\Delta(\varphi_{\mathbf{b}}\varphi_{\mathbf{c}}\pi_{\mathbf{a}})
	+\Delta(\varphi_{\mathbf{c}}\varphi_{\mathbf{a}}\pi_{\mathbf{b}})
	+\Delta(\varphi_{\mathbf{a}}\varphi_{\mathbf{b}}\pi_{\mathbf{c}}),
	\nonumber\\
	\dot{\Delta}(\varphi_{\mathbf{a}}\varphi_\mathbf{b}\pi_{\mathbf{c}})
	=&\Delta(\varphi_{\mathbf{b}}\pi_{\mathbf{a}}\pi_{\mathbf{c}})
	+\Delta(\varphi_{\mathbf{a}}\pi_{\mathbf{b}}\pi_{\mathbf{c}})
	-\omega_c^2
	\Delta(\varphi_{\mathbf{a}}\varphi_{\mathbf{b}}\varphi_{\mathbf{c}})
	\nonumber\\
	-&\mu\int_{p}
	\Big[\Delta(\varphi_{\mathbf{a}}\varphi_{\mathbf{b}}
	\varphi_{\mathbf{p}}\varphi_{\mathbf{c-p}})
	-\Delta(\varphi_{\mathbf{a}}\varphi_{\mathbf{b}})
	\Delta(\varphi_{\mathbf{p}}\varphi_{\mathbf{c-p}})
	\Big]
	\nonumber\\
	\dot{\Delta}(\varphi_{\mathbf{a}}\pi_\mathbf{b}\pi_{\mathbf{c}})
	=&
	\Delta(\pi_{\mathbf{a}}\pi_{\mathbf{b}}\pi_{\mathbf{c}})
	-\omega_b^2\Delta(\varphi_{\mathbf{a}}\varphi_{\mathbf{b}}\pi_{\mathbf{c}})
	-\omega_c^2\Delta(\varphi_{\mathbf{a}}\varphi_{\mathbf{c}}\pi_{\mathbf{b}})
	\nonumber\\
	-&\mu\int_{p}
	\Bigg\{\Big[
	\Delta(\varphi_{\mathbf{a}}\pi_{\mathbf{b}}
	\varphi_{\mathbf{p}}\varphi_{\mathbf{c-p}})
	-\Delta(\varphi_{\mathbf{a}}\pi_{\mathbf{b}})
	\Delta(\varphi_{\mathbf{p}}\pi_{\mathbf{c-p}})\Big] + (\mathbf{b\leftrightarrow c})\Bigg\}
	\nonumber\\
	\dot{\Delta}(\pi_{\mathbf{a}}\pi_\mathbf{b}\pi_{\mathbf{c}})
	=&
	-\omega_a^2\Delta(\varphi_{\mathbf{a}}\pi_{\mathbf{b}}\pi_{\mathbf{c}})
	-\omega_b^2\Delta(\pi_{\mathbf{a}}\varphi_{\mathbf{b}}\pi_{\mathbf{c}})
	-\omega_c^2\Delta(\pi_{\mathbf{a}}\pi_{\mathbf{b}}\varphi_{\mathbf{c}})
	\nonumber\\
	-&\mu\int_{p}
	\Bigg\{\Big[
	\Delta(\pi_{\mathbf{a}}\pi_{\mathbf{b}}
	\varphi_{\mathbf{p}}\varphi_{\mathbf{c-p}})
	-\Delta(\pi_{\mathbf{a}}\pi_{\mathbf{b}})
	\Delta(\varphi_{\mathbf{p}}\pi_{\mathbf{c-p}})\Big] + (\mathbf{a \to b\to c})\Bigg\}
	\nonumber\\
	+&\frac{\mu}{2}\tilde{\delta}_{\mathbf{a,b,c}}
	\,.
	\label{phi3-cubic-EoM}
\end{align}
Here, we set \(\hbar = 1\). The notation \((\mathbf{a \leftrightarrow b})\) denotes a term identical to the one preceding it but with the momenta \(\mathbf{a}\) and \(\mathbf{b}\) interchanged. Similarly, \((\mathbf{k \rightarrow p\rightarrow q})\) 
indicates two additional terms, obtained from the preceding term by cyclically permuting the momenta \(\mathbf{k}, \mathbf{p}\), and \( \mathbf{q} \) .

Finally, we note down the equations of motion of the classical variables \(\Eval{\varphi_\mathbf{a}}\) and \(\Eval{\pi_\mathbf{a}}\):
\begin{align}
\dot{\Eval{\varphi_\mathbf{a}}} &= \Eval{\pi_\mathbf{a}},\nonumber\\
\dot{\Eval{\pi_\mathbf{a}}} &= -\omega_a^2\Eval{\pi_\mathbf{a}} - \mu\int_k\Bigg[\Eval{\varphi_\mathbf{k}}\Eval{\varphi_\mathbf{a-k}} + \Delta(\varphi_\mathbf{k}\varphi_\mathbf{a-k})\Bigg].\label{eq:eomclassical}
\end{align}
These are obtained by the Poisson brackets
\[\{\Eval{\varphi_\mathbf{a}}, \Eval{\pi_\mathbf{b}}\} = \tilde\delta_\mathbf{a,b}, \qquad \{\Eval{\varphi_\mathbf{a}}, \Delta(\cdots)\} = \{\Eval{\pi_\mathbf{a}}, \Delta(\cdots)\} = 0,\]
in view of the effective Hamiltonian \(H_Q\) \eqref{eq:HamilQ}. 
Notably, as opposed to Eqs. \eqref{eq:eom} that rely on the assumption of \(\Eval{\varphi_\mathbf{a}} = \Eval{\pi_\mathbf{a}} = 0\), Eq. \eqref{eq:eomclassical} does not make this assumption. Nevertheless, when translational and rotational invariances are imposed, it will follow that \(\Eval{\varphi_\mathbf{a}} = \Eval{\pi_\mathbf{a}} = 0\)
for any momentum, such that the classical variables in Eq. \eqref{eq:eomclassical} will vanish. We will prove this in two steps, respectively in Section \ref{trans-rot-sec} and \ref{sec:quanteffect}.

\subsection{Translational and rotational invariance}
\label{trans-rot-sec}
Although specific realizations of field configurations might not always be homogeneous or isotropic (for instance, our universe has inhomogeneous structures),
we generally expect the quantum average of our system to exhibit translational 
and rotational invariance,
provided that the initial conditions adhere to these symmetries.
This assumption leads to significant simplifications in the effective equations of motion in three ways:
\begin{enumerate}
\label{list-symmetry}
\item At the leading order,
a given fluctuation couples to only a finite number of other fluctuations. In particular, $\delta$-functions arise in such a way that at $\mu^1$-order --- with no ``loop'' integrals
--- terms in the equations of motion will be proven to be local in the Fourier space despite objects like $\Delta(\varphi_{\mathbf{k}}\varphi_{\mathbf{p}}\varphi_{\mathbf{q}})$ originally triply integrated.
\item The Weyl-ordered expectation values within second and third order fluctuations remain real, although \(\varphi_{\mathbf{k}}\) and \(\pi_{\mathbf{k}}\) are not Hermitian operators.
\item For quantum averages associated with the inhomogeneous part of the field, terms proportional to
\(\Eval{\varphi_{\mathbf{k}}}\) can be disregarded in the equations of motion.
\end{enumerate}
The last simplification above leads to minimal
backreaction,
a crucial requirement for
(quantum) perturbation theory to be consistent
in cosmology.
Similar to the discussion in
Section \ref{sec:backreact},
this simplification enables us to decouple quantum
fluctuations from the classical background, while minimizing the growth of inhomogeneities
of the background,
thus, preserving a homogeneous background metric.
The second simplification will be important
for establishing a connection between our
effective Hamiltonian and traditional
effective potentials.

We showcase the above three simplifications
using our 
\(\varphi^3\)-model.
We first show that simplifications 
(1) to (3) 
are the consequences of translationally and rotationally invariant initial conditions. 
Then, we will argue that these simplifications are maintained throughout the evolution generated by \(H_{\text{Mink.},\Delta}\).

For any expectation value that is translationally and rotationally invariant, we find that for arbitrary products of Fourier-space operators \(A, B, C, \ldots\):
\begin{equation}
\Eval{A(\mathbf{k_1})B(\mathbf{k_2})C(\mathbf{k_3})\dots}
= (2\pi)^3\delta^3(\mathbf{k_1+k_2+k_3\dots})D(k_1,k_2,k_3,\dots; \{\mathbf{k}_i\cdot \mathbf{k}_j\})\,.
\label{homo-iso}
\end{equation}
For cubic-products, rotational invariance makes the function $D$ on the RHS depend only on the magnitudes of the three relevant momenta, i.e. \(D=D(k,p,q)\). Nonetheless,
for any higher power of product operators,
the function $D$ may depend also on the angles between certain momenta, as indicated by the collective notation \(\{\mathbf{k}_i\cdot \mathbf{k}_j\}\).
As a specific case of \eqref{homo-iso}, modes with non-vanishing momenta obey
\(\Eval{\varphi_{\mathbf{k}}}=0\)
for \(\mathbf{k} \neq \mathbf{0}\).
These modes represent the inhomogeneous contributions.
To carry our discussion over to inhomogeneities in cosmology,
we will proceed with the assumption that all external momenta 
\(\mathbf{a, b, c, d,}\ldots\) are non-zero
in any of the following analyses.

The first question one can ask is:
Do our effective equations of motion preserve translational
and rotational invariance in the sense of \eqref{homo-iso}?
Yes.
For linear operators,
\begin{align*}
\dot{\Eval{\varphi_{\mathbf{a}}}}=&\Eval{\pi_{\mathbf{a}}}\\
	\dot{\Eval{\pi_{\mathbf{a}}}}=& -\omega_a^2{\Eval{\varphi_{\mathbf{a}}}}
 -\mu\int_{p,q}\tilde{\delta}_{-\mathbf{a},\mathbf{p},\mathbf{q}}
 \Eval{\varphi_{\mathbf{p}}}
 \Eval{\varphi_{\mathbf{q}}}
	-\mu\int_{p,q}\tilde{\delta}_{\mathbf{-a,p,q}}
	\Delta(\varphi_{\mathbf{p}}\varphi_{\mathbf{q}})\,.
\end{align*}
If the initial conditions satisfy \eqref{homo-iso}, i.e., 
$\Eval{\varphi_\mathbf{a}}_{t = 0}$,
$\Eval{\pi_\mathbf{a}}_{t = 0}\propto\tilde{\delta}_\mathbf{a}=\tilde{\delta}_{\mathbf{a,0}}$
and $\Delta_{t=0}(\varphi_\mathbf{p}\varphi_\mathbf{q})\propto\tilde{\delta}_\mathbf{p,q}$, 
the background variables
$\Eval{\varphi_\mathbf{a}}$
and
$\Eval{\pi_{\mathbf{a}}}$
will not evolve for $\mathbf{a}\neq0$ (the inhomogeneous part of the field)
because its initial rate of change will be proportional to
$\tilde{\delta}_{\mathbf{-a,p,q}}\tilde{\delta}_{\mathbf{p,q}}\sim \tilde{\delta}_{\mathbf{a,0}}=0$.
Therefore,
we also have $\Eval{\varphi_{\mathbf{a}}}= 0$ for $\mathbf{a}\neq0$
at the next ``time-step''.
Physically, spatial symmetry guarantees
that there is no backreaction
from quantum fluctuations that
promote growth in background inhomogeneity ---
an important consistency condition for our
later application to cosmology. Furthermore, we will show that we also have $\Eval{\varphi_\mathbf{0}} = 0$ in Section \ref{sec:quanteffect}, so all terms in the equations of motion proportional to $\Eval{\varphi_\mathbf{0}}$ can be disregarded.

Let us now examine the quadratic and cubic fluctuations
based on equation sets \eqref{phi3-quad-EoM} and \eqref{phi3-cubic-EoM}.
We hope to see if they still maintain
their original delta-functions
such that the form \eqref{homo-iso}
is preserved.
We shall analyze, say,
$\dot{\Delta}(\varphi_{\mathbf{a}}\pi_{\mathbf{b}})$
and $\dot{\Delta}(\varphi_{\mathbf{a}}\varphi_{\mathbf{b}}\pi_{\mathbf{c}})$
as representatives.

Consider the equations of motion for
$\dot{\Delta}(\varphi_{\mathbf{a}}\pi_{\mathbf{b}})$
and $\dot{\Delta}(\varphi_{\mathbf{a}}\varphi_{\mathbf{b}}\pi_{\mathbf{c}})$ in
\eqref{phi3-quad-EoM} and \eqref{phi3-cubic-EoM}.
The terms without explicit
$\mu$ factors are all initially proportional to
$\tilde{\delta}_{\mathbf{a,b}}$
or $\tilde{\delta}_{\mathbf{a,b,c}}$
if they started from initial conditions \eqref{homo-iso}.
(The terms proportional to
$\Eval{\varphi}$ are absent
because we have argued that
linear expectation values vanish initially.)
Their integral terms 
(with the explicit $\mu$ factor)
are all proportional to delta-functions
of sums over external momenta:
\begin{align*}
	\text{Integral term of }
	\dot{\Delta}(\varphi_{\mathbf{a}}\pi_{\mathbf{b}})
	=&-\mu\int_{k,p,q}\tilde{\delta}_{\mathbf{k,p,q}}
	\tilde{\delta}_{\mathbf{b,k}}
	\Delta(\varphi_{\mathbf{a}}\varphi_{\mathbf{p}}\varphi_{\mathbf{q}})
	\sim
	\mu\int_{k,p,q}\tilde{\delta}_{\mathbf{k,p,q}} \tilde{\delta}_{\mathbf{b,k}}
	\tilde{\delta}_{\mathbf{a,p,q}}
	(\dots)\\
	&\sim \int_{k} \tilde{\delta}_{\mathbf{a,b}}(\dots),
\end{align*}
\begin{align*}
	\text{Integral term of }
	\dot{\Delta}(\varphi_{\mathbf{a}}\varphi_{\mathbf{b}}\pi_{\mathbf{c}})
	=&-\mu\int_{k,p,q}
	\tilde{\delta}_{\mathbf{k,p,q}}
	\tilde{\delta}_{\mathbf{c,k}}
	\Big(\Delta(\varphi_{\mathbf{a}}\varphi_{\mathbf{b}}
	\varphi_{\mathbf{p}}\varphi_{\mathbf{q}})
	-\Delta(\varphi_{\mathbf{a}}\varphi_{\mathbf{b}})
	\Delta(\varphi_{\mathbf{p}}\varphi_{\mathbf{q}})
	\Big)\\
	&\hspace{-80pt}\sim 
	\int_{k,p,q}\tilde{\delta}_{\mathbf{k,p,q}}
	\tilde{\delta}_{\mathbf{c,k}}
	(\tilde{\delta}_{\mathbf{a,b,p,q}}
	-\tilde{\delta}_{\mathbf{a,b}}\delta_{\mathbf{p,q}})
	(\dots)
	\sim \int_{k}\Big(\tilde{\delta}_{\mathbf{-a,-b,-c}}
	+\tilde{\delta}_{\mathbf{c,0}}\delta_{\mathbf{a,b}}\Big)
	(\dots)\\
	&\hspace{-80pt}\sim \int_k\tilde{\delta}_{\mathbf{-a,-b,-c}}(\dots)\,.
\end{align*}
The bracket $(\dots)$ consists of the functions $D$ and/or their products.
Thus, we see that the quantum fluctuations for inhomogeneities are proportional to the correct delta-functions
and \eqref{homo-iso} is preserved.

Note that in the above discussion, we have not made any assumptions about the fluctuations
aside from them initially obeying \eqref{homo-iso}.
If they satisfy the further assumption of obeying Wick's theorem (say, at the leading order in $O(\mu)$),
one can strip away some of the above Fourier-space integrals.
(Integrals of this type are undesirable in an equation of motion
for they make the evolution non-local in the Fourier space.)
For the evolution of cubic fluctuations,
one can get rid of these integrals if there is a hierarchy between
the fourth-order and second-order fluctuations,
such as $\Delta^{(4)}\sim \Delta^{(2)}\Delta^{(2)}$,
via Wick's theorem.
The integrand is then proportional to the product of two delta-functions
$\delta(\dots)\delta(\dots)$.
Such a simplification can be realized consistently
if one starts with states obeying Wick's theorem
and assumes it is preserved at leading order in $\mu^1$ throughout evolution.
For example,
for
$\dot{\Delta}(\varphi_{\mathbf{a}}\varphi_{\mathbf{b}}\pi_{\mathbf{c}})$,
applying Wick's theorem we have
\begin{align*}
	\text{Integral term of }
	\dot{\Delta}(\varphi_{\mathbf{a}}\varphi_{\mathbf{b}}\pi_{\mathbf{c}})
	=&-\mu\int_{k,p,q}
	\tilde{\delta}_{\mathbf{k,p,q}}
	\tilde{\delta}_{\mathbf{c,k}}
	\Big(\Delta(\varphi_{\mathbf{a}}\varphi_{\mathbf{b}}
	\varphi_{\mathbf{p}}\varphi_{\mathbf{q}})
	-\Delta(\varphi_{\mathbf{a}}\varphi_{\mathbf{b}})
	\Delta(\varphi_{\mathbf{p}}\varphi_{\mathbf{q}})
	\Big)\\
	\approx&-\mu\int_{k,p,q}\tilde{\delta}_{\mathbf{k,p,q}}
	\tilde{\delta}_{\mathbf{c,k}}
	\Big(
	\Delta(\varphi_{\mathbf{a}}\varphi_{\mathbf{b}})
	\Delta(\varphi_{\mathbf{p}}\varphi_{\mathbf{q}})
	+\Delta(\varphi_{\mathbf{a}}\varphi_{\mathbf{p}})
	\Delta(\varphi_{\mathbf{b}}\varphi_{\mathbf{q}})\\
	+&\Delta(\varphi_{\mathbf{a}}\varphi_{\mathbf{q}})
	\Delta(\varphi_{\mathbf{b}}\varphi_{\mathbf{p}})
	-\Delta(\varphi_{\mathbf{a}}\varphi_{\mathbf{b}})
	\Delta(\varphi_{\mathbf{p}}\varphi_{\mathbf{q}})
	\Big)+O(\mu^2)\\
	\approx&-\mu\int_{k,p,q}\tilde{\delta}_{\mathbf{k,p,q}}
	\tilde{\delta}_{\mathbf{c,k}}
	\Big(
	\tilde{\delta}_{\mathbf{a,p}}\tilde{\delta}_\mathbf{b,q}
	(\dots)
	+
	\tilde{\delta}_{\mathbf{a,q}}\tilde{\delta}_\mathbf{b,p}
	(\dots)
	\Big)\\
	\sim&-\mu
	\tilde{\delta}_{\mathbf{a,b,c}}(\dots)\,,
\end{align*}
which is local
in momentum space because the integration is eliminated. This result follows from the fact that our $\varphi^3$-model admits tree-level (i.e. $O(\mu)$ order) three-point vertices. It also follows that one cannot get rid of the
integral over internal momenta for
quadratic fluctuations in
\eqref{phi3-quad-EoM}.
The cubic fluctuations contribute to quadratic fluctuations
at the $O(\mu^2)$ order via ``loop''
integrals because cubic fluctuations
per se are already $O(\mu)$ if they evolved
from Gaussian-type initial conditions
(i.e. ones obeying Wick's theorem).
That is,
the $O(\mu)$-integrals in \eqref{phi3-quad-EoM}
represent quantum contributions to 2-pt functions,
which naturally come with integrals over internal momenta.
Finally, let us emphasize again that maintaining \eqref{homo-iso} does not necessarily require
Wick's theorem. Nevertheless, Wick's theorem does make the equations of motion simpler
by removing some of the Fourier-space integrals.

Let us recap what we have done and argue that spatial symmetries (in the sense of
\eqref{homo-iso}) are preserved throughout time evolution.
We have shown that if an initial configuration obeys
\eqref{homo-iso}: (1) the rate of change for the Fourier-space expectation values 
$\dot{\Eval{\varphi}}$ and $\dot{\Eval{\pi}}$ 
all vanish initially (for inhomogeneous modes), and (2) the rates of change of the fluctuations
are still proportional to the appropriate $\delta$-functions.
That is, if \eqref{homo-iso} is satisfied at some initial time $t=t_0$,
then it can be consistently satisfied at the next ``time step''
$t=t_0+\delta t$.
Since our equations are all first-order in $d/dt$,
iterating this logic for arbitrary time steps,
the preservation of \eqref{homo-iso}
is thus proven for any $t$.

In the end, let us briefly address rotational invariance.
Rotational invariance on top of translational invariance
guarantees that
the functions $D(k,p,q)$
are real functions for Weyl-ordered averages of
product operators that are Hermitian in position space.
Consider for example a Weyl-ordered cubic expectation value
\begin{align*}
	\Eval{A(\mathbf{k})B(\mathbf{p})C(\mathbf{q})}_W^*
	=&\int\int\int d^3x d^3y d^3z
	e^{+i(\mathbf{k}\mathbf{x}+\mathbf{p}\mathbf{y}+\mathbf{q}\mathbf{z})}
	\Eval{A(\mathbf{x})B(\mathbf{y})C(\mathbf{z})}_W\\
	=&\int\int\int d^3x' d^3y' d^3z'
	e^{-i(\mathbf{k}\mathbf{x}'+\mathbf{p}\mathbf{y}'+\mathbf{q}\mathbf{z}')}
	\Eval{A(-\mathbf{x}')B(-\mathbf{y}')C(-\mathbf{z}')}_W\\
	=&\int\int\int d^3x' d^3y' d^3z'
	e^{-i(\mathbf{k}\mathbf{x}'+\mathbf{p}\mathbf{y}'+\mathbf{q}\mathbf{z}')}
	\Eval{A(\mathbf{x}')B(\mathbf{y}')C(\mathbf{z}')}_W\\
	=&
	\Eval{A(\mathbf{k})B(\mathbf{p})C(\mathbf{q})}_W\,.
\end{align*}
In the first line, since the Weyl-ordering was symmetric to begin with, it ensures that a complex-conjugation
on the expectation value did not reshuffle the operator ordering.
We renamed our integration variables
in position-space in the second line and
used rotational and translational invariance in the third line.
Specifically, rotation together with translation can mimic the operation of parity,
inverting $(\mathbf{-x',-y',-z'})\rightarrow(\mathbf{x',y',z'})$,
for a triangle (whose vertices are located at $(\mathbf{x',y',z'})$)
representing a 3-point correlator. 
An analogous statement is true for an edge that represents a 2-pt correlator,
helping us prove the realness of $D$ for second-order fluctuations.
(To generalize the proof to $n$-pt correlators with $n\geq 4$,
one would have to impose symmetry under parity.)

We have therefore proven simplifications (1) to (3).
These simplifications
remove in the equations of motion terms linear in $\Eval{\varphi}$ and $\Eval{\pi}$ 
---
representing backreaction in our model ---
and strip away integrals over internal momenta at tree level.
Conversely,
classical variables associated with 
background inhomogeneities
will also decouple from quantum fluctuations,
retaining the fixed background geometry.
It is also worth noting the realness property of fluctuations will also help us find minimal-energy initial conditions in Section \ref{UP-IV-phi3}.

\subsection{Uncertainty principle and energy-minimization}
\label{UP-IV-phi3}
We need a set of initial conditions
for fluctuations. It is natural to consider
minimal-energy configurations 
as the prime candidates.
To find them, we look for the set of
fluctuation-values that minimize
$H_{\Delta}$.
Nevertheless, quantum fluctuations must obey
the uncertainty principle
(including its generalization to higher-order
fluctuations).
We must derive them in our context.
We will then see that for a system
with potential $V(\varphi)$,
the minimized effective Hamiltonian contains
the Coleman-Weinberg effective potential.
In particular,
the saturation of the uncertainty principle
will be a result of requiring the system
to have minimal energy.
This is true even for interacting fields
when quantum fluctuations are kept to quadratic
order.
For systems with no interactions,
the set of initial values for fluctuations
reduces to the familiar one in
flat-spacetime quantum field theory.

First, we derive the Weyl-ordered version of the well-known quadratic uncertainty principle. Given an arbitrary state \(|\psi\rangle\):
\begin{equation}
    \varphi_{\mathbf{-k}}|\psi\rangle = |u\rangle, \quad
    \pi_{\mathbf{-k}}|\psi\rangle = |v\rangle.
    \nonumber
    \label{}
\end{equation}
Using the Cauchy-Schwarz inequality for
\(|u\rangle\) and \(|v\rangle\), we have
\begin{align}
    \Eval{\varphi_{\mathbf{k}}\varphi_{\mathbf{-k}}}
    \Eval{\pi_{\mathbf{k}}\pi_{\mathbf{-k}}}
    \geq& |\Eval{\varphi_{\mathbf{k}}\pi_{\mathbf{-k}}}|^2 \nonumber\\
    =& \frac{1}{4}\Eval{\varphi_{\mathbf{k}}\pi_{\mathbf{-k}}
    +\pi_{\mathbf{-k}}\varphi_{\mathbf{k}}+i\hbar\tilde{\delta}(\mathbf{0})}
    \Eval{\varphi_{\mathbf{k}}\pi_{\mathbf{-k}}
    +\pi_{\mathbf{-k}}\varphi_{\mathbf{k}}+i\hbar\tilde{\delta}(\mathbf{0})}^*
    \nonumber\\
    =&|\Eval{\varphi_{\mathbf{k}}\pi_{\mathbf{-k}}}_W|^2
    +\mathrm{Im}[\Eval{\varphi_{\mathbf{k}}\pi_{\mathbf{-k}}}_W]
    +\frac{\hbar^2}{4}\tilde{\delta}(\mathbf{0})^2\nonumber\\
    =&\Eval{\varphi_{\mathbf{k}}\pi_{\mathbf{-k}}}_W^2
    +\frac{\hbar^2}{4}\tilde{\delta}(\mathbf{0})^2
    \text{ (if translationally and rotationally invariant)}.
    \label{Weyl-UP}
\end{align}
In the last line, we used the realness of 
\(\Eval{\varphi_{\mathbf{k}}\pi_{\mathbf{p}}}_W\),
which relies on rotational and translational invariance.
The appearance of \(\tilde{\delta}(\mathbf{0})\equiv(2\pi)^3\delta^3(\mathbf{0})\)
originates from the commutator
\([\varphi_{\mathbf{k}},\pi_{\mathbf{-k}}]=i(2\pi)^3\hbar\delta^3(\mathbf{k-k})\).

For inhomogeneous degrees of freedom
--- those with momenta \(\mathbf{k}\neq 0\) and thus
\(\Eval{\varphi_{\mathbf{k}}} = 0\)
--- the fluctuations coincide with Weyl-ordered expectation values (e.g., \(\Delta(\varphi_{\mathbf{k}}\varphi_{\mathbf{-k}}) = \Eval{\varphi_{\mathbf{k}}\varphi_{\mathbf{-k}}}_W\)). This generalizes to all combinations of product operators. Hence,
\begin{equation}
    \Delta(\varphi_{\mathbf{k}}\varphi_{\mathbf{-k}})
    \Delta(\pi_{\mathbf{k}}\pi_{\mathbf{-k}})
    \geq \Delta^2(\varphi_{\mathbf{k}}\pi_{\mathbf{-k}})
    +\frac{\hbar^2}{4}\tilde{\delta}(\mathbf{0})^2.
\end{equation}

\textit{Low-energy effective potential}.
Let us now demonstrate how one could derive the
Coleman-Weinberg effective potential as a low-energy
limit of our effective Hamiltonian.
Consider a scalar-field $\varphi$
with a canonical kinetic term and some arbitrary potential
$V(\varphi)$.
At second-order in fluctuations,
the effective Hamiltonian for our field theory is
\begin{align*}
	\Eval{H} &=
	H[\Eval{\varphi},\Eval{\pi}]
	+\int\frac{d^3k}{(2\pi)^3}
	\Big[\frac{1}{2}(
	\Delta(\pi_{\mathbf{k}}\pi_{\mathbf{-k}})
 +k^2 \Delta(\varphi_{\mathbf{k}}\varphi_{\mathbf{-k}}))
	+\frac{1}{2}V''(\Eval{\varphi})
	\Delta(\varphi_{\mathbf{k}}\varphi_{\mathbf{-k}})
	\Big],\\
	[\varphi_{\mathbf{k}},\pi_{\mathbf{p}}] &= i(2\pi)^3\hbar\delta^3(\mathbf{k+p})\,.
\end{align*}
We will write $V''(\Eval{\varphi})$ as $V''$ in what follows.

Consider the fluctuation part of the effective Hamiltonian
\begin{align}
	H_{\Delta}=&
	\int\frac{d^3k}{(2\pi)^3}
	\frac{1}{2}\Big[
	\Delta(\pi_{\mathbf{k}}\pi_{\mathbf{-k}})
	+(k^2+V'')
	\Delta(\varphi_{\mathbf{k}}\varphi_{\mathbf{-k}})
	\Big]\nonumber\\
	\geq&
	\int\frac{d^3k}{(2\pi)^3}
	\sqrt{k^2+V''}\sqrt{\Delta(\pi_{\mathbf{k}}\pi_{\mathbf{-k}})
	\Delta(\varphi_{\mathbf{k}}\varphi_{\mathbf{-k}})}\nonumber\\
	\geq&
	\int\frac{d^3k}{(2\pi)^3}
	\sqrt{k^2+V''}
	\sqrt{\Delta^2(\varphi_{\mathbf{k}}\pi_{\mathbf{-k}})+\frac{\hbar^2}{4}
	(2\pi)^6(\delta^3(\mathbf{k-k}))^2}\,.
 \label{min-Heff}
\end{align}
The first inequality uses 
$\Delta(\pi_{\mathbf{k}}\pi_{\mathbf{-k}}),(k^2+V'')\Delta(\varphi_{\mathbf{k}}\varphi_{\mathbf{-k}})\geq 0$,
which applies if our field is near
the true vacuum.
The second inequality is the uncertainty principle,
where the realness of
$\Delta(\varphi_{\mathbf{k}}\pi_{\mathbf{-k}})$ is guaranteed by spatial symmetries.
The second inequality also implies that to saturate the lower bound of $H_{\Delta}$,
the uncertainty principle must, too, be saturated.
In this sense,
the saturation of the uncertainty principle is a necessary consequence of minimizing the effective Hamiltonian when quantum fluctuations are kept to the second order.

The last line of \eqref{min-Heff} tells us that effective Hamiltonian is minimized
in the absence of field-momentum correlations, i.e.,
$\Delta(\varphi_{\mathbf{k}}\pi_{\mathbf{-k}})=0$,
analogous to \eqref{QM-min-EP} in quantum mechanics.
The minimized effective Hamiltonian $\Eval{H}$
then becomes
\begin{equation}
	\RM{Min}[\Eval{H}]
	=H[\Eval{\varphi},\Eval{\pi}]
	+(2\pi)^3\delta^3(\mathbf{0})
	\frac{\hbar}{2}\int \frac{d^3k}{(2\pi)^3}\sqrt{k^2+V''}\,.
	\label{CW-field}
\end{equation}
The $\delta^3(\mathbf{0})\sim \RM{Vol}[\mathbb{R}^3]$,
the volume in position space.
Divided by the volume,
the integral term of \eqref{CW-field}
can been shown \cite{bojowald2014canonical} to be equivalent to the covariant form
of the Coleman-Weinberg potential (density) up to constants of $\Eval{\varphi}$:
\begin{equation}
	V_{\RM{CW}}=\frac{i\hbar}{2}\int \frac{d^4k}{(2\pi)^4}
	\log\Big(1+\frac{V''}{\eta_{\mu\nu}k^{\mu}k^{\nu}}\Big)\,,\;
	\eta_{\mu\nu}=(-1,1,1,1)
	\label{}
\end{equation}
after integrating $\int dk^0$.

Let us remark on the comparison
with the past derivations of effective potentials
using fluctuations
\cite{bojowald2014canonical,baytacs2019effective,
baytacs2020faithful}. These works either make a priori assumptions of the saturation of the uncertainty 
principle and adiabatic approximations
in the evolution
or rely on canonical maps.
What is new here is that we make none of these assumptions nor
rely on any canonical maps (as they have not been found for fields).
The saturation of the uncertainty principle in our work is a \emph{consequence} of
requiring the effective Hamiltonian to be stable to fluctuations.
The uncertainty principle underpins how inequalities should be used
in a constrained minimization problem.
It, along with translational and rotational symmetries,
directly permits us to derive the low-energy effective potential.
The consistency between these symmetries and the evolution of fluctuations
is non-trivial and needs to be checked,
as we have done in
section \ref{trans-rot-sec}.
Physically, these symmetries lead to minimal backreaction in our framework
and legitimate using a fixed background metric in the cosmological application in
Section \ref{sec-dS-fields}.

\textit{Finding initial conditions}.
In an operator formalism, one often starts with an initial vacuum of the 
``free" Hamiltonian when the scattering particles are far away enough 
to not feel any interactions
\footnote{In cosmological applications,
one cannot assume initially vanishing interactions on physical grounds.
Nonetheless, in practice, this is often still assumed implicitly through an \(i\varepsilon\)-deformation of the contour in time. It is the opinion of the authors
that the assumption is in most applications
a reasonable compromise when 
the construction and calculations involving an interacting vacuum 
in curved spacetime is too difficult.
This difficulty is exacerbated by the gauge/constraint aspects of quantization when higher-order metric perturbations are involved.}. 
In our effective formalism,
initial conditions can be found in alignment with this philosophy.

To conform with the spirit of traditional quantum field theory,
we look for a set of initial values for fluctuations that minimize
the energy of the ``free'' part of our Hamiltonian,
defined as the part with quadratic field products.
For our $\varphi^3$ example,
it corresponds to the first line of
\eqref{eq:HamilDelta}.
We will denote the effective Hamiltonian of the ``free'' part as
\(H_{\mathrm{Mink.},\Delta^{(2)}}\).
To minimize
\(H_{\mathrm{Mink.},\Delta^{(2)}}\), we seek initial fluctuations that satisfy all inequalities while minimizing \(\sqrt{\Delta^2(\varphi_{\mathbf{k}}\pi_{\mathbf{-k}}) + \tilde{\delta}(\mathbf{0})^2/4}\). The set of fluctuations meeting this requirement is
\begin{align*}
    &\Delta(\pi_{\mathbf{k}}\pi_{\mathbf{-k}})
    =\omega_k^2\Delta(\varphi_{\mathbf{k}}\varphi_{\mathbf{-k}})
    =\omega_k\sqrt{\Delta^2(\varphi_{\mathbf{k}}\pi_{\mathbf{-k}})
    +\frac{1}{4}\tilde{\delta}(\mathbf{0})^2},\\
    &\Delta(\varphi_{\mathbf{k}}\pi_{\mathbf{-k}})=0.
\end{align*}
In more familiar terms:
\begin{equation}
    \Delta(\varphi_{\mathbf{k}}\varphi_{\mathbf{-p}})
    =\frac{1}{2\omega_k}\tilde{\delta}_\mathbf{k,p},\qquad
    \Delta(\pi_{\mathbf{k}}\pi_{\mathbf{-p}})
    =\frac{\omega_k}{2}\tilde{\delta}_\mathbf{k,p},\qquad
    \Delta(\varphi_{\mathbf{k}}\pi_{\mathbf{-p}})
    =0 \times \tilde{\delta}_\mathbf{k,p}.
    \label{Mink-IV}
\end{equation}
This corresponds to what one would have obtained
using the conventional operator formalism 
for a free-scalar vacuum in Minkowski spacetime, with
\begin{equation}
    \varphi_{\mathbf{k}}
    =\frac{1}{\sqrt{2\omega_k}}e^{-i\omega_kt} a_{\mathbf{k}}
    +\frac{1}{\sqrt{2\omega_k}}e^{+i\omega_kt} a_{\mathbf{-k}}^{\dagger},\;
    \pi_{\mathbf{k}}=\dot{\varphi}_{\mathbf{k}}.
    \label{oprt-soln-free}
\end{equation}
Here, the time-independent annihilation operator \(a_{\mathbf{k}}\) annihilates the Minkowski vacuum \(|0\rangle\) associated with the \(t\)-slicing.

Note that the initial condition \eqref{Mink-IV}, obtained solely through the minimal energy condition, exhibits spatial homogeneity and isotropy. Utilizing the findings we derived in Section \ref{trans-rot-sec}, the field will maintain its spatial homogeneity and isotropy throughout its evolution. In other words, the minimal energy condition implies spatial symmetry.

Appending initial conditions \eqref{Mink-IV} with an initial Gaussian hierarchy, where Wick's theorem applies to the fluctuations, 
one obtains a complete set of initial conditions for evolving arbitrary orders of fluctuations
\(\Delta^{(n)}\).

\subsection{Solutions}\label{flat-solution-sec}

Having established the equations of motion \eqref{phi3-quad-EoM} and \eqref{phi3-cubic-EoM}, along with the homogeneous and isotropic initial conditions \eqref{Mink-IV}, we are now equipped to derive effective solutions for the $\varphi^3$-system perturbatively. We
will focus on the $\mu^1$-order solutions for 
quadratic and cubic fluctuations, $\Delta^{(2)}$ and $\Delta^{(3)}$. We'll also touch upon the quantum contributions in this section. A thorough analysis of effective solutions at various orders will be reported elsewhere.

To determine the equations of motion for $\Delta^{(n)}$, which involves $n$ operators of $\varphi$ and/or $\pi$ fields, we assume a perturbative expansion of $\Delta^{(n)}$ based on powers of $\mu$:
$$\Delta^{(n)} = \Delta^{(n)}_0 + \Delta^{(n)}_1 + \Delta^{(n)}_2 + \cdots,$$
where $\Delta^{(n)}_k$ denotes the component of $\Delta^{(n)}$ that is of the order $\mu^k$. Following our previous discussion, once all $\Delta_0^{(n)}$ values --- the quantum fluctuations of a free scalar field at $O(\mu^0)$ --- have been determined, it becomes feasible to systematically compute $\Delta^{(n)}_k$ at any given order
in $\mu$.

\subsubsection{Quantum fluctuations for free theory}\label{sec:freelevel}

For $\mu^0$-order, $\Delta^{(n)}_0$'s adhere to the following free equations of motion:
\begin{align}
\dot\Delta_0(\varphi_{\mathbf{a}_1}\cdots\varphi_{\mathbf{a}_m}\pi_{\mathbf{b}_1}\cdots\pi_{\mathbf{b}_n}) &= \sum_{i = 1}^m\Delta_0(\varphi_{\mathbf{a}_1}\cdots\varphi_{\mathbf{a}_{i-1}}\pi_{\mathbf{a}_i}\varphi_{\mathbf{a}_{i+1}}\cdots\varphi_{\mathbf{a}_m}\pi_{\mathbf{b}_1}\cdots\pi_{\mathbf{b}_n})\nonumber\\
&-\sum_{j = 1}^n\omega_{b_j}^2\Delta_0(\varphi_{\mathbf{a}_1}\cdots\varphi_{\mathbf{a}_m}\pi_{\mathbf{b}_1}\cdots\pi_{\mathbf{b}_{j-1}}\varphi_{\mathbf{b}_j}\pi_{\mathbf{b}_{j+1}}\cdots\pi_{\mathbf{b}_n}),\label{eq:eomfree}
\end{align}
The initial conditions \eqref{Mink-IV} lead to static solutions for both quadratic and cubic fluctuations
such that at any time $t$:
\begin{equation}\label{eq:solfreequadratic}
    \Delta_0(\varphi_{\mathbf{a}}\varphi_{\mathbf{b}}; t)
    = \frac{1}{2\omega_a}\tilde{\delta}_{\mathbf{a,b}},\qquad
    \Delta_0(\varphi_{\mathbf{a}}\pi_{\mathbf{b}}; t)
    = 0,\qquad
    \Delta_0(\pi_{\mathbf{a}}\pi_{\mathbf{b}}; t)
    = \frac{\omega_a}{2}\tilde{\delta}_{\mathbf{a,b}}.
\end{equation}
These static solutions are a natural consequence of the initial conditions \eqref{Mink-IV}, corresponding to the vacuum state $\ket{0}$ of the free scalar field, which is an eigenstate of the free Hamiltonian and hence remains static.

Given the vacuum state $\ket{0}$, the higher-order fluctuations at \(\mu^0\)-order can be derived from $\Delta_0^{(2)}$ \eqref{eq:solfreequadratic} using Wick's theorem. For instance, the cubic fluctuations $\Delta_0^{(3)}$ are all $0$:
\begin{equation}\label{eq:solfreecubic}
\Delta_0(\varphi_\mathbf{a}\varphi_\mathbf{b}\varphi_\mathbf{c}; t) 
= \Delta_0(\varphi_\mathbf{a}\varphi_\mathbf{b}\varphi_\mathbf{c}; t)
= \Delta_0(\varphi_\mathbf{a}\varphi_\mathbf{b}\varphi_\mathbf{c}; t)
= \Delta_0(\varphi_\mathbf{a}\varphi_\mathbf{b}\varphi_\mathbf{c}; t) = 0,
\end{equation}
and the quartic ones are:
\begin{align}
	\Delta_0\left(\varphi_\mathbf{a}\varphi_\mathbf{b}\varphi_\mathbf{c}\varphi_\mathbf{d}; t\right) &= \Delta_0\left(\varphi_\mathbf{a}\varphi_\mathbf{b}\right)\Delta_0\left(\varphi_\mathbf{c}\varphi_\mathbf{d}\right) + \Delta_0\left(\varphi_\mathbf{a}\varphi_\mathbf{c}\right)\Delta_0\left(\varphi_\mathbf{b}\varphi_\mathbf{d}\right) + \Delta_0\left(\varphi_\mathbf{a}\varphi_\mathbf{d}\right)\Delta_0\left(\varphi_\mathbf{b}\varphi_\mathbf{c}\right)\nonumber\\
	&= \frac{1}{4\omega_a\omega_c}\tilde\delta_\mathbf{a,b}\tilde\delta_\mathbf{c,d} + \frac{1}{4\omega_a\omega_b}\tilde\delta_\mathbf{a,b}\tilde\delta_\mathbf{c,d} + \frac{1}{4\omega_a\omega_d}\tilde\delta_\mathbf{a,b}\tilde\delta_\mathbf{c,d},\nonumber\\ 
	\Delta_0\left(\varphi_\mathbf{a}\varphi_\mathbf{b}\pi_\mathbf{c}\pi_\mathbf{d}; t\right) &= \Delta_0\left(\varphi_\mathbf{a}\varphi_\mathbf{b}\right)\Delta_0\left(\pi_\mathbf{c}\pi_\mathbf{d}\right) = \frac{\omega_c}{4\omega_a}\tilde\delta_\mathbf{a,b}\tilde\delta_\mathbf{c,d},\nonumber\\ 
	\Delta_0\left(\pi_\mathbf{a}\pi_\mathbf{b}\pi_\mathbf{c}\pi_\mathbf{d}; t\right) &= \Delta_0\left(\pi_\mathbf{a}\pi_\mathbf{b}\right)\Delta_0\left(\pi_\mathbf{c}\pi_\mathbf{d}\right) + \Delta_0\left(\pi_\mathbf{a}\pi_\mathbf{c}\right)\Delta_0\left(\pi_\mathbf{b}\pi_\mathbf{d}\right) + \Delta_0\left(\pi_\mathbf{a}\pi_\mathbf{d}\right)\Delta_0\left(\pi_\mathbf{b}\pi_\mathbf{c}\right)\nonumber\\
	&= \frac{\omega_a\omega_c}{4}\tilde\delta_\mathbf{a,b}\tilde\delta_\mathbf{c,d} + \frac{\omega_a\omega_b}{4}\tilde\delta_\mathbf{a,b}\tilde\delta_\mathbf{c,d} + \frac{\omega_a\omega_d}{4}\tilde\delta_\mathbf{a,b}\tilde\delta_\mathbf{c,d}\nonumber \\
\Delta_0(\varphi_\mathbf{a}\varphi_\mathbf{b}\varphi_\mathbf{c}\pi_\mathbf{d}; t) &= \Delta_0(\varphi_\mathbf{a}\pi_\mathbf{b}\pi_\mathbf{c}\pi_\mathbf{d}; t) =  0.\label{eq:solfreequartic}
\end{align}
These constants are indeed static solutions to the free equations of motion \eqref{eq:eomfree}.

\subsubsection{Quantum fluctuations at tree-level}\label{sec:treelevel}

The equations of motion \eqref{phi3-quad-EoM} indicate that the evolutions of quadratic fluctuations
\(\Delta_1^{(2)}\)\
of order $\mu$ depend on the free cubic fluctuations
\(\Delta_0^{(3)}\). Fortunately, the free cubic \(\Delta^{(3)}_0 = 0\) at \(\mu^0\)-order. Therefore, the subleading terms of $\Delta^{(2)}$ are at least of order \(\mu^2\):
\begin{equation*}
    \Delta^{(2)} = \Delta^{(2)}_0 + O(\mu^2).
\end{equation*}
Therefore, for tree-level solutions of \(\Delta^{(2)}\) up to $\mu$-order, focusing on the \(\mu^0\)-order --- the free solution $\Delta^{(2)}_0$ \eqref{eq:solfreequadratic} --- is sufficient.

Subsequently, with the \(\mu^0\)-order solutions for \(\Delta_0^{(2)}\)\,s and \(\Delta_0^{(4)}\)\,s determined in \eqref{eq:solfreequadratic} and \eqref{eq:solfreequartic}, we can now apply them to \eqref{phi3-cubic-EoM} to solve for \(\Delta_1^{(3)}\). The equations of motion for $\Delta^{(3)}_1$, ignoring $O(\mu^2)$ terms, are:
\begin{align*}
\dot\Delta_1(\varphi_\mathbf{a}\varphi_\mathbf{b}\varphi_\mathbf{c}) &= \Delta_1(\pi_\mathbf{a}\varphi_\mathbf{b}\varphi_\mathbf{c}) + \Delta_1(\varphi_\mathbf{a}\pi_\mathbf{b}\varphi_\mathbf{c}) + \Delta_1(\varphi_\mathbf{a}\varphi_\mathbf{b}\pi_\mathbf{c}), \\
    \dot\Delta_1(\varphi_\mathbf{a}\varphi_\mathbf{b}\pi_\mathbf{c}) &= \Delta_1(\pi_\mathbf{a}\varphi_\mathbf{b}\pi_\mathbf{c}) + \Delta_1(\varphi_\mathbf{a}\pi_\mathbf{b}\pi_\mathbf{c}) - \omega_c^2 \Delta_1(\varphi_\mathbf{a}\varphi_\mathbf{b}\varphi_\mathbf{c}) - \frac{\mu}{2\omega_a\omega_b}\tilde\delta_{k,p,q}, \\
    \dot\Delta_1(\varphi_\mathbf{a}\pi_\mathbf{b}\pi_\mathbf{c}) &= \Delta_1(\pi_\mathbf{a}\pi_\mathbf{b}\pi_\mathbf{c}) - \omega_b^2\Delta_1(\varphi_\mathbf{a}\varphi_\mathbf{b}\pi_\mathbf{c}) - \omega_c^2\Delta_1(\varphi_\mathbf{a}\pi_\mathbf{b}\varphi_\mathbf{c}), \\
    \dot\Delta_1(\pi_\mathbf{a}\pi_\mathbf{b}\pi_\mathbf{c}) &= -\omega_a^2\Delta_1(\varphi_\mathbf{a}\pi_\mathbf{b}\pi_\mathbf{c}) - \omega_b^2\Delta_1(\pi_\mathbf{a}\varphi_\mathbf{b}\pi_\mathbf{c}) - \omega_c^2\Delta_1(\pi_\mathbf{a}\pi_\mathbf{b}\varphi_\mathbf{c}) + \frac{\mu}{2}\tilde\delta_\mathbf{a, b, c}.
\end{align*}

\begin{figure}\centering
\includegraphics[width=15cm]{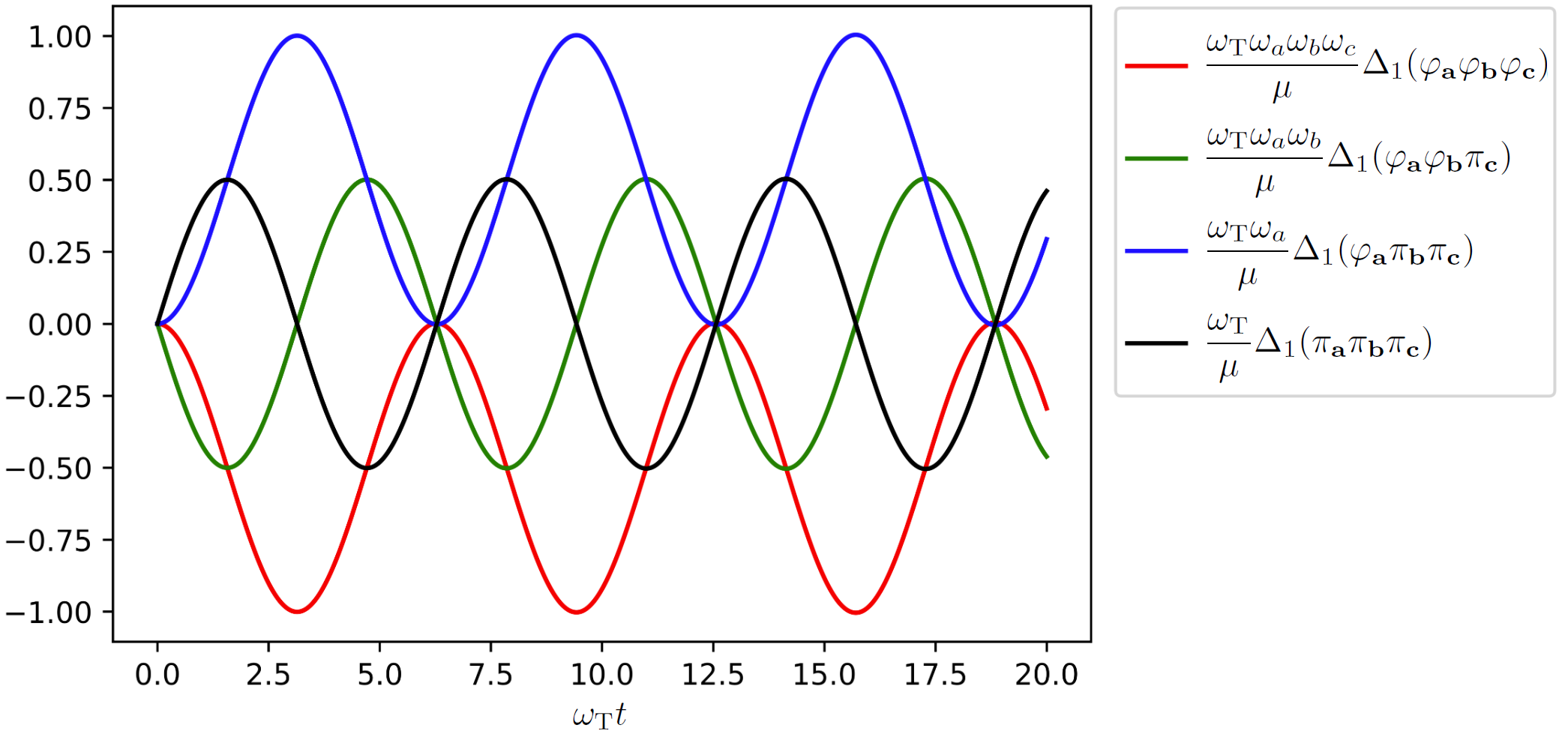}
\caption{Evolutions of $\Delta_1(\varphi_\mathbf{a}\varphi_\mathbf{b}\varphi_\mathbf{c})$, $\Delta_1(\varphi_\mathbf{a}\varphi_\mathbf{b}\pi_\mathbf{c})$, $\Delta_1(\varphi_\mathbf{a}\pi_\mathbf{b}\pi_\mathbf{c})$, and $\Delta_1(\pi_\mathbf{a}\pi_\mathbf{b}\pi_\mathbf{c})$.}
\label{fig:cubicdelta}
\end{figure}

The above set forms a non-homogeneous linear ordinary differential equation system involving eight \(\Delta_1^{(3)}\) equations with the initial condition
\begin{equation*}
    \Delta_1^{(3)}(t = 0) = 0.
\end{equation*}
For this initial condition, the solution is:
\begin{align}
\omega_a\omega_b\omega_c\Delta_1(\varphi_\mathbf{a}\varphi_\mathbf{b}\varphi_\mathbf{c}) &= \frac{\mu\left(\cos\omega_{\rm{T}}t - 1\right)}{2\omega_{\rm{T}}}\tilde\delta_\mathbf{a, b, c}, \nonumber\\
    \omega_b\omega_c\Delta_1(\pi_\mathbf{a}\varphi_\mathbf{b}\varphi_\mathbf{c}) &= \omega_c\omega_a\Delta_1(\varphi_\mathbf{a}\pi_\mathbf{b}\varphi_\mathbf{c}) = \omega_a\omega_b\Delta_1(\varphi_\mathbf{a}\varphi_\mathbf{b}\pi_\mathbf{c}) = - \frac{\mu\sin\omega_{\rm{T}}t}{2\omega_{\rm{T}}}\tilde\delta_\mathbf{a, b, c}, \nonumber\\
    \omega_a\Delta_1(\varphi_\mathbf{a}\pi_\mathbf{b}\pi_\mathbf{c}) &= \omega_b\Delta_1(\varphi_\mathbf{b}\pi_\mathbf{c}\pi_\mathbf{a}) = \omega_c\Delta_1(\varphi_\mathbf{c}\pi_\mathbf{a}\pi_\mathbf{b}) =  \frac{\mu(1 -\cos\omega_{\rm{T}}t)}{2\omega_{\rm{T}}}\tilde\delta_\mathbf{a, b, c}, \nonumber\\
    \Delta_1(\pi_\mathbf{a}\pi_\mathbf{b}\pi_\mathbf{c}) &= \frac{\mu\sin\omega_{\rm{T}}t}{2\omega_{\rm{T}}}\tilde\delta_\mathbf{a, b, c},\label{eq:soltreecubic}
\end{align}
where $\omega_{\rm{T}} = \omega_a + \omega_b + \omega_c$ is the inherent angular frequency of $\Delta^{(3)}$\,s. The evolutions of these cubic fluctuations are depicted in Fig. \ref{fig:cubicdelta}.

These solutions demonstrate that, at the tree level in Minkowski spacetime, all $\Delta^{(3)}$ fluctuations are bounded and evolve harmonically. This nature is a characteristic feature of a Minkowski spacetime. In Section \ref{sec-dS-fields}, we will observe that in a cosmological de Sitter spacetime, \(\Delta^{(3)}_1\) fluctuations diverge in a \(\varphi^3\)-model as the conformal time \(\eta\) approaches \(0\), corresponding to the far-future of an exponentially expanding
spacetime.

\subsubsection{Quantum fluctuations at quantum-level}\label{sec:quanteffect}

To conclude this section, we outline the typical quantum effects on the effective dynamics of the $\varphi^3$ model in the Minkowski spacetime. A more exhaustive study is ongoing.

We first examine the quantum effect on the classical variables $\Eval{\varphi_\mathbf a}$ and $\Eval{\pi_\mathbf{a}}$ at the $\mu^1$-order. Given the initial condition $\Eval{\varphi_\mathbf{a}} = \Eval{\pi_\mathbf{a}} = 0$, we have proved in Section \ref{sec-interacting-field} that $\Eval{\varphi_\mathbf{a}} = 0$ for all $\mathbf{a}\ne 0$ at any time. We now prove $\Eval{\varphi_\mathbf{0}} = 0$. The equations of motion \eqref{eq:eomclassical} are
\begin{align*}
\dot{\Eval{\varphi_{\mathbf{a}}}}_1 &= \Eval{\pi_{\mathbf{a}}}_1,\\
\dot{\Eval{\pi_{\mathbf{a}}}}_1 &= -\omega_a^2\Eval{\varphi_{\mathbf{a}}}_1
-\mu\int_{p}\Bigg[\Eval{\varphi_\mathbf{p}}_0\Eval{\varphi_\mathbf{a-p}}_0 +  \Delta_0(\varphi_{\mathbf{p}}\varphi_{\mathbf{a-p}})\Bigg] = -\omega_a^2\Eval{\varphi_{\mathbf{a}}}_1 - \frac{\mu}{2m}\tilde\delta_\mathbf{a}\int\frac{d^3k}{(2\pi)^3}.
\end{align*}
These equations imply that $\Eval{\varphi_\mathbf{a}}\ne 0$ only for $\mathbf{a} = 0$. This scenario represents a homogeneous field strength $\Eval{\varphi(\mathbf{x})}$, uniform across all positions $\mathbf{x}$. This expectation value is negative (under the assumption that \(\mu > 0\)) and exhibits UV divergence. This divergence is anticipated, given that the \(\mu\varphi^3\) term in the Hamiltonian, being unbounded from below, naturally pushes the field \(\varphi\) towards infinite descent. 

In quantum field theory, such a divergent field strength $\Eval{\varphi}$ correlates with tadpole diagrams. Drawing from textbook quantum field theory, we know that tadpole terms break Lorentz invariance. If we ignore the general relativistic effect of $\Eval{\varphi}$ on the spacetime curvature and keep the spacetime Minkovski, it is conventional to redefine the field operator \(\varphi\) by re-expanding it around its true vacuum \(\Eval{\varphi}\):
$$\varphi \to \varphi - \Eval{\varphi}.$$
With this redefined field, the equations of motion for quantum fluctuations remain unchanged, resulting in
$$\Eval{\varphi_\mathbf{a}}(t) = \Eval{\pi_\mathbf{a}}(t) = 0.$$
This is the reason for omitting the term $\Delta(\varphi_\mathbf{k}\varphi_\mathbf{p})\Eval{\varphi_\mathbf{q}}$ in the fluctuation-part of the effective Hamiltonian, $H_{\text{Mink.}, \Delta}$, as indicated in equation \eqref{eq:HamilDelta}. This omission leads to the simplified form of the equations of motion presented in equation \eqref{eq:eom}.

Then, as a simple yet nontrivial example, we consider an observable quantum effect --- the $\mu^2$-order correction, $\Delta^{(2)}_2$, to the quadratic quantum fluctuations. Having determined all $\Delta_1^{(3)}$ in Eqs. \eqref{eq:soltreecubic}, we can now express the $\mu^2$-order equations of motion for quadratic quantum fluctuations as follows:
\begin{align*}
\dot\Delta_2(\varphi_\mathbf{a}\varphi_\mathbf{b}) &= \Delta_2(\pi_\mathbf{a}\varphi_\mathbf{b}) + \Delta_2(\varphi_\mathbf{a}\pi_\mathbf{b}),\\
\dot\Delta_2(\varphi_\mathbf{a}\pi_\mathbf{b}) &= \Delta_2(\pi_\mathbf{a}\pi_\mathbf{b}) - \omega_b^2\Delta_2(\varphi_\mathbf{a}\varphi_\mathbf{b}) - \mu\int_p{(2\pi)^3}\Delta_1(\varphi_\mathbf{a}\varphi_\mathbf{p}\varphi_\mathbf{b - p}),\\ 
&= \Delta_2(\pi_\mathbf{a}\pi_\mathbf{b}) - \omega_b^2\Delta_2(\varphi_\mathbf{a}\varphi_\mathbf{b}) + \frac{\mu^2}{2}\tilde\delta_{\mathbf{a, b}}\int_p\frac{1 - \cos\omega_{\rm{T}}t}{\omega_{\rm{T}}\omega_a\omega_p\omega_{|\mathbf{a+p}|}}\\
\dot{\Delta}_2(\pi_\mathbf{a}\pi_\mathbf{b}) &= -\omega_a^2\Delta_2(\varphi_\mathbf{a}\pi_\mathbf{b}) - \omega_b^2\Delta_2(\varphi_\mathbf{a}\pi_\mathbf{b}) - \mu\int_p\left[\Delta_1(\pi_\mathbf{a}\varphi_\mathbf{p}\varphi_\mathbf{b-p}) + \Delta_1(\pi_\mathbf{b}\varphi_\mathbf{p}\varphi_\mathbf{a-p})\right]\\
&=-\omega_a^2\Delta_2(\varphi_\mathbf{a}\pi_\mathbf{b}) - \omega_b^2\Delta_2(\varphi_\mathbf{a}\pi_\mathbf{b}) + \mu^2\tilde\delta_\mathbf{a,b}\int_p
\frac{\sin\omega_{\rm{T}}t}{\omega_{\rm{T}}\omega_a\omega_{|\mathbf{a+p}|}},
\end{align*}
where 
\[\omega_{\rm{T}} = \omega_a + \omega_p + \omega_\mathbf{|\mathbf{a+p}|}.\]
The RHS of these equations exhibit \(\ln\Lambda\) UV divergence, where \(\Lambda\) is the UV cutoff. Renormalization is necessary to address this divergence, resulting in $\mu^2$-order corrections to quadratic $\Delta^{(2)}_0$ in \eqref{eq:solfreequadratic}. The renormalization procedure remains a subject of ongoing investigations.

\section{Fields on cosmological spacetimes}
\label{sec-dS-fields}

We have seen in the previous section that our framework can provide a consistent
effective description of quantum fields in flat spacetime.
The description allowed for analytical solutions to the $\varphi^3$-model
without relying on any operator-method input..

For cosmological spacetimes, 
such as the exponentially expanding spacetime of inflation,
solving the effective equations for an interacting field is not easy.
Results from the in-in method \cite{chen2010primordial} already suggest that
for cubic fluctuations,
the solution can at best only be written in terms of
non-elementary complex integrals that integrate to the (conformal)
time of interest
$\eta$.
It is only in the superhorizon limit can one hope to express the solutions
in terms of elementary functions.

In our effective framework,
we will soon see the obstruction 
to obtaining analytical solutions
manifest in the form of time-dependent
interaction strengths.
In the $\varphi^3$-model on de Sitter,
these will appear as functions $\Omega_k^2(\eta)$
for the quadratic interaction
and $\lambda(\eta)$ for the cubic interaction.
The analogue of \eqref{phi3-cubic-EoM}
in an expanding spacetime
will typically not be solvable analytically when these 
interaction strengths depend on $\eta$.

Fortunately, following the strategy outlined in
Section \ref{flat-solution-sec},
one could still numerically
describe the evolution of 3-pt correlators.
One simply needs to organize the solution in
cubic-interaction strength
and use the uncertainty principle to find
the desired minimal-energy initial conditions.
At the zeroth-order in cubic-interaction strength,
the analogue of \eqref{phi3-quad-EoM}
can still be solved analytically
despite the quadratic-interaction strength being time-dependent. As we will see, however, this process sees additional subtleties.
The expansion of spacetime
permits different choices of canonical variables
that correspond to different minimal-energy initial conditions,
even at the zeroth-order in cubic-interaction strength
\footnote{This issue is separate from the ambiguity of vacuum in
curved-spacetime quantum field theory.
Here, we are not comparing vacua associated with different time-slicings
but minimal energies associated with different canonical variables of a fixed
time-slicing.}.
The differences vanish if one sets initial conditions at
past-infinity.
But for inflation, it has been debated whether too much expansion could lead to the
``trans-Planckian'' problem \cite{martin2001trans}.
Settling the debate is beyond the scope of this work.
For demonstration purposes,
we will instead simply pick a set of canonical variables
to show that our effective framework is self-consistent
and that one could arrive at the correct value for
the cubic fluctuation
$\Delta(\delta\varphi\delta\varphi\delta\varphi)$
(in the superhorizon limit)
when compared with the in-in method outlined in
\cite{weinberg2005quantum}.

\subsection{Free field}
Let us warm up by analyzing a free field on a fixed de Sitter background.
The action of such a system is
\begin{equation}
	S_{2}=\int d\eta d^3x\frac{1}{2}
	\sqrt{|g|}\Big[
	-g^{\mu\nu}\partial_{\mu}\varphi\partial_{\nu}\varphi
	\Big]\,.
	\label{dS-free-field}
\end{equation}
Here, the metric $g_{\mu\nu}$ is the textbook FLRW metric
in conformal time with scale factor $a(\eta)$.
We can split the field into a background (homogeneous) part
and a perturbation (inhomogeneous, not containing the zero mode) part
\begin{equation}
	\varphi(\eta,\mathbf{x})=\bar{\varphi}(\eta)+\delta\varphi(\eta,\mathbf{x})
	\nonumber\,.
	\label{}
\end{equation}

Substituting this separation into action
\eqref{dS-free-field},
and upon imposing the background equations of motion
(for $\bar{\varphi}(\eta)$),
the $O(\delta\varphi)$ term vanishes.
We thus have an action where perturbation decouples from the background
system
\begin{equation}
	S_2|_{\RM{OS}}=
	\int d\eta d^3x a^2\frac{1}{2}(\bar{\varphi}')^2
	+\int d\eta d^3x a^2\frac{1}{2}
	\Big[
	(\delta\varphi')^2-\frac{1}{2}\delta^{ij}\partial_i\delta\varphi
	\partial_j\delta\varphi\Big]\nonumber\,,
	\label{}
\end{equation}
where the subscript ``$\RM{OS}$'' means on-shell for the background system.

As an aside, we note that
on-shell equations of motion are in general not valid in the presence of quantum corrections. So a vanishing $O(\delta\varphi)$-contribution to the action should not be taken for granted. Nevertheless, in the context of cosmological perturbation theory, the homogeneous part of the field system is typically not quantized as they are strongly coupled to the background spacetime, for which there is no good quantum description yet. In our framework, this implies that we assume no quantum fluctuations of $\bar{\varphi}(\eta)$.
Thus, the $O(\delta\varphi)$-term of \eqref{dS-free-field} will (in Fourier-space) be proportional to $\Eval{\delta\varphi_{\mathbf{k\neq 0}}}$
after quantization
and vanishes consistently based on our discussion in Section \eqref{trans-rot-sec}.
(Our framework can in principle be used to discuss the case
where quantum fluctuations of homogeneous fields are allowed
\cite{bojowald2021canonical}.
When gravity is involved, however, one must then commit to some description of quantum gravity for it to work consistently.
See \cite{brizuela2019moment, brizuela2021mode} for attempts in this direction.)

Now consider the $O(\delta\varphi^2)$-part
of the action
\begin{equation}
	S_{\RM{Free}}
	=\int d\eta d^3x a^2
	\Big[
	\frac{1}{2}
	(\delta\varphi')^2-\frac{1}{2}\delta^{ij}\partial_i\delta\varphi
	\partial_j\delta\varphi\Big]\,,
	\label{free-action-cos}
\end{equation}
which we call the ``free'' part to separate it from the interacting
part consisting of $O(\delta\varphi^3)$-order and higher.
(We wish to emphasize that
the notion of the ``free'' part is ambiguous when there
are $\delta\varphi'$-type interactions, but it is not an issue here.)

The prescription for obtaining an effective Hamiltonian
and its equations of motion is similar to the one outlined in Section
\ref{sec-Mink-fields}.
But there is a nuance in the case of a cosmological
spacetime: There are at least three ``natural'' sets of canonical variables
\cite{grain2020canonical}.
We must choose one before we can begin canonical quantization.
Different choices cannot be expected to be physically equivalent
\cite{fulling1973nonuniqueness}, as we now address.

\subsubsection{Different choices of variables and initial conditions}
We claim that there are three ``natural'' sets of configuration variables
because of two observations.
First, canonically quantizing \eqref{free-action-cos}
``as is'' and imposing an initial vacuum 
(associated with $\eta$-time slicing)
lead to divergences at early times,
which may make quantization inconsistent.
Second,
action \eqref{free-action-cos}
is not one that collects the familiar simple harmonic oscillators
in Fourier space due to the $a^2$ factor.
If one ignores these two issues,
one could insist that the variables $\delta\varphi$ and $\pi=a^2\delta\varphi'$
are natural and carry out the standard canonical quantization.
Otherwise, one could find two other sets of variables,
which we now introduce.

At the present level,
the two issues highlighted above are mostly aesthetic,
assuming that one imposes initial conditions at past infinity
($\eta\rightarrow -\infty$).
The observables will not be affected.
(Nevertheless,
the situation will worsen a lot for generic non-Gaussian systems.)
One might be tempted to make the system appear more like
harmonic oscillators
and its initial conditions more convergent with a change of variables
to $f=a\delta\varphi$.
The ``$f$-variable''
enables us to do two things.
One, we can trade initial divergences for late-time divergences.
The late time divergences are not as problematic because the system will have
classicalized and one would not have to deal with infinities in Hilbert space.
The correlators associated with $\delta\varphi$ will also remain finite.
Two, we may refurbish our system into simple harmonic oscillators
with one catch: We need to integrate by parts.
The act of integration by parts affects nothing if one imposes
quantum initial conditions at past infinity.
But the situation will be different otherwise.

Let us be more specific.
The first of the three natural variables we can pick is the original
$\delta\varphi$ and its canonical momentum
$\pi_{\delta\varphi}=a^2\delta\varphi'$.
The other two options arise after changing to the variable $f=a\delta \varphi$.
After switching to $f$-variables,
one is faced with two possible actions,
differentiated by our willingness to ignore a boundary term
\begin{align}
	&\text{Option 1: }
	S_{\RM{free},1}=\int d\eta d^3x
	\frac{1}{2}\Big[ (f')^2
	-2\mathcal{H}ff'+\mathcal{H}^2f^2
	-\delta^{ij}\partial_if\partial_jf\Big]\nonumber\\
	&\text{Canonical momentum: }
	\pi_{1}(\eta,\mathbf{x})= f'(\eta,\mathbf{x})-\mathcal{H}f\nonumber\\
	&\text{Option 2: }
	S_{\RM{free},2}=
	\int d\eta d^3x
	\frac{1}{2}\Big[
	(f')^2+\frac{a''}{a}f^2-\delta^{ij}\partial_if\partial_jf\Big]
	-\cancel{\frac{1}{2}\int d\eta d^3x
	(\mathcal{H}f^2)'}\nonumber\\
	&\text{Canonical momentum: }
	\pi_{2}(\eta,\mathbf{x})= f'(\eta,\mathbf{x})\,,
	\label{IBP-or-not}
\end{align}
where $\mathcal{H}=a'/a$ is the conformal time Hubble parameter.
The two options will introduce two distinct canonical momenta
due to differences in terms with ``velocity'' $f'$ in the Lagrangian.

The action from option 1 is completely equivalent to the original one
\eqref{free-action-cos},
whereas the action from option 2
--- which ignores the boundary term ---
resembles a set of simple harmonic oscillators,
albeit with time-dependent ``frequencies''.
So in terms of picking a ``natural'' starting point for canonical quantization,
neither of the options offers a significant merit over the other.

Does this mean one can pick any one of the two options?
No, unless one sets initial conditions at past infinity.
If one follows the standard canonical quantization
and assumes an initial minimal-energy or Gaussian state
(which includes the usual Bunch-Davies vacuum),
what matters are the initial values for the quadratic expectation values
\begin{equation}
	\Eval{ff}\,,\;\Eval{f\pi}\,,\;\Eval{\pi\pi}\,,
	\label{}
\end{equation}
where the $\pi$ above can be either $\pi_1$ or $\pi_2$.
Clearly, since $\pi_2-\pi_1=\mathcal{H}f\sim -\eta^{-1}$
on de Sitter, the differences in quadratic expectation values
for the two choices of momentum
vanish only when $\eta\rightarrow -\infty$ at past infinity.
(The situation will worsen when there is a ``velocity'' type
non-Gaussian terms in the Lagrangian
--- the differences in initial expectation values may even diverge
for different choices of canonical momentum.)

Our effective framework offers a more intuitive way to understand the discrepancy:
Different canonical pairs correspond to 
different effective Hamiltonians,
and hence, different minimal-energy conditions.
In our case, the differences in minimal energy states vanish
in the infinite-past limit.

In Fourier space, by applying our prescription for obtaining
the effective Hamiltonian, we have
\begin{align}
	&\text{Option 1: }
	H_{1,\Delta}=\int_k \Big[\frac{1}{2}\Delta(\pi_{1,\mathbf{k}}
	\pi_{1,\mathbf{-k}})
	+\mathcal{H}\Delta(f_{\mathbf{k}}\pi_{1,\mathbf{-k}})
	+\frac{1}{2}k^2\Delta(f_{\mathbf{k}}f_{\mathbf{-k}})\Big]\nonumber\\
	&\text{Option 2: }
	H_{2,\Delta}=\int_k
	\Big[\frac{1}{2}\Delta(\pi_{2,\mathbf{k}}\pi_{2,\mathbf{-k}})
	+\frac{1}{2}(k^2+\frac{a''}{a})\Delta(f_{\mathbf{k}}f_{\mathbf{-k}})\Big]
	\nonumber\,.
\end{align}
In de Sitter, both $a''/a$ and $\mathcal{H}$
vanish in the infinite-past limit ($\eta\rightarrow -\infty$),
so the difference between the two effective Hamiltonians vanishes.
One can use our uncertainty-principle prescription
in Section \ref{UP-IV-phi3} to find
a common set of initial values for the two options
--- they correspond to the Minkowski vacuum
in the operator/Hilbert-space Language.
This is what modes feel in the subhorizon limit under Bunch-Davies initial conditions
because their scales are too small to feel any effects from curvature.

But for a finite initial time $|\eta_i|<\infty$,
there is no common set of initial values for the two options anymore.
To find their minimal-energy conditions,
one should use the uncertainty principle,
per Section \ref{UP-IV-phi3}.
The prescription outlined there tells us that the minima of the two Hamiltonians
are
\begin{align*}
	&\text{Option 1: }
	H_{1,\Delta}\geq \int_k\Big[
	\mathcal{H}\Delta(f_{\mathbf{k}}\pi_{1,\mathbf{-k}})
	+k\sqrt{\Delta^2(f_{\mathbf{k}}\pi_{1,\mathbf{-k}})
	+\frac{1}{4}\tilde{\delta}(\mathbf{0})^2}\Big]\,.\\
	&\text{Option 2: }
	H_{2,\Delta}\geq \int_k\Big[
	(k^2-\frac{2}{\eta^2})^{1/2}
	\sqrt{\Delta^2(f_{\mathbf{k}}\pi_{2,\mathbf{-k}})
	+\frac{1}{4}\tilde{\delta}(\mathbf{0})^2}\Big]\,.\\
\end{align*}
Note that $1/4\tilde{\delta}
(\mathbf{0})^2\rightarrow \hbar^2/4\tilde{\delta}(\mathbf{0})^2$
if we restore $\hbar$,
and $\tilde{\delta}(\mathbf{0})$ comes from $\tilde{\delta}(\mathbf{k-k})$.
The smallest values of the two lower bounds are reached when 
\begin{align}
	\text{Option 1: }
	&k^2\Delta(f_{\mathbf{k}}f_{\mathbf{-k}})=
	\Delta(\pi_{1,\mathbf{k}}\pi_{1,\mathbf{-k}})=
	k\sqrt{\Delta^2(f_{\mathbf{k}}\pi_{1,\mathbf{-k}})
	+\frac{1}{4}\tilde{\delta}(\mathbf{0})^2}
	\,,\;\nonumber\\
	 &\Delta^2(f_{\mathbf{k}}\pi_{1,\mathbf{-k}})=
	\frac{1}{4}\frac{1}{(k/\mathcal{H})^2-1}\tilde{\delta}(\mathbf{0})^2
	\,.\nonumber\\
	\text{Option 2: }
	&(k^2-\frac{2}{\eta^2})\Delta(f_{\mathbf{k}}f_{\mathbf{-k}})=
	\Delta(\pi_{2,\mathbf{k}}\pi_{2,\mathbf{-k}})=
	(k^2-\frac{2}{\eta^2})^{1/2}
	\sqrt{\Delta^2(f_{\mathbf{k}}\pi_{2,\mathbf{-k}})
	+\frac{1}{4}\tilde{\delta}(\mathbf{0})^2}
	\,,\;\nonumber\\
	&\Delta^2(f_{\mathbf{k}}\pi_{2,\mathbf{-k}})=0\,.
 \label{min-Ham-IV}
\end{align}
In particular, when the effective Hamiltonians are minimized,
the initial 2-pt functions
\begin{equation}
	\Delta(f_{\mathbf{k}}f_{\mathbf{-k}})|_{\text{option 1}}=
	\frac{1}{2}(k^2-\frac{1}{\eta_i^2})^{-1/2}\tilde{\delta}(\mathbf{0})\,,\;
	\Delta(f_{\mathbf{k}}f_{\mathbf{-k}})|_{\text{option 2}}=
	\frac{1}{2}(k^2-\frac{2}{\eta_i^2})^{-1/2}\tilde{\delta}(\mathbf{0})\,,
	\label{2-pt-func-diff}
\end{equation}
differ significantly for long-wavelength modes.
Nevertheless, they coincide when the initial time is taken to
the $\eta_i\rightarrow -\infty$ limit.

Suppose one demands that the effective system
(for either option 1 or option 2)
rests in its minimal-energy configuration initially
according to \eqref{min-Ham-IV}.
Not all modes ``feel" a Minkowski vacuum,
as is commonly assumed in an operator/Hilbert-space
framework \cite{maldacena2003non}.
The modes are actually in squeezed states
\cite{danielsson2002note},
where the squeezing parameter depends on both
the wavenumber $k$ and initial time 
$\eta_i$ in a non-polynomial way.

The non-Bunch-Davies initial conditions
has observational consequences.
This was discussed in \cite{grain2020canonical}
by employing different canonical variables and
using an operator/Hilbert-space language.
Alternatively, our effective framework
provides an easier parameterization of initial conditions
using quantum fluctuations as dynamical quantities.
As a potential application in this direction,
we would like to point out the possibility of
explaining the large-angle anomaly in the CMB
\cite{copi2010large}.
We see, for example in \eqref{2-pt-func-diff},
that large-wavelength modes
deviate from the Minkowski vacuum the most.
To explain the weakness of large-angle temperature
correlations, we would require modes with large wavelengths
to contribute weaker power spectra compared with the standard
Bunch-Davies result.
In \cite{grain2020canonical},
squeezed initial conditions akin to \eqref{min-Ham-IV}
lead to a modified power spectrum that is ambiguous in whether
it amplifies or weakens the large-angle sector of the
power spectrum.
So a squeezed initial state may not be enough to explain the anomaly.
Using our method, however,
one could parameterize more general hierarchies
between fluctuations in an $\eta_i$- or $k$-dependent way.
The ratio between the fluctuations
$\Delta(ff)$ and $\Delta(\pi\pi)$,
such as the ones in \eqref{min-Ham-IV},
are among the simplest parameterizations of 
a fluctuation-hierarchy ---
Our method provides a way to probe more complicated ones,
potentially involving higher-order fluctuations.
The examination of CMB anomalies
using our effective framework is the subject of an ongoing
investigation, which will be reported elsewhere.

\subsubsection{The equations of motion and its solution}
To demonstrate our effective framework without having to worry about
ambiguity in variables,
we will assume that initial conditions can be set at past infinity
so that we are free to pick any one of the aforementioned canonical momenta.
We will use option 2 in \eqref{IBP-or-not} for easier comparison with the
flat-spacetime case of Section \ref{sec-Mink-fields}.
That is, our effective Hamiltonian in Fourier space will be
\begin{equation}
	H_{\RM{free},\Delta}=\int_{k}\frac{1}{2}
	\Big[\Delta(\pi_{\mathbf{k}}\pi_{\mathbf{-k}})
	+\Omega_k^2(\eta)\Delta(f_{\mathbf{k}}f_{\mathbf{-k}})\Big]\,,\;
	\pi_{\mathbf{k}}=f'_{\mathbf{k}}\,,\;
	\Omega_k^2(\eta)=k^2-\frac{2}{\eta^2}\,,
	\label{}
\end{equation}
where we have used a fixed de Sitter solution for $a''/a$.

To obtain initial conditions we again assume
that the initial fluctuations follow a Gaussian hierarchy
(where Wick's theorem applies)
and minimize $H_{\RM{free},\Delta}$.
The only initial free parameters we must find are the quadratic expectation
values,
which can be found with the uncertainty-principle prescription
as before
\begin{align}
	\Delta(f_{\mathbf{p}}\pi_{\mathbf{q}})=&0\times
	\tilde{\delta}_{\mathbf{p,q}}\nonumber\\
	\Delta(f_{\mathbf{p}}f_{\mathbf{q}})
	=& \frac{1}{2p}\tilde{\delta}_{\mathbf{p,q}}\nonumber\\
	\Delta(\pi_{\mathbf{p}}\pi_{\mathbf{q}})
	=&\frac{p}{2}\tilde{\delta}_{\mathbf{p,q}}\,.
	\label{}
\end{align}
They look exactly like the Minkowski version,
as is expected from a Bunch-Davies initial condition in the subhorizon limit.

With the Poisson bracket rules for the fluctuations,
we can derive the equations of motion for
$H_{\RM{free},\Delta}$
to be
\begin{align}
	\Delta'(f_{\mathbf{p}}f_{\mathbf{q}})=&
	\Delta(f_{\mathbf{p}}\pi_{\mathbf{q}})
	+\Delta(f_{\mathbf{q}}\pi_{\mathbf{p}})\nonumber\\
	\Delta'(f_{\mathbf{p}}\pi_{\mathbf{q}})=&
	\Delta(\pi_{\mathbf{p}}\pi_{\mathbf{q}})
	-\Omega_q^2(\eta)\Delta(f_{\mathbf{p}}f_{\mathbf{q}})\nonumber\\
	\Delta'(\pi_{\mathbf{p}}\pi_{\mathbf{q}})
	=&-\Omega_q^2(\eta)\Delta(\pi_{\mathbf{p}}f_{\mathbf{q}})
	-\Omega_p^2(\eta)\Delta(\pi_{\mathbf{q}}f_{\mathbf{p}})\,.
	\label{free-cos-EoM}
\end{align}

When one assumes statistical homogeneity and isotropy,
the quadratic expectation values above are delta-functions
times real functions of $\eta$.
Imposing our initial conditions,
these equations of motion can be solved by hand to be
\begin{align}
	\Delta(f_{\mathbf{p}}f_{\mathbf{q}})
	=&\Big(\frac{1}{2p}+\frac{1}{2p^3\eta^2}\Big)\tilde{\delta}_{\mathbf{p,q}}
	\nonumber\\
	\Delta(f_{\mathbf{p}}\pi_{\mathbf{q}})
	=&-\frac{1}{2(p\eta)^3}\tilde{\delta}_{\mathbf{p,q}}\nonumber\\
	\Delta(\pi_{\mathbf{p}}\pi_{\mathbf{q}})=&
	\Big(\frac{p}{2}-\frac{1}{2p\eta^2}+\frac{1}{2p^3\eta^4}\Big)
	\tilde{\delta}_{\mathbf{p,q}}\,.
	\label{dS-free-sol}
\end{align}
These solutions coincide with the expectation values obtained in an operator formalism,
assuming Bunch-Davies initial conditions. Utilizing the findings we derived in Section \ref{trans-rot-sec}, the field in dS spacetime will maintain its spatial homogeneity and isotropy throughout its evolution. This  spatial symmetry is also the consequence of minimal energy condition.

In particular, in the spatially-flat gauge,
the comoving curvature perturbation $\mathcal{R}$ and the perturbation $\delta\varphi$
are related by
\begin{equation}
	\mathcal{R}=\frac{\mathcal{H}}{\bar{\varphi}'}\delta\varphi\nonumber\,.
	\label{}
\end{equation}
This gives the power spectrum of $\mathcal{R}$
to be
\begin{equation}
	\Eval{\mathcal{R}_{\mathbf{p}}\mathcal{R}_{\mathbf{q}}}
	=(\frac{\mathcal{H}}{a\bar{\varphi}'})^2
	\Eval{f_{\mathbf{p}}f_{\mathbf{q}}}
	\longrightarrow \frac{H^2}{\dot{\bar{\varphi}}^2}
	\frac{H^2}{2p^3}(2\pi)^3\delta^3(\mathbf{p+q})\,,
	\label{}
\end{equation}
where the arrow indicates the superhorizon limit.

\subsection{Interacting field: non-Gaussianities}
\label{sec-interacting-field}
In this section, we apply our formalism to the case of an interacting field
on fixed de Sitter background.
This system serves as a good approximation if one considers up to leading order in
slow roll and assumes sufficient decoupling between metric 
and matter perturbations.
Our goal is to compute the 3-pt function,
or non-Gaussianity, of the theory.
In passing, we demonstrate that our method is self-contained
in the sense that nothing (from equations of motion to initial values to solution schemes) necessarily requires any operator-formalism input.
The techniques developed in Section \ref{sec-Mink-fields}
suffice.

Let us first consider a generic matter action with potential $V(\varphi)$
\begin{equation}
	S[\varphi]=\int d\eta d^3x
	\sqrt{|g|}\Big[
	-\frac{1}{2}g^{\mu\nu}\partial_{\mu}\varphi\partial_{\nu}\varphi
	-V(\varphi)\Big]\,.
	\label{}
\end{equation}
We will again assume the following splitting and change of variables
\begin{equation}
	\varphi(\eta,\mathbf{x})=\bar{\varphi}(\eta)+\delta\varphi(\eta,\mathbf{x})
	\,,\;
	f=a\delta\varphi(\eta,\mathbf{x})\,.
	\label{}
\end{equation}
The action can then be split into a background part, containing only $\bar{\varphi}$, and a perturbation part containing terms $O(f^2,f^3)$ or higher.
\begin{equation}
	S[\varphi]|_{\RM{OS}}=S_0[\bar{\varphi}]
	+\int d\eta d^3x
	[\frac{1}{2}(f')^2+\frac{1}{2}(\frac{2}{\eta^2}f^2
	-\delta^{ij}\partial_if\partial_jf)
	-a^4\sum_{n=2}\frac{1}{n!}(\partial_{\phi}^{(n)}V)|_{\phi_0}
	(\frac{f}{a})^n]
	\label{de-Sitter-scalar}
\end{equation}
To arrive at the above action,
we have imposed the equation of motion for the background field
$\bar{\varphi}$, as implied by the subscript ``$\RM{OS}$'' for on-shell.
This makes terms linear in $\delta\varphi$ vanish
and explains why the sum in the integration starts at $n=2$.
Additionally, integration by parts is performed.

Now, consider the case for $V(\varphi)=\lambda_0\varphi^3/3$
where $\lambda_0$
is a time-independent small parameter controlling the magnitude of non-Gaussianity.
(The smallness is also necessary from the point of view of
background spacetime stability.)
Up to leading order in slow-roll,
the $V_{\varphi\varphi}$ term in
\eqref{de-Sitter-scalar}
is negligible compared to other terms,
giving us the action
\begin{equation}
	S[\varphi]|_{\RM{OS}}\approx S_0[\bar{\varphi}]
	+\int d\eta d^3x
	\Bigg[ \frac{1}{2}(f')^2+\frac{1}{2}\Big(\frac{2}{\eta^2}f^2
	-\delta^{ij}\partial_if\partial_jf\Big)
	+\frac{\lambda_0}{3H}\frac{1}{\eta}f^3\Bigg]\,.
	\label{de-Sitter-phi3}
\end{equation}
Viewed in Fourier space, the action resembles that of a 
$\varphi^3$-model on flat spacetime but
with an effective time-dependent mass and cubic-interaction
(for a mode with wavenumber $k=|\mathbf{k}|$)
\begin{equation}
	\Omega_k^2(\eta)=k^2-\frac{2}{\eta^2}\,,\;
	\lambda(\eta)=-\frac{\lambda_0}{H\eta}.
	\label{}
\end{equation}
The sign change for the coupling strength
is due to the translation from the scale factor to conformal time,
$a(\eta)=-1/(\eta H)$, on fixed de Sitter.

To apply our formalism to the system \eqref{de-Sitter-phi3},
we first notice that its $O(f^2,f^3)$-part
does not ``backreact" on
$S_0[\bar{\varphi}]$
in the sense that $\bar{\varphi}$ and $f$ do not mix in the Lagrangian
and the background spacetime is fixed regardless of how $f$ evolves.
So in a Hamiltonian formalism,
we only need to examine the $O(f^2,f^3)$-part of the action
if we are only interested in the evolution of $f$.

Thus the Lagrangian and Hamiltonian governing variable $f$
are (in the Fourier space)
\begin{align*}
	S_{f^3}\equiv& S[\varphi]|_{\RM{OS}}-S_0[\bar{\varphi}]\\
	=&\int d\eta\int_k  \Big[
	\frac{1}{2}f'_{\mathbf{k}}f'_{\mathbf{-k}}
	+\frac{1}{2}\Omega_k^2(\eta)f_{\mathbf{k}}f_{\mathbf{-k}}\Big]\\
	+&\int d\eta\int_{k,p,q}
	\frac{\lambda(\eta)}{3}f_{\mathbf{k}}f_{\mathbf{p}}f_{\mathbf{q}}
	\tilde{\delta}_{\mathbf{k,p,q}}\,,\\
	H_{f^3}=&\int d\eta
	\int_k\Big[
	\frac{1}{2}\pi_{\mathbf{k}}\pi_{\mathbf{-k}}
	+\frac{1}{2}\Omega_k^2(\eta)f_{\mathbf{k}}f_{\mathbf{-k}}\Big]\\
	+&\int d\eta \int_{k,p,q}
	\frac{\lambda(\eta)}{3}f_{\mathbf{k}}f_{\mathbf{p}}f_{\mathbf{q}}
	\tilde{\delta}_{\mathbf{k,p,q}}\,.
\end{align*}
Since there are no velocity-type interactions,
the canonical conjugate to $f_{\mathbf{k}}$ is simply
$\pi_{\mathbf{k}}=f'_{\mathbf{k}}$,
with Fourier-space normalization
$\{f_{\mathbf{k}},\pi_{\mathbf{p}}\}=\tilde{\delta}_{\mathbf{k,p}}$.

Upon employing our effective formalism,
the fluctuation part of the effective Hamiltonian associated with
$H_{f^3}$ is
\begin{align}
	H_{\Delta}[H_{f^3}]
	\equiv& H_Q[H_{f^3}]-H_{\rm classical}[H_{f^3}]\nonumber\\
	=&
	\frac{1}{2}\int_k
	\Big[
	\Delta(\pi_{\mathbf{k}}\pi_{\mathbf{-k}})
	+\Omega_k^2(\eta)\Delta(f_{\mathbf{k}}f_{\mathbf{-k}})
	\Big]\nonumber\\
	+&\int_{k,p,q}\frac{\lambda(\eta)}{3}
	\Big[
	3\Delta(f_{\mathbf{k}}f_{\mathbf{p}})\Eval{f_{\mathbf{k}}}
	+\Delta(f_{\mathbf{k}}f_{\mathbf{p}}f_{\mathbf{q}})
	\Big]\tilde{\delta}_{\mathbf{k,p,q}}\,,
	\label{eff-Ham-phi3}
\end{align}
where we have re-shuffled the integration variables to get the
factor of the $O(\Eval{f})$ term in the third line.

For solutions respecting translational and rotational invariance
on the level of quantum averages, terms
in $H_{\Delta}$ proportional to
linear-operator expectation values like 
$\Eval{f_{\mathbf{k}}}$ and $\Eval{\pi_{\mathbf{k}}}$
will vanish.
The cubic-$\Delta$ equations of motion are therefore
\begin{align}
	\Delta'(f_{\mathbf{b}}f_{\mathbf{c}}f_{\mathbf{d}})
	=&\Delta(f_{\mathbf{b}}f_{\mathbf{c}}\pi_{\mathbf{d}})
	+\Delta(f_{\mathbf{c}}f_{\mathbf{d}}\pi_{\mathbf{b}})
	+\Delta(f_{\mathbf{d}}f_{\mathbf{b}}\pi_{\mathbf{c}})
	\nonumber\\
	\Delta'(f_{\mathbf{b}}f_{\mathbf{c}}\pi_{\mathbf{d}})=&
	\Delta(f_{\mathbf{b}}\pi_{\mathbf{c}}\pi_{\mathbf{d}})
	+\Delta(f_{\mathbf{c}}\pi_{\mathbf{b}}\pi_{\mathbf{d}})
	-\Omega_d^2(\eta)
	\Delta(f_{\mathbf{b}}f_{\mathbf{c}}f_{\mathbf{d}})
	\nonumber\\
	+&\lambda(\eta)\int_{k,p,q}\tilde{\delta}_{\mathbf{k,p,q}}
	(-\tilde{\delta}_{\mathbf{d},\mathbf{k}})
	[
	\Delta(f_{\mathbf{b}}f_{\mathbf{c}}
	f_{\mathbf{p}}f_{\mathbf{q}})
	-\Delta(f_{\mathbf{p}}f_{\mathbf{q}})
	\Delta(f_{\mathbf{b}}f_{\mathbf{c}})
	]\nonumber\\
	\Delta'(f_{\mathbf{b}}\pi_{\mathbf{c}}\pi_{\mathbf{d}})
	=&
	\Delta(\pi_{\mathbf{b}}\pi_{\mathbf{c}}\pi_{\mathbf{d}})
	-[\Omega_{c}^2(\eta)
	\Delta(f_{\mathbf{b}}f_{\mathbf{c}}\pi_{\mathbf{d}})
	+(\mathbf{c}\leftrightarrow \mathbf{d})]\nonumber\\
	+&\lambda(\eta)\int_{k,p,q}
	\tilde{\delta}_{\mathbf{k,p,q}}
	[-\tilde{\delta}_{\mathbf{c,k}}
	(\Delta(f_{\mathbf{b}}f_{\mathbf{p}}f_{\mathbf{q}}
	\pi_{\mathbf{d}})
	-\Delta(f_{\mathbf{b}}\pi_{\mathbf{d}})
	\Delta(f_{\mathbf{p}}f_{\mathbf{q}}))
	+(\mathbf{c}\leftrightarrow \mathbf{d})
	]\nonumber\\
	\Delta'(\pi_{\mathbf{b}}\pi_{\mathbf{c}}\pi_{\mathbf{d}})
	=&-\Omega_{b}^2(\eta)
	\Delta(f_{\mathbf{b}}\pi_{\mathbf{c}}\pi_{\mathbf{d}})
	-\Omega_{c}^2(\eta)
	\Delta(f_{\mathbf{c}}\pi_{\mathbf{d}}\pi_{\mathbf{b}})
	-\Omega_{d}^2(\eta)
	\Delta(f_{\mathbf{d}}\pi_{\mathbf{b}}\pi_{\mathbf{c}})
	\nonumber\\
	+&\lambda(\eta)\int_{k,p,q}
	\tilde{\delta}_{\mathbf{k,p,q}}
	\Big[-\tilde{\delta}_{\mathbf{b,k}}
	(\Delta(\pi_{\mathbf{c}}\pi_{\mathbf{d}}
	f_{\mathbf{p}}f_{\mathbf{q}})
	-\Delta(\pi_{\mathbf{c}}\pi_{\mathbf{d}})
	\Delta(f_{\mathbf{p}}f_{\mathbf{q}}))\nonumber\\
	+&(\text{Two more terms cycling }
	\mathbf{b\rightarrow c\rightarrow d})
	\Big]\nonumber\\
	+&\frac{1}{2}\lambda(\eta)
	\tilde{\delta}_{\mathbf{b,c,d}}\,.
	\label{EoM-phi3}
\end{align}
These equations,
upon integrating delta-functions,
look formally like the equations of motion \eqref{phi3-cubic-EoM}
for the
$\varphi^3$-model on Minkowski
but with a time-dependent mass and cubic-interaction parameter.

\subsubsection{Finding the solution}
We will try to solve the cubic fluctuations perturbatively in
$O(\lambda)$.
Given a Gaussian initial condition,
the cubic fluctuations vanish in $O(\lambda^0)$.
So to find the leading order non-trivial solution,
we only need to go to $O(\lambda)$.
What this means for the equation set \eqref{EoM-phi3}
is that the even-order fluctuations inside integrals only need to be kept
up to $O(\lambda^0)$, i.e., its ``free'' solution.

Previously, the free solution was found to be
\begin{align*}
	\Delta(f_{\mathbf{k}}f_{\mathbf{p}})=&\Big(
	\frac{1}{2k}+\frac{1}{2k^3\eta^2}\Big)
	\tilde{\delta}_{\mathbf{k,p}}\\
	\Delta(f_{\mathbf{k}}\pi_{\mathbf{p}})=&-\frac{1}{2(k\eta)^3}
	\tilde{\delta}_{\mathbf{k,p}}\\
	\Delta(\pi_{\mathbf{k}}\pi_{\mathbf{p}})=&
	\Big(\frac{k}{2}-\frac{1}{2k\eta^2}+\frac{1}{2k^3\eta^4}\Big)
	\tilde{\delta}_{\mathbf{k,p}}\,,
\end{align*}
with all other fluctuations obtained via a Gaussian hierarchy,
which in particular means that all odd-order fluctuations vanish.

\begin{figure}
\centering
\includegraphics[height=6cm]{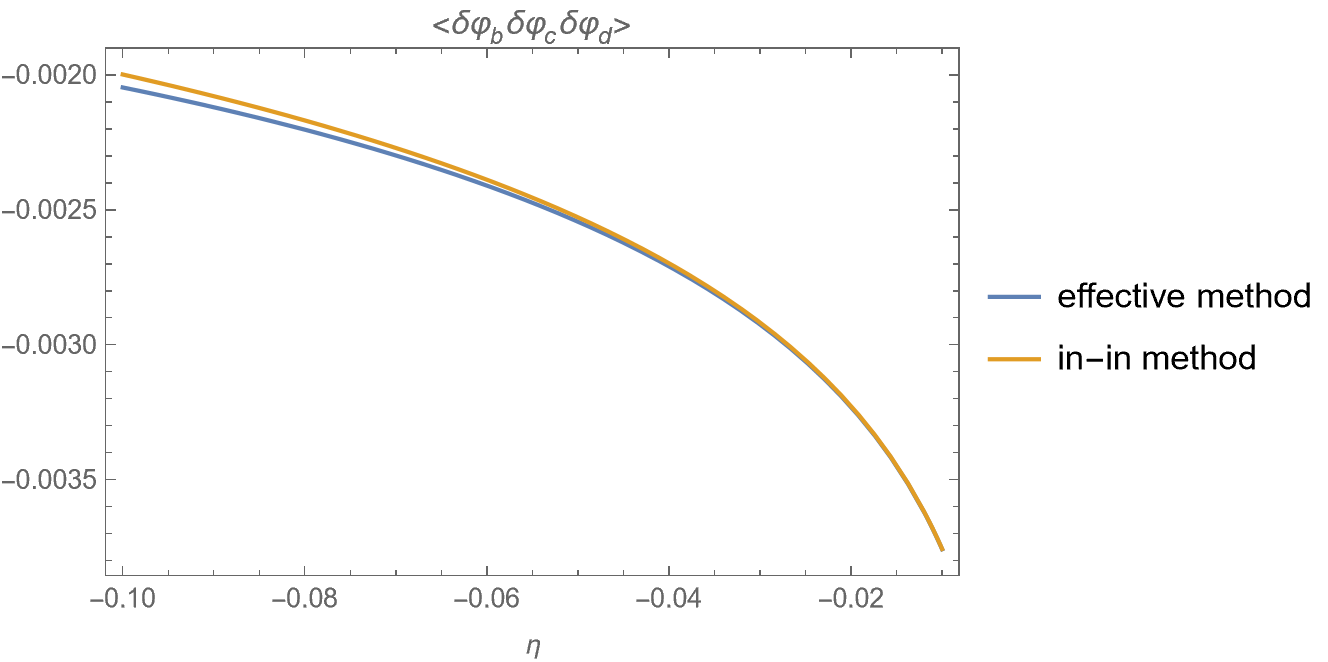}
\caption{The comparison of the 3-pt correlator between the solution to
\eqref{EoM-phi3} (blue)
and the superhorizon limit of the ``in-in" result (yellow).
Fields and time are in units of 
$H$ and $H^{-1}$,
where $H=\dot{a}/a$ is the Hubble parameter.
The external momenta used were
$(\mathbf{|b|,|c|,|d|})=(H,2H,1.5H)$.
(In our context,
a limit is superhorizon for the correlator
when the largest momenta in the correlator is much smaller than $aH$.)
The $\varphi^3$-coupling is taken to be $\lambda_0= 0.01 H$.
We have checked that other sets of external momenta also exhibit excellent
agreement with the ``in-in" result.
The free solution used in \eqref{EoM-phi3}
is that of \eqref{dS-free-sol},
which reduces in the far-past to the Bunch-Davies initial condition
commonly used for the ``in-in" method,
making our comparison fair.}
\label{dS-comparison}
\end{figure}

With the initial values determined and free solutions specified,
one can solve the cubic order fluctuations up to $O(\lambda)$.
Because of the time dependence in
$\Omega_b^2(\eta)$ and $\lambda(\eta)$,
the solutions cannot be obtained analytically.
We therefore use numerical evolution.
The result for the 3-pt correlator
$\Eval{\delta\varphi_{\mathbf{b}}\delta\varphi_{\mathbf{c}}
\delta\varphi_{\mathbf{c}}}$
is shown in Figure \ref{dS-comparison},
where we also make the comparison with the superhorizon-limit
expression obtained
from the ``in-in'' method.
While the comparison is made in the superhorizon limit,
we hope to emphasize that an advantage of our framework is its applicability to
intermediate time-regions that are not in the super/sub-horizon limit.
In particular, initial conditions can be set at finite $\eta$.

Our numerical evolution shows excellent agreement with the ``in-in'' method in the superhorizon limit,
which exhibits a $\log[-(a+b+c)\eta]$ 
divergence originating from the fixed de Sitter background.
For more realistic cosmological applications,
we would treat gauge symmetry and the weak time dependence of the Hubble parameter
more seriously.
The 3-pt correlator of gauge-invariant variables
would then be finite and slow-roll suppressed
for most single-field inflation models.

\section{Conclusions and discussions}
\label{sec-conclusions}
We have introduced a systematic framework for describing a quantum-field system
in an effective way, without invoking an operator-Hilbert space language.
The effective description, originally developed in \cite{bojowald2006effective}
for quantum mechanical systems,
treats fluctuations $\Delta$
of canonical pairs as messengers of quantum effects.
Via effective Poisson brackets \eqref{method-rules},
one obtains an effective Hamiltonian that governs the evolution of fluctuations.
This provides us with a time-dependent description of how quantum fluctuations
backreact on the classical part of the system.
At second order in fluctuations,
minimizing the effective Hamiltonian while respecting the uncertainty principle
leads to the Coleman-Weinberg effective potential,
which can be understood as a low-energy limit 
of the $H_{\Delta}$ that we constructed.
We have shown that this way of obtaining the low-energy effective potential 
works for both quantum mechanics and quantum fields.
In particular, the saturation of minimal uncertainty becomes a result
rather than an assumption when one wishes to find the low-energy
effective potential.

The novel part of our work lies in showing how one could go beyond the low-energy
effective potential for interacting field systems
by allowing fluctuations to evolve freely
instead of taking fixed values that minimize energy.
At first glance, a field system would give rise to infinitely many equations
of motion for the fluctuations.
By arguing from the perspective of minimal backreaction,
we have motivated the use of
translational and rotational symmetries
and shown how to make finite the number of equations.
Physically, we have thus shown that
minimal backreaction helps maintain spatial symmetries,
which decouple the evolution of fluctuations
from both themselves and the 
inhomogeneous part of the classical background.

In flat space, one could even find analytical solutions by hand
following the strategy outlined in 
Section \ref{sec-Mink-fields}.
The key elements used were: (1) perturbation theory
(in interaction strength);
(2) a set of minimal-energy initial conditions
found using the uncertainty principle;
(3) an initial Gaussian hierarchy where Wick's theorem applies.
These three elements are also manifest in the core techniques of
textbook quantum field theory
--- perturbation around a Gaussian system and the assumption of initial ``free''
vacuum, which together imply Wick's theorem.
In our framework, the three key elements are enough 
to completely fix the system of equations,
allowing us to solve for the fluctuations by hand.
We have demonstrated this with a $\varphi^3$-model.
But the strategy also applies to systems with more generic interactions,
even on curved spacetimes.

In particular, we have shown that for a $\varphi^3$-model on a de Sitter background
our result for the 3-pt correlator (or bispectrum)
agrees with the ``in-in'' result in the superhorizon limit.
This gives us confidence to apply it to more realistic inflationary systems,
with more complicated interactions and additional subtleties arising from
gauge invariance.
The systematic application of our method to the calculation of non-Gaussianities
will be left for future work.
We end by mentioning some opportunities and challenges in this direction.

\textit{Backreaction and stability of the background}.
Our framework naturally incorporates how time-dependent quantum fluctuations
backreact on classical field variables.
A preliminary investigation in this direction,
using toy models and an approximate canonical map
for second-order fluctuations,
was carried out in \cite{bojowald2021canonical}.
In our work, we have extended the investigation to higher-order fluctuations
for interacting fields on static and expanding backgrounds.
We have shown that backreaction is minimized when spatially symmetric
initial conditions are used.
Nevertheless,
the stability of spatial symmetries to 
quantum backreaction remains to be explored.

For inflationary cosmology, one could envision a scenario where a small
deviation from initial homogeneity and isotropy causes the inhomogeneous background
modes to couple to quantum fluctuations.
The growth of these modes could further enhance the coupling
and consequently also enhance the backreaction, causing more inhomogeneity growth.
A positive feedback loop would then ensue to produce too much structure
when inflation ends.
This scenario is of course incompatible with the observation of a
Gaussian primordial power spectrum.
Therefore it did not occur in our universe.
One would like to know why.
Does its absence impose constraints on the allowed inflaton models
or on initial quantum states?
Is there some other unknown mechanism that suppresses the coupling between
quantum fluctuations and the classical inhomogeneous background,
perhaps related to the quantum-to-classical transition?
The systematic framework developed in our paper can be used in search of answers
to these questions.

\textit{Non-Gaussianities and the choice of canonical variables}.
As we have seen in Section \ref{sec-dS-fields},
if one imposes initial conditions at large but finite times,
the minimal-energy state would be different for canonical variables
that seem equally ``natural''.
In particular, the initial values for a given fluctuation, such as $\Delta(ff)= a^2\Delta(\delta\varphi\delta\varphi)$, would be different.
In this case,
different choices of canonical variables, as pointed out in 
\cite{grain2020canonical}, 
may give rise to different predictions.
This is true even when the canonical variables 
are related by only linear canonical transformations.
Therefore,
we expect the issue to be worse when one considers non-Gaussianities.
The cubic interaction terms will open up the possibility of choices
of canonical variables that are non-linearly related,
as can be the case when there are ``velocity-type'' interactions, such as $\delta\varphi'(\partial_{\mu}\delta\varphi\partial^{\mu}\delta\varphi)$
interaction terms that can appear in the $P(X,\varphi)$ theories
\cite{garriga1999perturbations}.
When deriving the form of cubic interactions,
the terms ignored by integration-by-parts may also complicate our choice of
variables.

Our effective framework offers a way of understanding the difference
arising from different variable choices.
In the case of quadratic systems, different choices of variables
give rise to different effective Hamiltonians.
The uncertainty principle then selects out different minimal-energy
configurations for the fluctuations,
which eventually grow and classicalize to seed primordial perturbations.
On the other hand, when non-Gaussian interactions are present,
different choices of canonical variables can define different notions
of a ``free'' Hamiltonian, around which one applies perturbation theory.
What is considered to be ``free'' in one Hamiltonian, viz those composed of quadratic terms, might contain cubic interactions in a different Hamiltonian.
This can happen 
when the two Hamiltonians are related by non-linear canonical transformation.
The ambiguity existing in the notion of a ``free''  Hamiltonian may be mitigated
by assigning initial values based on a minimization of 
the full effective Hamiltonian instead of just the free one.

\textit{Minimal-energy state and the uncertainty principle}.
If one wishes to minimize the full effective Hamiltonian,
which contains cubic fluctuations or higher,
one requires the uncertainty principle of higher-order fluctuations.
These will be inequalities similar to the celebrated
Heisenberg uncertainty principle for position and momentum,
but instead contain $\Delta^{(n)}$ with $n\geq 3$.
An interesting thing is:
If one again uses the Cauchy-Schwarz inequality,
the cubic moments would be on the RHS of
the analogue of \eqref{Weyl-UP}.
That is, cubic fluctuations would be bounded from above
by products of quartic and quadratic fluctuations,
as opposed to having a lower bound.
This conforms with intuition from classical analysis,
where systems with only $\varphi^3$-interaction has no global minima,
unless one includes a quartic interaction.
If no other inequalities are constraining cubic fluctuations,
it would seem to suggest that one should consider up to at least a fourth order
in fluctuations if one wishes to find a minimal-energy state via the
uncertainty principle.
Fortunately, the inequalities constraining cubic and quartic fluctuations
only need to be derived once
because they are model-independent.

\textit{Quantum corrections to the power spectrum}.
Lastly, we wish to comment on quantum corrections to second-order fluctuations,
which in our context are defined as correction terms that are,
for example, $O(\mu^2)$ or higher in
\eqref{phi3-quad-EoM}.
Taking \eqref{phi3-quad-EoM} for instance,
the leading order correction to the 2-pt correlators comes from
the cubic fluctuations. To obtain the $O(\mu^2)$ correction,
one should solve the ``tree-level''
solution of cubic fluctuations, namely to $O(\mu)$.
For higher order corrections, say to $O(\mu^4)$,
one ought to obtain the $O(\mu)$ solution for the quintic fluctuations,
giving us $O(\mu^2)$ solutions of the quartic fluctuations,
further giving us the $O(\mu^3)$ solutions of the cubic fluctuations
and hence finally giving us the $O(\mu^4)$ solutions to the quadratic fluctuations.
For arbitrarily higher order, one simply iterates this process more,
starting with the tree-level solution of the highest required order,
which is also the easiest one to solve.
See Section \ref{flat-solution-sec}.

While the strategy is clear, as we substitute higher-order fluctuations
into lower order ones, we encounter our version of ``loop'' integrals,
defined as integrals over internal momenta that cannot be eliminated with
delta-functions.
One would need to resort to certain regularization schemes and renormalize
the theory. This is difficult even at $O(\mu^2)$ order for a
$\varphi^3$-model on Minkowski,
as shown in Section \ref{flat-solution-sec}.
For generic orders in $\mu$,
we expect two main difficulties.
First, different orders of fluctuations would be mixed in the equations
of motion. 
Unless one can find a way to decouple equations of motion
for different orders of fluctuations,
we might not be able to obtain analytical expressions for the integrand
inside a loop-integral.
Second,
even if one obtains an analytical expression for the integrand,
the loop-integral is still not guaranteed to be expressible using
simple functions.
Renormalization would remain difficult and deserve future study.

\section*{Acknowledgements}
DD is supported by China Postdoctoral Science Foundation (Certificate No.2023M730704). YW is supported by the General Program of Science and Technology of Shanghai No. 21ZR1406700, and Shanghai Municipal Science and Technology Major Project (Grant No. 2019SHZDZX01). YW is grateful for the Hospitality of the Perimeter Institute during his visit, where the main part of this work is done. We thank Martin Bojowald for discussions and useful suggestions.

\clearpage
\appendix

\section{Poisson bracket for fluctuations of arbitrary order}
\label{BCH-PB-appendix}
Here, we give a prescription for finding arbitrary but specified fluctuations
$\Delta^{(n)}$.
The prescription is most
useful for brackets between $\Delta^{(n)}$
that are at least both of cubic order or higher.
(Quadratic-cubic brackets and quadratic-quadratic brackets can be
easily ``read out'' using the Leibniz rule
and intuition from classical Poisson brackets.)
The key technique used here is the Baker-Campbell-Haussdorf
(BCH)
formula.
We will therefore call the prescription derived in this section
the BCH-Poisson bracket prescription --- or BHC-PB prescription for short.

A naive adaptation of the quantum mechanical
derivation in
\cite{bojowald2006effective}
--- by turning the weighted sums of canonical variables
into weighted integrals
---
does not work well in practice.
This is because, with this naive implementation,
one obtains integrated versions of Poisson brackets over momenta,
as opposed to the form of \eqref{phiphiphipipipi}.
There is no good way to get rid of the integrals,
especially when there are constant terms like
the last line of the last equation in \eqref{phiphiphipipipi}.

So, we will use an alternative strategy
to obtain Poisson brackets without integrations over momenta.
The only price we pay is that we have to repeat the
prescription for each distinct bracket.
Fortunately,
once the Poisson brackets are derived for a given order,
they can be used for any physical system up to that order.
Additionally, there are not that many of them.
For example, for any system with arbitrary cubic interactions,
there are only eight non-trivial cubic-cubic Poisson brackets.
(For $\varphi^3$-interactions, 
only three of these cubic-cubic brackets are needed.)

\subsection{The derivation}
Let us now proceed with the derivation.
For simplicity of demonstration,
we derive the prescription for the
$\varphi_1\pi_2\pi_3-\pi_a\varphi_b\varphi_c$
bracket
in $1+1$ dimensions
\footnote{The results of this section can be generalized to
$3+1$ dimensions by replacing all $(2\pi)\delta(\dots)$
with $(2\pi)^3\delta^3(\dots)$.}
\begin{align*}
	\{\Delta(\varphi_1\pi_2\pi_3),\Delta(\pi_a\varphi_b\varphi_c)\}
\end{align*}
and then apply the prescription in a few examples.
(The subscripts $1,2,3,a,b,c$ label the momenta $k_1,k_2,k_3,k_a,k_b,k_c$.)

To utilize the BCH formula,
we further introduce the following notation
\begin{align*}
	A(\alpha)=&\Eval{e^{\alpha_1\delta\varphi_1+\alpha_2\delta\pi_2
	+\alpha_3\delta\pi_3}}\\
	B(\beta)=&\Eval{e^{\beta_a\delta\pi_a+\beta_b\delta\varphi_b
	+\beta_c\delta\varphi_c}}\\
	e^X=&e^{\alpha_1\hat{\varphi}_1+\alpha_2\hat{\pi}_2+\alpha_3\hat{\pi}_3}\\
	e^Y=&e^{\beta_a\hat{\pi}_a+\beta_b\hat{\varphi}_b+\beta_c\hat{\varphi}_c}
	\\
	\delta\varphi_i=&\hat{\varphi}_i-\Eval{\hat{\varphi}_i}
	\equiv \hat{\varphi}_i-\bar{\varphi}_i\,,\;
	\delta\pi_i=\hat{\pi}_i-\Eval{\hat{\pi}_i}
	\equiv \hat{\pi}_i-\bar{\pi}_i\,.
\end{align*}
To clarify, there is no sum over repeated indices,
and the objects $\delta\varphi_i$ and $\delta\pi_i$
here have nothing to do with the inhomogeneities in the main text.
Note also that our definition of 
$A(\alpha),B(\beta),X,Y$ is specific to the brackets
we wish to compute.
This is what we meant previously when we claimed that the prescription
needs to be repeated
(with a different $A(\alpha)$ and $B(\beta)$)
for each distinct bracket.

By construction,
the quantities $A(\alpha)$ and $B(\beta)$
contain respectively the two Weyl-ordered cubic moments
involved in the
bracket that we want to compute
\begin{align*}
	\alpha_1\alpha_2\alpha_3\Eval{\delta\varphi_1\delta\pi_2\delta\pi_3}_W
	\subset &\Eval{\sum_{n=0}\frac{1}{n!}(\alpha_1\delta\varphi_1
	+\alpha_2\delta\pi_2+\alpha_3\delta\pi_3)^n}=A(\alpha)\\
	\beta_a\beta_b\beta_c\Eval{\delta\pi_a\delta\varphi_b\delta\varphi_c}_W
	\subset &\Eval{\sum_{n=0}\frac{1}{n!}
	(\beta_a\delta\pi_a+\beta_b\delta\varphi_b+\beta_c\delta\varphi_c)^n}
	=B(\beta)\,,
\end{align*}
where the left-hand sides arise from picking out the $n=3$ 
terms of the sum.
These terms naturally form a symmetrically ordered sum,
and the overall coefficients are exactly $1$ when we write them into a 
Weyl-ordered notation $\Eval{\dots}_W$.
Thus the $O(\alpha_1\alpha_2\alpha_3\beta_a\beta_b\beta_c)$
term (with overall coefficient 1) in $\{A(\alpha),B(\beta)\}$
will be exactly the bracket we want
\begin{align*}
	\alpha_1\alpha_2\alpha_3\beta_a\beta_b\beta_c
	\{\Delta(\varphi_1\pi_2\pi_3),\Delta(\pi_a\varphi_b\varphi_c)\}
	\subset_1 \{A(\alpha),B(\beta)\}\,.
\end{align*}
where the subscript of $\subset_1$ indicates the overall coefficient is $1$.

Now we wish to expand the exponential products in the
bracket $\{A(\alpha),B(\beta)\}$
via the BCH formula.
We first split the bracket between the generating functions
$A(\alpha)$ and $B(\beta)$
into four parts
\begin{align*}
	\{A(\alpha),B(\beta)\}
	=&
	\{\Eval{e^Xe^{-\bar{X}}},\Eval{e^Ye^{-\bar{Y}}}\}\\
	=&\{\Eval{e^{X}},\Eval{e^{Y}}\}e^{-\bar{X}}e^{-\bar{Y}}\\
	+&\{\Eval{e^{X}},e^{-\bar{Y}}\}e^{-\bar{X}}\Eval{e^Y}\\
	+&\{e^{-\bar{X}},\Eval{e^{Y}}\}\Eval{e^X}e^{-\bar{Y}}\\
	+&\{e^{-\bar{X}},e^{-\bar{Y}}\}\Eval{e^X}\Eval{e^Y}\,,
\end{align*}
where the notation $\bar{X},\bar{Y}$ should be obvious
--- they are $\bar{X}=\alpha_1\bar{\varphi}_1+\alpha_2\bar{\pi}_2+\dots$
and analogously for $\bar{Y}$.

We then use the BCH formula
to calculate the part that contains averages of exponentials.
\begin{align*}
	\{\Eval{e^X},\Eval{e^Y}\}=&
	\frac{1}{i\hbar}\Eval{e^Xe^Y-e^Ye^X}\\
	=&\frac{1}{i\hbar}
	\Eval{e^{X+Y+[X,Y]/2}-e^{Y+X+[Y,X]/2}}\\
	=&\frac{1}{i\hbar}\Eval{e^{X+Y}(e^{i(-i[X,Y]/2)}-e^{-i(-i[X,Y]/2)})}\\
	=&\Eval{e^{X+Y}}
	\frac{2}{\hbar}\sin(-\frac{i}{2}[X,Y])\\
	\sin(-\frac{i}{2}[X,Y])=&
	\sin(\pi\hbar(\alpha_1\beta_a\delta_{1a}
	-\alpha_2\beta_b\delta_{2b}-\alpha_2\beta_c\delta_{2c}
	-\alpha_3\beta_b\delta_{3b}-\alpha_3\beta_c\delta_{3c}))\,.
\end{align*}
The notation of delta functions is $\delta_{1a}=\delta(k_1+k_a)$.
Every line but the last applies to any linear sum of canonical variables
as $X, Y$. The last line is specific to the $\varphi\pi\pi-\pi\varphi\varphi$
bracket.

The two mixed brackets
--- containing an average of exponentials 
and exponentials of averages --- 
can be calculated via \eqref{method-rules}.
The first of such brackets gives
\begin{align*}
	\{\Eval{e^{X}},e^{-\bar{Y}}\}
	=&\{\Eval{e^X},(-\Eval{Y})\}e^{-\bar{Y}}\\
	=&\frac{1}{i\hbar}\Eval{e^X[X,-Y]}e^{-\bar{Y}}\\
	=&\frac{1}{i\hbar}[X,-Y]\Eval{e^X}e^{-\bar{Y}}\,.
\end{align*}
In the second and third lines, we used the fact that $Y, X$ 
are linear in canonical variables.
Similarly, the second mixed bracket gives
\begin{align*}
	\{e^{-\bar{X}},\Eval{e^Y}\}
	=&e^{-\bar{X}}\{-\Eval{X},\Eval{e^Y}\}\\
	=&\frac{1}{i\hbar}e^{-\bar{X}}\Eval{[-X,Y]e^Y}\\
	=&\frac{1}{i\hbar}[-X,Y]\Eval{e^Y}e^{-\bar{X}}\,.
\end{align*}

Finally, the bracket between two exponentials of averages is most direct
\begin{align*}
	\{e^{-\bar{X}},e^{-\bar{Y}}\}=&
	e^{-\bar{X}}e^{-\bar{Y}}\{\Eval{X},\Eval{Y}\}
	=\frac{1}{i\hbar}[X,Y]e^{-\bar{X}}e^{-\bar{Y}}\,.
\end{align*}

Combining these brackets,
we arrive at a key result of the BCH-PB prescription
\begin{align}
	\{A(\alpha),B(\beta)\}=&
	\{\Eval{e^{X-\bar{X}}},\Eval{e^{Y-\bar{Y}}}\}\nonumber\\
	=&\Eval{e^{X-\bar{X}+Y-\bar{Y}}}\frac{2}{\hbar}
	\sin(\frac{1}{2i}[X,Y])\nonumber\\
	&-\frac{1}{i\hbar}[X,Y]A(\alpha)B(\beta)
	\label{BCH-PB-general}\\
	[X,Y]=&2\pi i\hbar(\alpha_1\beta_a\delta_{1a}\nonumber\\
	&-\alpha_2\beta_b\delta_{2b}-\alpha_2\beta_c\delta_{2c}\nonumber\\
	&-\alpha_3\beta_b\delta_{3b}-\alpha_3\beta_c\delta_{3c})\,.\nonumber
\end{align}

We stress that the formula \eqref{BCH-PB-general}
is still general, in the sense that as long as we adjust
how the canonical variables in $X$ and $Y$ are summed, we can use \eqref{BCH-PB-general} to compute any bracket.
The specific form of $[X, Y]$ here
is merely a consequence of our current intention of computing
a $\varphi\varphi\pi-\pi\varphi\varphi$ type bracket.
Let us now present some examples of applying the
BCH-PB prescription.

\subsection{Examples of using the generic BCH-PB formula}
Here we use \eqref{BCH-PB-general}
to calculate some examples.
For simplicity of bookkeeping orders of $\alpha$ and $\beta$
coefficients,
we will write $[X,Y]=2\pi i\hbar \sum O(\alpha\beta)$
to indicate that it is linear in the coefficient-product
$\alpha\times\beta$.

\textbf{Example 1}:
The first example is the original $\varphi\pi^2-\pi\varphi^2$
bracket that we hoped to compute. This will be a very non-trivial test 
as the bracket is one of the most difficult cubic-cubic brackets to compute.
It contains both $\pi$ and $\varphi$ in both slots of the bracket
and has a counterintuitive $O(\hbar^2)$ constant term.

Here $X, Y$ are the ones specified in the previous section
\begin{align*}
	A(\alpha)=&\Eval{e^{\alpha_1\delta\varphi_1+\alpha_2\delta\pi_2
	+\alpha_3\delta\pi_3}}\\
	B(\beta)=&\Eval{e^{\beta_a\delta\pi_a+\beta_b\delta\varphi_b
	+\beta_c\delta\varphi_c}}\\
	e^X=&e^{\alpha_1\hat{\varphi}_1+\alpha_2\hat{\pi}_2+\alpha_3\hat{\pi}_3}\\
	e^Y=&e^{\beta_a\hat{\pi}_a+\beta_b\hat{\varphi}_b+\beta_c\hat{\varphi}_c}\,.
\end{align*}

As explained before, we already know that the 
$O(\alpha_1\dots\alpha_3\beta_a\dots\beta_3)$
term in the direct expansion of the generating function in the bracket
$\{A(\alpha),B(\beta)\}$
is multiplied by the
$\varphi\varphi\pi-\pi\varphi\varphi$
bracket that we want (with an overall coefficient \(1\)).
So now to compute said bracket in terms of fluctuations $\Delta$,
we only need to find all the $O(\alpha_1\dots\alpha_3\beta_a\dots\beta_c)$
term on the RHS of \eqref{BCH-PB-general}.

Consider the first line of \eqref{BCH-PB-general}
\begin{align*}
	\sin(\frac{1}{2i}[X,Y])\Eval{e^{X-\bar{X}+Y-\bar{Y}}}
	=&\left[ 
	\pi\hbar\sum O(\alpha\beta)
	-\frac{1}{3!}(\pi\hbar\sum O(\alpha\beta))^3+\dots\right]
	\Eval{e^{X-\bar{X}+Y-\bar{Y}}}\\
	=&[\pi\hbar\sum O(\alpha\beta)
	\frac{1}{4!}\Eval{(X-\bar{X}+Y-\bar{Y})^4}+\dots]\\
	&-\frac{1}{3!}
	[\pi^3\hbar^3(\sum O(\alpha\beta))^3\times 1+\dots]\,.
\end{align*}
We immediately see the constant term arise
in the last line.
However, not every term of the last line should be included in the
$\varphi\pi^2-\pi\varphi^2$ bracket,
but only the ones with different indices,
as we are looking for $O(\alpha_1\dots\alpha_3\beta_a\dots\beta_c)$
terms.
We will get back to this in a brief moment.
Let us first consider the
$1/4 !$-line above.

For the $1/4!$-line, not all the terms in the expectation value fit the 
indices-requirement
--- it depends on which term in $\sum O(\alpha\beta)$
multiplies with the expectation value.
For example, for
$\alpha_1\beta_a\delta_{1a}\subset \sum O(\alpha\beta)$,
we would obtain one of the desired fourth-order fluctuations
\begin{align*}
	\sum O(\alpha\beta)\frac{1}{4!}
	\Eval{(X-\bar{X}+Y-\bar{Y})^4}\supset_1&\frac{1}{4!}
	\alpha_1\beta_a\delta_{1a}\Eval{[(\alpha_2\delta\pi_2+\alpha_3\delta\pi_3\\
	&+\beta_b\delta\varphi_b+\beta_c\delta\varphi_c)+
	(\alpha_1\delta\varphi_1+\beta_a\delta\pi_a)]^4}\\
	=&\alpha_1\beta_a\delta_{1a}\frac{1}{4!}
	\Eval{(\alpha_2\delta\pi_2+\alpha_3\delta\pi_3
	+\beta_b\delta\varphi_b+\beta_c\delta\varphi_c)^4}+\dots\\
	\supset_1&\alpha_1\beta_a\delta_{1a}
	\alpha_2\alpha_3\beta_b\beta_c
	\Eval{(\delta\pi_2\delta\pi_3\delta\varphi_b\delta\varphi_c)}_W\,,
\end{align*}
where all the ``$\supset_1$''\,s imply the relation
that the left-hand side contains the right-hand side 
with an overall coefficient of $1$.
Repeating this logic,
the terms in the $1/4!$-line relevant to our bracket are
\begin{align*}
	&\frac{2}{\hbar}[\pi\hbar\sum O(\alpha\beta)
	\frac{1}{4!}\Eval{(X-\bar{X}+Y-\bar{Y})^4}+\dots]\\
	&\supset
	2\pi\alpha_1\alpha_2\alpha_3\beta_a\beta_b\beta_c\times\\
	&[\delta_{1a}\Delta(\pi_2\pi_3\varphi_a\varphi_b)
	-\delta_{2b}\Delta(\varphi_1\pi_3\pi_a\varphi_c)
	-\delta_{2c}\Delta(\varphi_1\pi_3\pi_a\varphi_b)
	-(2\leftrightarrow 3)]
\end{align*}
The above contains all the $\Delta^{(4)}$ terms in the
$\varphi\pi^2-\pi\varphi^2$ bracket
with the correct overall coefficients
(modulo the $\alpha_1\dots\beta_c$ factor).

Now consider the $(\sum O(\alpha\beta))^3$ line.
These terms will help us find the constant terms.
Again, we do not need all the terms in this line.
The sum contains $5$ terms 
and once cubed we only need the resulting product terms that contain all distinct indices.
The $\alpha_1\beta_a$ 
factor must appear and always come from 
$\alpha_1\beta_a\delta_{1a}$,
whereas the $\alpha_2\alpha_3\beta_b\beta_c$ can come in different ways.
There are $3\times2\times1$ ways of getting
$\alpha_1\beta_a\delta_{1a}\alpha_2\beta_b(-\delta_{2b})
\alpha_3\beta_c(-\delta_{3c})$
and $\alpha_1\beta_a\alpha_{1a}\alpha_2\beta_c(-\delta_{2c})
\alpha_3\beta_b(-\delta_{3b})$ respectively.
Namely, we need 12 terms in $(\sum O(\alpha\beta)^3)$.
Specifically, the relevant terms in the $O(\sum O(\alpha\beta)^3)$
line are
\begin{equation}
	-\frac{2}{\hbar}\frac{1}{3!}
	[\pi^3\hbar^3(\sum O(\alpha\beta))^3\times 1+\dots]\supset
	-2\pi^3\hbar^2\alpha_1\dots\alpha_3\beta_a\dots\beta_c
	(\delta_{1a}\delta_{2b}\delta_{3c}+
	\delta_{1a}\delta_{2c}\delta_{3b})\,.
	\label{}
\end{equation}
This concludes the search for relevant terms in the first line of
\eqref{BCH-PB-general}.

Now consider the second line of \eqref{BCH-PB-general},
which gives us the $\Delta^{(2)}\Delta^{(2)}$ terms.
Specifically, the relevant terms are
\begin{align*}
	-\frac{1}{i\hbar}[X,Y]
	A(\alpha)B(\beta)&\supset_1
	-\frac{2\pi}{2!}\sum O(\alpha\beta)
	\Eval{(\alpha_1\varphi_1+\alpha_2\pi_2+\alpha_3\pi_3)^2}
	\Eval{(\beta_a\pi_a+\beta_b\varphi_b+\beta_c\varphi_c)^2}\\
	=&-2\pi\alpha_1\alpha_2\alpha_3\beta_a\beta_b\beta_c\\
	&\times[ \delta_{1a}\Delta(\pi_2\pi_3)\Delta(\varphi_b\varphi_c)
	-(\delta_{2b}\Delta(\varphi_1\pi_3)\Delta(\pi_a\varphi_c)
	+\delta_{2c}\Delta(\varphi_1\pi_3)\Delta(\pi_a\varphi_b)\\
	&+(2\leftrightarrow 3))]\,.
\end{align*}

Now we have collected all the possible 
$\alpha_1\alpha_2\alpha_3\beta_a\beta_b\beta_c$ terms of the RHS of
\eqref{BCH-PB-general}.
Matching this with the LHS
--- the $\alpha_1\alpha_2\alpha_3\beta_a\beta_c\beta_b
\{\Delta(\varphi_1\pi_2\pi_c),\Delta(\pi_a\varphi_b\varphi_c)\}$
term ---
we obtain the full result for the bracket
\begin{align*}
	&\{\Delta(\varphi_1\pi_2\pi_3),\Delta(\pi_a\varphi_b\varphi_c)\}\\	
	=&2\pi[\delta_{1a}\Delta(\pi_2\pi_3\varphi_a\varphi_b)
	-\delta_{2b}\Delta(\varphi_1\pi_3\pi_a\varphi_c)
	-\delta_{2c}\Delta(\varphi_1\pi_3\pi_a\varphi_b)
	-(2\leftrightarrow 3)]\\
	&-2\pi^3\hbar^2
	(\delta_{1a}\delta_{2b}\delta_{3c}+
	\delta_{1a}\delta_{2c}\delta_{3b})\\
	&-2\pi\times[ \delta_{1a}\Delta(\pi_2\pi_3)\Delta(\varphi_b\varphi_c)\\
	&-\delta_{2b}\Delta(\varphi_1\pi_3)\Delta(\pi_a\varphi_c)
	-\delta_{2c}\Delta(\varphi_1\pi_3)\Delta(\pi_a\varphi_b)
	-(2\leftrightarrow 3)]
\end{align*}
One can check that this reduces to the quantum mechanical result
derived in \cite{bojowald2006effective} 
if we set all $2\pi\delta_{ij}$ to one.

\textbf{Example 2}:
The second example is the $\varphi^3-\pi^3$ bracket,
but we will only look for 
the constant (and counterintuitive) term
of the last equation in \eqref{phiphiphipipipi}
to show how it arises.

The direct expansion of the exponentials in $\{A(\alpha),B(\beta)\}$
contains the desired bracket between the fluctuations
\begin{equation}
	\{A(\alpha),B(\beta)\}\supset_1
	\alpha_1\dots\alpha_3\beta_a\dots\beta_c
	\{\Eval{\delta\varphi_1\delta\varphi_2\delta\varphi_3}_W,
	\Eval{\delta\pi_a\delta\pi_b\delta\pi_c}_W\}\,.
	\label{}
\end{equation}
where ``$\supset_1$'' means ``contains up to an overall coefficient $1$''.

The constant term of the bracket originates from the
$\sin(\dots)$ term in
\eqref{BCH-PB-general}.
The relevant $O(\alpha_1\dots\alpha_3\beta_a\dots\beta_c)$ terms are
\begin{align*}
	\frac{2}{\hbar}\sin[\pi\hbar\sum O(\alpha\beta)]
	\supset_1&
	-\frac{2}{\hbar}(\pi\hbar)^3\frac{1}{3!}\\
	&\times[\alpha_1\beta_a\delta_{1a}+\alpha_1\beta_b\delta_{1b}
	+\alpha_1\beta_c\delta_{1c}\\
	&+\alpha_2\beta_a\delta_{2a}+\alpha_2\beta_b\delta_{2b}
	+\alpha_2\beta_c\delta_{2c}\\
	&+\alpha_3\beta_a\delta_{3a}
	+\alpha_3\beta_b\delta_{3b}+
	\alpha_3\beta_c\delta_{3c}]^3\\
	\supset_1&
	-\frac{2}{\hbar}(\pi\hbar)^3\frac{1}{3!}\\
	&\times[3!\alpha_1\beta_a\delta_{1a}\alpha_2\beta_b\delta_{2b}
	\alpha_3\beta_c\delta_{3c}\\
	&+(\text{five more possible pairings between } 
	1,2,3 \leftrightarrow a,b,c)]\,.
\end{align*}
Generalizing this to $3+1$ dimensions,
via $(2\pi)\delta(\dots)\rightarrow (2\pi)^3\delta^3(\dots)$,
we see that this constant term matches the one in
\eqref{phiphiphipipipi} when $\hbar=1$.

\textbf{Example 3}.
This example shows how ``mirror-brackets'' are computed, such as $\{\Delta(\varphi\varphi\pi),\Delta(\varphi\varphi\pi)\}$.
These brackets vanish trivially in the quantum mechanical case
due to the lack of Fourier-space labels.
In the field case, the
$\{\Delta(\varphi_1\varphi_2\pi_3),\Delta(\varphi_a\varphi_b\pi_c)\}$
bracket is one of the two mirror brackets of the cubic-cubic type.
(The other one is the $\varphi\pi\pi-\varphi\pi\pi$ type bracket.)

The linear sums $X$ and $Y$ in this case are
\begin{align*}
	X=&\alpha_1\varphi_1+\alpha_2\varphi_2+\alpha_3\pi_3\,,\;
	Y=\beta_a\varphi_b+\beta_b\varphi_b+\beta_c\pi_c\,,\;\\
	[X,Y]=& 2\pi i\hbar(\alpha_1\beta_c\delta_{1c}+\alpha_2\beta_c\delta_{2c}
	-\alpha_3\beta_a\delta_{3a}-\alpha_3\beta_b\delta_{3b})\,.
\end{align*}

The bracket is contained in
\begin{equation}
	\{A(\alpha),B(\beta)\}\supset_1
	\alpha_1\alpha_2\alpha_3\beta_a\beta_b\beta_c
	\{\Eval{\delta\varphi_1\delta\varphi_2\delta\pi_3}_W,
	\Eval{\delta\varphi_a\delta\varphi_b\pi_c}_W\}.
	\label{}
\end{equation}
We therefore need to look for 
$O(\alpha_1\alpha_2\alpha_3\beta_a\beta_b\beta_c)$ terms.

The $O(\alpha_1\alpha_2\alpha_3\beta_a\beta_b\beta_c)$
terms in the first line of the BCH-PB formula \eqref{BCH-PB-general}
are
\begin{align*}
	\Eval{e^{X-\bar{X}+Y-\bar{Y}}}\frac{2}{\hbar}\sin
	\left( \frac{1}{2i}[X,Y] \right)
	=&\Eval{e^{X-\bar{X}+Y-\bar{Y}}}\frac{2}{\hbar}\\
	&\times\frac{2}{\hbar}[\pi\hbar\sum O(\alpha\beta)
	-\frac{1}{3!}(\pi\hbar)^3\sum O(\alpha\beta)^3+O(\alpha^5\beta^5)]\,.
\end{align*}
It turns out that the $\sum O(\alpha^3\beta^3)$
terms above
excludes
$O(\alpha_1\alpha_2\alpha_3\beta_a\beta_b\beta_c)$
terms
because in the sum the product $\alpha_1\alpha_2$ always comes with
a factor of $\beta_c^2$,
which is not the desired power of $\beta_c$.
This is ultimately a consequence of the specific form of
$[X,Y]$ for this bracket.
So we only have
\begin{align*}
	\Eval{e^{X-\bar{X}+Y-\bar{Y}}}\frac{2}{\hbar}\sin
	\left( \frac{1}{2i}[X,Y] \right)
	\supset_1&\Eval{e^{X-\bar{X}+Y-\bar{Y}}}\frac{2}{\hbar}\\
	&\times\frac{2}{\hbar}[\pi\hbar\sum O(\alpha\beta)]\,.
\end{align*}
Here the $\sum O(\alpha\beta)\propto [X,Y]$,
so we simply need to pick the remaining $O(\alpha^2\beta^2)$
terms from $\Eval{e^{X-\bar{X}+Y-\bar{Y}}}$
(the $n=4$ term to be specific)
for $O(\alpha_1\alpha_2\alpha_3\beta_a\beta_b\beta_c)$.
The terms in the exponential are naturally completely symmetrized and readily give
$\Delta^{(4)}$'s.

In the end, we obtain the relevant terms
\begin{align*}
	\Eval{e^{X-\bar{X}+Y-\bar{Y}}}\frac{2}{\hbar}\sin
	\left( \frac{1}{2i}[X,Y] \right)
	\supset_1
	&2\pi \alpha_1\alpha_2\alpha_3\beta_a\beta_b\beta_c 
	[\Delta(\varphi_2\pi_3\varphi_a\varphi_b)\delta_{1c}
	+\Delta(\varphi_1\pi_3\varphi_a\varphi_b)\delta_{2c}\\
	&-\Delta(\varphi_1\varphi_2\varphi_b\pi_c)\delta_{3a}
	-\Delta(\varphi_1\varphi_2\varphi_a\pi_c)\delta_{3b}]\,.
\end{align*}

The remaining $O(\alpha_1\dots\beta_c)$ terms are in the second line of
\eqref{BCH-PB-general}.
The factor $[X,Y]$ gives us $O(\alpha\beta)$.
We thus need $O(\alpha^2)\subset A(\alpha)$
and $O(\beta^2)\subset A(\beta)$.
Picking out the right terms is straightforward.
One eventually finds the remaining relevant terms to be
\begin{align*}
	-\frac{1}{i\hbar}[X,Y]A(\alpha)B(\beta)\supset_1
	-\alpha_1\alpha_2\alpha_3\beta_a\beta_b\beta_c&(2\pi)
	[\Delta(\varphi_2\pi_3)\Delta(\varphi_a\varphi_b)\delta_{1c}
	+\Delta(\varphi_1\pi_3)\Delta(\varphi_a\varphi_b)\delta_{2c}\\
	&-\Delta(\varphi_1\varphi_2)\Delta(\varphi_b\pi_c)\delta_{3a}
	-\Delta(\varphi_1\varphi_2)\Delta(\varphi_a\varphi_c)\delta_{3b}]\,.
\end{align*}
Combining the relevant terms in
$\Eval{e^{X-\bar{X}+Y-\bar{Y}}}\frac{2}{\hbar}\sin
	\left( \frac{1}{2i}[X,Y] \right)$
 and $-\frac{1}{i\hbar}[X,Y]A(\alpha)B(\beta)$,
 we obtain the
 results for
 $\{\Delta(\varphi_1\varphi_2\pi_3),\Delta(\varphi_a\varphi_b\pi_c)\}$.
 
\clearpage
\bibliographystyle{JHEP}
 \bibliography{refs-Qcos-draft-combined.bib}

\providecommand{\href}[2]{#2}\begingroup\raggedright\begin{thebibliography}{10}

\bibitem{bojowald2006effective}
M.~Bojowald and A.~Skirzewski, \emph{{Effective equations of motion for quantum systems}}, \href{https://doi.org/10.1142/S0129055X06002772}{\emph{Rev. Math. Phys.} {\bfseries 18} (2006) 713} [\href{https://arxiv.org/abs/math-ph/0511043}{{\ttfamily math-ph/0511043}}].

\bibitem{planck2018const}
{\scshape Planck} collaboration, \emph{{Planck 2018 results. VI. Cosmological parameters}}, \href{https://doi.org/10.1051/0004-6361/201833910}{\emph{Astron. Astrophys.} {\bfseries 641} (2020) A6} [\href{https://arxiv.org/abs/1807.06209}{{\ttfamily 1807.06209}}].

\bibitem{guth1981inflationary}
A.H.~Guth, \emph{{The Inflationary Universe: A Possible Solution to the Horizon and Flatness Problems}}, \href{https://doi.org/10.1103/PhysRevD.23.347}{\emph{Phys. Rev. D} {\bfseries 23} (1981) 347}.

\bibitem{mukhanov1992theory}
V.F.~Mukhanov, H.A.~Feldman and R.H.~Brandenberger, \emph{{Theory of cosmological perturbations. Part 1. Classical perturbations. Part 2. Quantum theory of perturbations. Part 3. Extensions}}, \href{https://doi.org/10.1016/0370-1573(92)90044-Z}{\emph{Phys. Rept.} {\bfseries 215} (1992) 203}.

\bibitem{kofman1994reheating}
L.~Kofman, A.D.~Linde and A.A.~Starobinsky, \emph{{Reheating after inflation}}, \href{https://doi.org/10.1103/PhysRevLett.73.3195}{\emph{Phys. Rev. Lett.} {\bfseries 73} (1994) 3195} [\href{https://arxiv.org/abs/hep-th/9405187}{{\ttfamily hep-th/9405187}}].

\bibitem{krotov2011infrared}
D.~Krotov and A.M.~Polyakov, \emph{{Infrared Sensitivity of Unstable Vacua}}, \href{https://doi.org/10.1016/j.nuclphysb.2011.03.025}{\emph{Nucl. Phys. B} {\bfseries 849} (2011) 410} [\href{https://arxiv.org/abs/1012.2107}{{\ttfamily 1012.2107}}].

\bibitem{polyakov2012infrared}
A.M.~Polyakov, \emph{{Infrared instability of the de Sitter space}},  \href{https://arxiv.org/abs/1209.4135}{{\ttfamily 1209.4135}}.

\bibitem{danielsson2019quantum}
U.~Danielsson, \emph{{The quantum swampland}}, \href{https://doi.org/10.1007/JHEP04(2019)095}{\emph{JHEP} {\bfseries 04} (2019) 095} [\href{https://arxiv.org/abs/1809.04512}{{\ttfamily 1809.04512}}].

\bibitem{jona1964relativistic}
G.~Jona-Lasinio, \emph{{Relativistic field theories with symmetry breaking solutions}}, \href{https://doi.org/10.1007/BF02750573}{\emph{Nuovo Cim.} {\bfseries 34} (1964) 1790}.

\bibitem{coleman1973radiative}
S.R.~Coleman and E.J.~Weinberg, \emph{{Radiative Corrections as the Origin of Spontaneous Symmetry Breaking}}, \href{https://doi.org/10.1103/PhysRevD.7.1888}{\emph{Phys. Rev. D} {\bfseries 7} (1973) 1888}.

\bibitem{jackiw1979time}
R.~Jackiw and A.~Kerman, \emph{{Time Dependent Variational Principle and the Effective Action}}, \href{https://doi.org/10.1016/0375-9601(79)90151-8}{\emph{Phys. Lett. A} {\bfseries 71} (1979) 158}.

\bibitem{mulryne2010moment}
D.J.~Mulryne, D.~Seery and D.~Wesley, \emph{{Moment transport equations for non-Gaussianity}}, \href{https://doi.org/10.1088/1475-7516/2010/01/024}{\emph{JCAP} {\bfseries 01} (2010) 024} [\href{https://arxiv.org/abs/0909.2256}{{\ttfamily 0909.2256}}].

\bibitem{mulryne2011moment}
D.J.~Mulryne, D.~Seery and D.~Wesley, \emph{{Moment transport equations for the primordial curvature perturbation}}, \href{https://doi.org/10.1088/1475-7516/2011/04/030}{\emph{JCAP} {\bfseries 04} (2011) 030} [\href{https://arxiv.org/abs/1008.3159}{{\ttfamily 1008.3159}}].

\bibitem{vachaspati2018classical}
T.~Vachaspati and G.~Zahariade, \emph{{Classical-quantum correspondence and backreaction}}, \href{https://doi.org/10.1103/PhysRevD.98.065002}{\emph{Phys. Rev. D} {\bfseries 98} (2018) 065002} [\href{https://arxiv.org/abs/1806.05196}{{\ttfamily 1806.05196}}].

\bibitem{vachaspati2019classical}
T.~Vachaspati and G.~Zahariade, \emph{{Classical-Quantum Correspondence for Fields}}, \href{https://doi.org/10.1088/1475-7516/2019/09/015}{\emph{JCAP} {\bfseries 09} (2019) 015} [\href{https://arxiv.org/abs/1807.10282}{{\ttfamily 1807.10282}}].

\bibitem{bojowald2021canonical}
M.~Bojowald and D.~Ding, \emph{{Canonical description of cosmological backreaction}}, \href{https://doi.org/10.1088/1475-7516/2021/03/083}{\emph{JCAP} {\bfseries 03} (2021) 083} [\href{https://arxiv.org/abs/2011.03018}{{\ttfamily 2011.03018}}].

\bibitem{bojowald2021multi}
M.~Bojowald, S.~Brahma, S.~Crowe, D.~Ding and J.~McCracken, \emph{{Multi-field inflation from single-field models}}, \href{https://doi.org/10.1088/1475-7516/2021/08/047}{\emph{JCAP} {\bfseries 08} (2021) 047} [\href{https://arxiv.org/abs/2011.02843}{{\ttfamily 2011.02843}}].

\bibitem{bojowald2014canonical}
M.~Bojowald and S.~Brahma, \emph{{Canonical derivation of effective potentials}},  \href{https://arxiv.org/abs/1411.3636}{{\ttfamily 1411.3636}}.

\bibitem{brizuela2019moment}
D.~Brizuela and U.~Muniain, \emph{{A moment approach to compute quantum-gravity effects in the primordial universe}}, \href{https://doi.org/10.1088/1475-7516/2019/04/016}{\emph{JCAP} {\bfseries 04} (2019) 016} [\href{https://arxiv.org/abs/1901.08391}{{\ttfamily 1901.08391}}].

\bibitem{brizuela2021mode}
D.~Brizuela and I.~de~Leon, \emph{{Mode coupling on a geometrodynamical quantization of an inflationary universe}}, \href{https://doi.org/10.1088/1475-7516/2021/07/054}{\emph{JCAP} {\bfseries 07} (2021) 054} [\href{https://arxiv.org/abs/2105.03138}{{\ttfamily 2105.03138}}].

\bibitem{brizuela2022quantum}
D.~Brizuela and T.~Pawlowski, \emph{{Quantum fluctuations and semiclassicality in an inflaton-driven evolution}}, \href{https://doi.org/10.1088/1475-7516/2022/10/080}{\emph{JCAP} {\bfseries 10} (2022) 080} [\href{https://arxiv.org/abs/2107.04342}{{\ttfamily 2107.04342}}].

\bibitem{grain2020canonical}
J.~Grain and V.~Vennin, \emph{{Canonical transformations and squeezing formalism in cosmology}}, \href{https://doi.org/10.1088/1475-7516/2020/02/022}{\emph{JCAP} {\bfseries 02} (2020) 022} [\href{https://arxiv.org/abs/1910.01916}{{\ttfamily 1910.01916}}].

\bibitem{arickx1986gaussian}
F.~Arickx, J.~Broeckhove, W.~Coene and P.~Van~Leuven, \emph{Gaussian wave-packet dynamics}, \href{https://doi.org/10.1002/qua.560300741}{\emph{International Journal of Quantum Chemistry} {\bfseries 30} (1986) 471}.

\bibitem{prezhdo2006quantized}
O.V.~Prezhdo, \emph{Quantized hamilton dynamics}, \href{https://doi.org/10.1007/s00214-005-0032-x}{\emph{Theoretical Chemistry Accounts} {\bfseries 116} (2006) 206}.

\bibitem{mukhopadhyay2019rolling}
M.~Mukhopadhyay and T.~Vachaspati, \emph{{Rolling classical scalar field in a linear potential coupled to a quantum field}}, \href{https://doi.org/10.1103/PhysRevD.100.096018}{\emph{Phys. Rev. D} {\bfseries 100} (2019) 096018} [\href{https://arxiv.org/abs/1907.03762}{{\ttfamily 1907.03762}}].

\bibitem{vachaspati2019HR}
T.~Vachaspati and G.~Zahariade, \emph{{Classical-Quantum Correspondence and Hawking Radiation}}, \href{https://doi.org/10.1088/1475-7516/2019/04/013}{\emph{JCAP} {\bfseries 04} (2019) 013} [\href{https://arxiv.org/abs/1803.08919}{{\ttfamily 1803.08919}}].

\bibitem{baytacs2019effective}
B.~Bayta\c{s}, M.~Bojowald and S.~Crowe, \emph{{Effective potentials from semiclassical truncations}}, \href{https://doi.org/10.1103/PhysRevA.99.042114}{\emph{Phys. Rev. A} {\bfseries 99} (2019) 042114} [\href{https://arxiv.org/abs/1811.00505}{{\ttfamily 1811.00505}}].

\bibitem{baytacs2020faithful}
B.~Baytas, M.~Bojowald and S.~Crowe, \emph{{Faithful realizations of semiclassical truncations}}, \href{https://doi.org/10.1016/j.aop.2020.168247}{\emph{Annals Phys.} {\bfseries 420} (2020) 168247} [\href{https://arxiv.org/abs/1810.12127}{{\ttfamily 1810.12127}}].

\bibitem{cametti1999comparison}
F.~Cametti, G.~Jona-Lasinio, C.~Presilla and F.~Toninelli, \emph{{Comparison between quantum and classical dynamics in the effective action formalism}},  in \emph{{International School of Physics, 'Enrico Fermi', Course 143: New Directions in Quantum Chaos}}, pp.~431--448, 7, 1999, \href{https://doi.org/10.3254/978-1-61499-228-8-431}{DOI} [\href{https://arxiv.org/abs/quant-ph/9910065}{{\ttfamily quant-ph/9910065}}].

\bibitem{chen2010primordial}
X.~Chen, \emph{{Primordial Non-Gaussianities from Inflation Models}}, \href{https://doi.org/10.1155/2010/638979}{\emph{Adv. Astron.} {\bfseries 2010} (2010) 638979} [\href{https://arxiv.org/abs/1002.1416}{{\ttfamily 1002.1416}}].

\bibitem{martin2001trans}
J.~Martin and R.H.~Brandenberger, \emph{{The TransPlanckian problem of inflationary cosmology}}, \href{https://doi.org/10.1103/PhysRevD.63.123501}{\emph{Phys. Rev. D} {\bfseries 63} (2001) 123501} [\href{https://arxiv.org/abs/hep-th/0005209}{{\ttfamily hep-th/0005209}}].

\bibitem{weinberg2005quantum}
S.~Weinberg, \emph{{Quantum contributions to cosmological correlations}}, \href{https://doi.org/10.1103/PhysRevD.72.043514}{\emph{Phys. Rev. D} {\bfseries 72} (2005) 043514} [\href{https://arxiv.org/abs/hep-th/0506236}{{\ttfamily hep-th/0506236}}].

\bibitem{fulling1973nonuniqueness}
S.A.~Fulling, \emph{{Nonuniqueness of canonical field quantization in Riemannian space-time}}, \href{https://doi.org/10.1103/PhysRevD.7.2850}{\emph{Phys. Rev. D} {\bfseries 7} (1973) 2850}.

\bibitem{maldacena2003non}
J.M.~Maldacena, \emph{{Non-Gaussian features of primordial fluctuations in single field inflationary models}}, \href{https://doi.org/10.1088/1126-6708/2003/05/013}{\emph{JHEP} {\bfseries 05} (2003) 013} [\href{https://arxiv.org/abs/astro-ph/0210603}{{\ttfamily astro-ph/0210603}}].

\bibitem{danielsson2002note}
U.H.~Danielsson, \emph{{A Note on inflation and transPlanckian physics}}, \href{https://doi.org/10.1103/PhysRevD.66.023511}{\emph{Phys. Rev. D} {\bfseries 66} (2002) 023511} [\href{https://arxiv.org/abs/hep-th/0203198}{{\ttfamily hep-th/0203198}}].

\bibitem{copi2010large}
C.J.~Copi, D.~Huterer, D.J.~Schwarz and G.D.~Starkman, \emph{{Large angle anomalies in the CMB}}, \href{https://doi.org/10.1155/2010/847541}{\emph{Adv. Astron.} {\bfseries 2010} (2010) 847541} [\href{https://arxiv.org/abs/1004.5602}{{\ttfamily 1004.5602}}].

\bibitem{garriga1999perturbations}
J.~Garriga and V.F.~Mukhanov, \emph{{Perturbations in k-inflation}}, \href{https://doi.org/10.1016/S0370-2693(99)00602-4}{\emph{Phys. Lett. B} {\bfseries 458} (1999) 219} [\href{https://arxiv.org/abs/hep-th/9904176}{{\ttfamily hep-th/9904176}}].

\end{thebibliography}\endgroup
\end{document}